\documentclass[a4paper,fleqn]{cas-dc}

\usepackage[numbers]{natbib}
\usepackage{amsmath,amssymb,amsfonts}
\usepackage{algorithmic}
\usepackage{graphicx}
\usepackage{enumitem}
\usepackage{booktabs}
\usepackage{textcomp}
\usepackage{multirow}
\usepackage{array}

\def\tsc#1{\csdef{#1}{\textsc{\lowercase{#1}}\xspace}}
\tsc{WGM}
\tsc{QE}
\tsc{EP}
\tsc{PMS}
\tsc{BEC}
\tsc{DE}
\begin{document}
\let\WriteBookmarks\relax
\def\floatpagepagefraction{1}
\def\textpagefraction{.001}
\shorttitle{Machine Learning for Software Engineering: A Systematic Mapping}
\shortauthors{Saad Shafiq et al.}

\title [mode = title]{Machine Learning for Software Engineering: A Systematic Mapping}                      
\tnotemark[1]

\tnotetext[1]{The research reported in this paper has been partly funded by the Linz Institute of Technology, and the Austrian Ministry for Transport, Innovation and Technology, the Federal Ministry for Digital and Economic Affairs, and the Province of Upper Austria in the frame of the COMET - Competence Centers for Excellent Technologies - program managed by FFG.}

%\tnotetext[2]{The second title footnote which is a longer text matter to fill through the whole text width and overflow into another line in the footnotes area of the first page.}

\hyphenation{every-where spec-ifically sch-eme table}

\author[1]{Saad Shafiq}[type=editor,
%                        auid=000,bioid=1,
%                        prefix=Sir,
%                        role=Researcher,
                        orcid=0000-0002-5901-1420
]
% \cormark[1]
\fnmark[1]
\ead{saad.shafiq@jku.at}
%\ead[url]{www.cvr.cc, cvr@sayahna.org}

\credit{Data curation, Writing - Original draft preparation}

\address[1]{Johannes Kepler University, Linz, Austria}

\author[1,2]{Atif Mashkoor}[type=editor,
%                        auid=000,bioid=1,
%                        prefix=Sir,
%                        role=Researcher,
                        orcid=0000-0003-1210-5953
]
% \cormark[1]
\fnmark[2]
\ead{atif.mashkoor@jku.at}

\credit{Writing, Conceptualization of this study, Methodology}

\address[2]{Software Competence Center Hagenberg GmbH, Hagenberg, Austria}

\author[1]{Christoph Mayr-Dorn}[type=editor,
%   role=Researcher,
%   prefix=Sir,
  orcid=0000-0001-9791-6442
]
% \cormark[1]
\fnmark[3]
\ead{christoph.mayr-dorn@jku.at}

\credit{Writing, Conceptualization of this study, Methodology}

\author[1]{Alexander Egyed}[type=editor,
%   role=Researcher,
%   prefix=Sir,
  orcid=0000-0003-3128-5427
]
% \cormark[1]
\fnmark[4]
\ead{alexander.egyed@jku.at}

\credit{Writing, Conceptualization of this study, Methodology}

%\address[2]{Johannes Kepler University, 4040 Linz, Austria}

%\cortext[cor2]{Principal corresponding author}
%\fntext[fn1]{This is the first author footnote. but is common to third author as well.}
%\fntext[fn2]{Another author footnote, this is a very long footnote and it should be a really long footnote. But this footnote is not yet   sufficiently long enough to make two lines of footnote text.}

%\nonumnote{This note has no numbers. In this work we demonstrate $a_b$ the formation Y\_1 of a new type of polariton on the interface  between a cuprous oxide slab and a polystyrene micro-sphere placed on the slab.
%   }

\begin{abstract}
\noindent\textit{Context:} The software development industry is rapidly adopting machine learning for transitioning modern day software systems towards highly intelligent and self-learning systems. However, the full potential of machine learning for improving the software engineering life cycle itself is yet to be discovered, i.e., up to what extent machine learning can help reducing the effort\slash{}complexity of software engineering and improving the quality of resulting software systems. To date, no comprehensive study exists that explores the current state-of-the-art on the adoption of machine learning across software engineering life cycle stages.

\noindent\textit{Objective:} This article addresses the aforementioned problem and aims to present a state-of-the-art on the growing number of uses of machine learning in software engineering.

\noindent\textit{Method:} We conduct a systematic mapping study on applications of machine learning to software engineering following the standard guidelines and principles of empirical software engineering.

\noindent\textit{Results:} This study introduces a machine learning for software engineering (MLSE) taxonomy classifying the state-of-the-art machine learning techniques according to their applicability to various software engineering life cycle stages. Overall, 227 articles were rigorously selected and analyzed as a result of this study.

\noindent\textit{Conclusion:} From the selected articles, we explore a variety of aspects that should be helpful to academics and practitioners alike in understanding the potential of adopting machine learning techniques during software engineering projects.

%\noindent\texttt{\textbackslash begin{abstract}} \dots 
%\texttt{\textbackslash end{abstract}} and
%\verb+\begin{keyword}+ \verb+...+ \verb+\end{keyword}+ 
%which
%contain the abstract and keywords respectively. 
%
%\noindent Each keyword shall be separated by a \verb+\sep+ command.
\end{abstract}

%\includegraphics{figs/grabs.pdf}
%\item Research highlights item 1
%\item Research highlights item 2
%\item Research highlights item 3


\maketitle

\section{Introduction}

The software engineering (SE) industry is always looking for better and efficient ways of building higher quality software systems. However, in practice, the strong emphasis on time to market tends to ignore many, well-known SE practices. That is, practitioners often focus more on programming as compared to requirements gathering, planning, specification, architecture, design and documentation~\cite{Meinke2018b} – all of which are ultimately known to greatly benefit the cost effectiveness and quality of software systems. Lack of human resources is often cited as the main reason for doing so. Herein lies the great potential for machine learning (ML) since its algorithms are proven to be most befitting to problem domains that aim to replicate human behavior. Hence, it stands to reason that human-centric SE activities should also benefit from ML~\cite{Harman2012}.

The growing demand on agility and ability to solve complex problems in SE has already lead researchers to explore the potential of ML in this field. To date, ML has many demonstrated benefits in SE. Applications of ML for SE range from resolving ambiguous requirements to predicting software defects~\cite{Tsai2002}. For example, Sultanov et al.~\cite{Sultanov2013} used reinforcement learning (a type of ML) on understanding the relationships among software requirements at different levels of abstraction. Their approach shows how ML can automatically generate traceable links between high-level and low-level requirements. However, ML is not a single technique but rather an assortment of techniques. The challenge of using ML for SE is thus not only finding the right way of modeling the problem but also comparing various ML techniques and their potential. For example, several researchers have explored software projects predictions in order to better estimate the time to market their software products. For this purpose, various ML techniques were used and compared, e.g., artificial neural networks (ANN), rule induction (RI), case-based reasoning, support vector machines (SVM), and regression-based trees~\cite{Braga2007,Cuadrado-gallego2010,Silva2010}.

%However, the published applications of ML for SE have not yet established the same level of dominance. ML has proven very useful but rarely game changing. To date, there are few specialized conferences on ML for SE but it is important to note that the exponential growth in the number of articles on ML for SE being published year-to-year suggests that this research field is still in the process of establishing itself. Hence, we have yet to see whether ML will revolutionize SE.
In many areas of science and engineering, such as image recognition or autonomous driving, ML has already revolutionized development. The applications of ML to SE is increasing in significance, which is evident through the exponential growth in the number of articles on ML for SE being published every year. Consequently, it is of interest to understand which SE life cycle stages benefit the most from this trend; or even to understand which ML techniques are most suitable for which SE life cycle stage(s). This leads to the motivation of conducting this systematic mapping study. This systematic mapping study provides a bird’s-eye view on the current state-of-the-art of the field and suggests the open areas of research where more primary studies are needed. This study also provides a classification scheme as a MLSE (machine learning for software engineering) taxonomy highlighting the key areas of SE where ML has proven to be promising. In terms of scope, it is important to note that we are not interested in the application of ML in software projects in general (this is already well established) but in the application of ML in support of SE life cycle stages, e.g., requirements engineering, specification, analysis, design, testing, or maintenance. While this article presents a first, comprehensive study on the general use of ML for SE, it should be noted that some specialized studies already exist, e.g., ML for automated software testing~\cite{Durelli2019}.

The rest of the article is organized as follows. Section~\ref{RM} explains the research methodology and protocol followed in the study. Results of the study are discussed in Section~\ref{Discussion}. In the end, the study is concluded with addressing the threats to validity of the study and conclusion in Sections~\ref{TTV} and \ref{Conclusion}, respectively.

\section{Research Methodology}\label{RM}
This section describes how we obtained articles for our study, the key research questions, and how we systematically addressed them. We obtained the most relevant articles by employing an appropriate search strategy, formulating insightful goals and research questions, and devising a strong data extraction process. For this purpose, we have followed the research methodology described below, which is based on the updated guidelines provided by Petersen et al.~\cite{Petersen2015} for the research protocol and the creation of the classification scheme. The guidelines represent the basic principles of conducting systematic mapping studies in the domain of SE. We used Mendeley\footnote{https://www.mendeley.com} as our primary article management tool in this study. The timeline of this study is from the start of 1991 (the oldest relevant article we could found in the search was from year 1991) to the end of 2019 (we started writing this article in the start of 2020). 

\subsection{Goals, Questions and Metrics}

It stands to reason that a systematic study is always directed and kept on track by following a strict research protocol in order to improve the quality and impact of the study. To achieve this, we followed the Goal, Question and Metric (GQM) paradigm suggested by Basili et al.~\cite{Basili}. The aim was to guide the study by specifying its goals, formulating its research questions and identifying potential metrics in order to have a systematic data extraction process. The metrics are later used as attributes (keywords) in the data extraction process (described in Step 6 of Section~\ref{researchprotocol}). In the following, we summarize the goals, research questions and metrics (underlined) of the study.

\subsubsection{Goals}\label{Goals}
\begin{enumerate}[font={\bfseries},label={G\arabic*.}]
\item To identify the susceptibility of various ML techniques to SE life cycle stages
\item To understand the maturity of research in this area
\item To identify the demographics of this area
\item To understand the challenges, limitations and future directions for upcoming research in this area
\end{enumerate}

The first three goals lead to the research questions discussed in the following subsection. Due to the descriptive and elaborative nature of the fourth goal, we decided to thoroughly discuss it in Section~\ref{Discussion}.

\subsubsection{Questions}\label{RQ}
\begin{enumerate}[font={\bfseries},label={G\arabic*.}]
\item The susceptibility of various ML techniques to SE life cycle stages
\begin{enumerate}[font={\bfseries},label={Q\arabic*.}]
\item[]
	\begin{enumerate}[label*=\arabic*.\\,leftmargin=0\parindent]
	\setcounter{enumii}{1}
	\item What \textbf{\underline{SE life cycle stages}} are being focused on by the academic and industrial researchers in the area?
	
\textbf{Rationale:}
Our interest is to understand what SE life cycle stage the researchers tend to focus on, whether, a particular SE life cycle stage or the amalgamation of two or more. The SE life cycle stages are based on, but not limited to, the knowledge areas mentioned in SWEBOK~\cite{Swebok} characterizing the practice of SE, e.g., Software Requirements, Software Design or Software Maintenance.
	\item What are the \textbf{\underline{applications of ML}} in SE?

\textbf{Rationale:}
We are interested to know about the specific applications of ML that exist in SE, e.g., a ML technique was used to automate the test case generation or to predict potential bugs in the system.
	\item What \textbf{\underline{type of ML and its techniques}} are being employed for SE?
	
\textbf{Rationale:}
We are interested to know whether a particular type/technique consistently employed for a specific life cycle stage. Type of ML refers to how the models have been trained, e.g., supervised, semi-supervised or unsupervised. Whereas the ML techniques are the  algorithms used for classification or clustering problems, e.g., support vector machine (SVM), random forests (RF) or neural networks (NN).
	\end{enumerate}
\end{enumerate}

\item The maturity of research in the area
\begin{enumerate}[font={\bfseries},label={Q\arabic*.}]
\item[]
	\begin{enumerate}[label*=\arabic*.\\,leftmargin=0\parindent]
	\setcounter{enumii}{2}
	\item What is the \textbf{\underline{contribution facet}} of the articles?

\textbf{Rationale:}
The contribution facet partially corroborates the attributes provided by~\cite{Banerjee2013,Petersen2008} and are supplemented by our own views obtained by analysing the extracted articles. The attributes are defined as follows:

\begin{itemize}
\item \textbf{Tool:} Article proposing a new tool or improving an existing one and describing its evaluation.

\item \textbf{Approach/Method:} Article proposing a new approach or improving the existing one.

\item \textbf{Model/Framework:} Article introducing a new approach or a framework.

\item \textbf{Algorithm/Process:} Article proposing a new algorithm or describing a SE process.

\item \textbf{Comparative Analysis:} Article evaluating different approaches and reporting results of the comparative study.
\end{itemize}
	\item What is the \textbf{\underline{research facet}} of the articles?

\textbf{Rationale:}
The research facet of an article refers to the maturity of the research in terms of empirical evidence provided in the article or whether an article was proposing a solution or evaluating an existing approach. The research facet is defined as follows:
\begin{itemize}
\item \textbf{Evaluation:} Article evaluating or validating the proposed approach using empirical methods.
\item \textbf{Knowledge:} Article describing the experiences and opinions of authors on the existing approaches.
\item \textbf{Solution:} Article proposing a new solution and describing its applicability with the help of examples and arguments.
\end{itemize}

\item What \textbf{\underline{datasets}} are commonly employed in the articles?

\textbf{Rationale:}
We are interested to know about the datasets that are most commonly used to evaluate the research results in the domain of ML for SE.
	\end{enumerate}
\end{enumerate}
	
\item The demographics of research in the area
\begin{enumerate}[font={\bfseries},label={Q\arabic*.}]
\item[]
	\begin{enumerate}[label*=\arabic*.\\,leftmargin=0\parindent]
	\setcounter{enumii}{3}
	\item What are the \textbf{\underline{trends in terms of years}} of publications in the area?
	
\textbf{Rationale:}
Trends in terms of years refers to the number of publications varying from a year to another. Here, we want to assess how active this research area is.
	\item What are the \textbf{\underline{highest publishing venues}} of the area?

\textbf{Rationale:}
We are interested to know about the venues which have highest publications with respect to the area of ML for SE.
	\end{enumerate}
\end{enumerate}
%\item What are the future prospects of ML for SE?
%	\begin{enumerate}[label*=\arabic*.]
%	\item What are the \textbf{\underline{challenges}} highlighted by the researchers?
%
%\textbf{Rationale:}
%The major and key challenges highlighted by the authors in their studies which were/weren't later overcome by them.
%	\item What are the \textbf{\underline{limitations}} highlighted by the researchers?
%	
%\textbf{Rationale:}
%The limitations reported by the authors in their studies that they did not manage to address.
%	\item What are the \textbf{\underline{future directions}} implicated by the researchers?
%
%\textbf{Rationale:}
%The future areas of research reported by the authors to further investigate upon.
%	\end{enumerate}
\end{enumerate}

\subsection{Research Protocol}\label{researchprotocol}
\begin{figure*}
\centering
\includegraphics[width=0.75\linewidth]{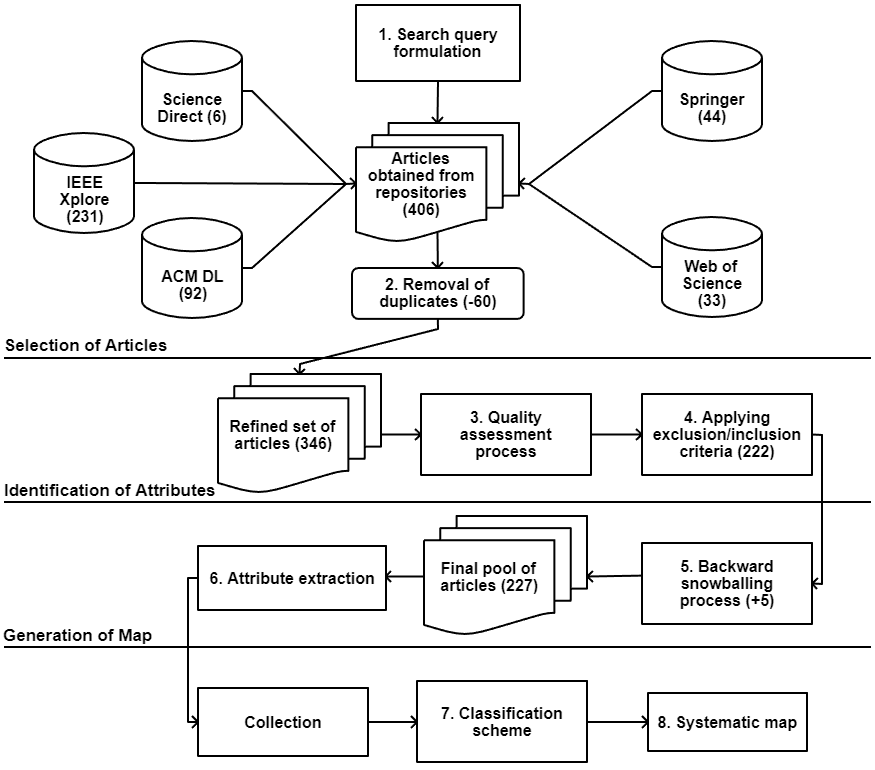}
% \centerline{}
\caption{Research Protocol followed in the study}
\label{RP}
\end{figure*}

A research protocol is essential to conduct an independent, objective study. It regulates the flow of research and maximizes the meaningful outcomes from the study. For this purpose, we have designed a research protocol that describes the elements of the study and is illustrated in Fig.~\ref{RP}. Following are the main steps of the research protocol.
%illustrates the sequence of steps involved in the search and the overall protocol of this study.

\textbf{1. Search query formulation:} Our search query uses two element PICO search as advised by Petersen et al.~\cite{Petersen2008}. The two elements of the PICO framework are the following:
%\begin{enumerate}
%\item \textbf{Problem 'P':} (requirement, specification, design, model, analysis, architecture, implementation, code, test, verification, validation, maintenance)
%\item \textbf{Intervention 'I':} (ML, deep learning)
%\item \textbf{Comparison 'C':} At this stage, no empirical comparison has been made among the interventions, alternate intervention terms have been identified in the study.
%\item \textbf{Outcome 'O':} Exploring measurable outcomes is one of the goal of the study thus not considered in the study at this stage.
%\end{enumerate}

\textbf{Problem `P':} (requirement, specification, design, model, analysis, architecture, implementation, code, test, verification, validation, maintenance) and
\textbf{Intervention `I':} (ML, deep learning). We have not considered Comparison `C' and Outcome `O' in order to limit the search spectrum and the broad scope of the study.

The search query was formulated in an iterative fashion in order to ensure highest evidence-based retrieval of articles. The query was applied to titles and abstracts of articles in five well known digital repositories: IEEEXplore\footnote{\url{https://ieeexplore.ieee.org/}}, ACM Digital Library\footnote{\url{https://dl.acm.org/}}, ScienceDirect\footnote{\url{https://www.sciencedirect.com/}}, Springer\footnote{\url{https://www.springer.com/}} and Web of Science\footnote{\url{https://apps.webofknowledge.com/}}. The search yielded a total of 406 articles. The search string used in all repositories was:

 \textbf{\textit{("machine learning" OR "deep learning") AND software AND requirement* OR specification* OR design* OR model* OR analysis OR architecture OR implementation OR code OR test* OR verification OR validation OR maintenance}}\footnote{Asterisk (*) is a wildcard that refers to zero or more characters in a word}

All repositories, except Springer, returned the number of articles as shown in Fig.~\ref{RP} corresponding to the search query applied only on titles. Springer initially yielded 4502 articles as a result of the query; however, most of these articles were quite irrelevant to the scope of our study even after applying filters such as "Computer Science" as discipline and "SE" and "Artificial Intelligence" as sub-disciplines to reduce the search space. The first author then went through the titles and abstracts of the articles (if the goal of the article is unclear from the title) and stopped the search process when the first page with all irrelevant articles was reached. This resulted into 44 articles.

%Since the Articles were further gone through quality assessment process (described in~\cite{Budgen2007,KITCHENHAM20097}) performed by the authors of this study.

% This threat was reduced by the selected backward snowballing process~\cite{Wohlin2014a} (mentioned below) conducted by the first author.

\textbf{2. Removal of duplicates:} In Step 2, we removed the duplicate articles from the database. After removal of duplicates (60 articles), the remaining pool of articles was left with a tally of 346.

\textbf{3. Quality assessment process (QAP):} In Step 3, the articles underwent the defined quality assessment process in order to maximize the overall authenticity and quality of the study. The quality assessment process consists of a multi filtration method based on the guidelines provided by Kitchenham et al.~\cite{KITCHENHAM20097}. In this method, random and equal set of articles are distributed among the participants of a study in order to mitigate any bias. The method comprises of a four-questions checklist, where each question is answered using a defined scale as described in Table~\ref{QAP}. The sum of scores for all questions can vary from 3 to 10, while 10 being the highest quality. Similarly, all participants of this study evaluated their particular set of articles by rating each article based on the questions mentioned in the checklist. The resultant scores were then accumulated and utilized in the following exclusion/inclusion criteria. The questions of the checklist and the scales used in this study are shown in Table~\ref{QAP}.
% Once the quality assessment process concluded, the articles adhering to the check-list were included in the final pool and the rest (69 articles) was discarded.

\begin{table}
\caption{Quality assessment process}
\begin{tabular}{p{0.3cm}p{3cm}p{2cm}p{1.5cm}}
\toprule
Sr. no. & Questions& Scale & Rationale \\
\midrule
1st& Does relevance and appropriateness of the article correspond to the research goals of the study? & low=1, medium=2, high=3 & Relevance of the article\\
\hline
2nd& Does the Primary study evaluated empirically? & yes=1,no=0 & Empirical Evaluation\\
\hline
3rd& Is there a certain level of bias?& low=3, medium=2, high=1 &  Risk of Bias\\
\hline
4th& Results reporting quality and authenticity?& low=1, medium=2, high=3 & Quality of Results\\
\bottomrule
\end{tabular}
\label{QAP}
\end{table}

\textbf{4. Applying exclusion/inclusion criteria:} In Step 4, we apply exclusion and inclusion criteria to the pool of articles in order to further refine their quality. This process yielded 222 articles.

Exclusion Criteria: %Articles obtained from the query results underwent a screening process and were scrutinized based on the below mentioned exclusion criteria.

\begin{enumerate}
\item Articles that were not relevant to the scope (i.e., Articles that were not addressing the context of applications of ML for SE (negating Q1 in QAP checklist)) of the study were excluded
\item Articles that were not available in full text format were excluded
\item Articles demonstrating poor empirical soundness, i.e., score lower than 5 (refer to the QAP in Step 3) were excluded
%\item Articles that are not written in English were excluded
%\item Duplicate articles are excluded
\end{enumerate}

Inclusion Criteria: Articles were then selected based on the following inclusion criteria.

\begin{enumerate}
\item Articles of more than a single page were included
\item Articles assigned with a minimum score of 5 or more out of 10 in the QAP were included
\item Articles that were peer reviewed were included
%\item Articles having defined contributions and adhering to the principles of EBSE (Evidence based SE) are included
%\item Articles comparing and evaluating different ML techniques are included
\item Articles that were entirely written in English were included
%\item Articles obtained from forward and backward snowballing are included
\end{enumerate}

\textbf{5. Backward snowballing process:} In Step 5, we applied backward snowballing~\cite{Wohlin2014a} (further searches based on references in the existing articles of the pool) in order to ensure a broad spectrum of articles relevant to the scope. The process yielded five additional articles suggesting that the initial search and exclusion/inclusion criteria covered the scope of our study well. The tally now stands at 227.

\textbf{6. Attribute extraction:} In Step 6, the first author of this study went through the abstracts and derived the main attributes from each article. If the discussion in an abstract was not conclusive, the author investigated the conclusion or even the full text of the article. Once, the attributes are extracted, the authors established initial set of categories, which were refined iteratively and then generalized in order to broadly cover the research area. The generalized attributes along with article references are maintained in MS Excel sheets referred to as the collection in this study.

\begin{figure}
\centering
\includegraphics[width=0.75\linewidth]{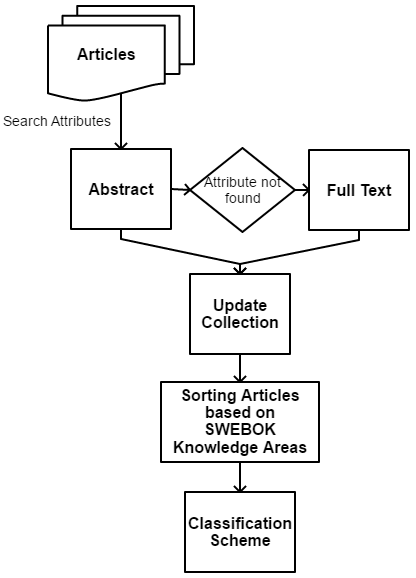}
% \centerline{}
\caption{Attribute extraction and classification scheme}
\label{CS}
\end{figure}

\textbf{7. Classification scheme} In Step 7, we define a classification scheme to ensure accurate assessment of attributes. The generalized attributes obtained were then sorted by the participants of the study based on the knowledge areas provided in~SWEBOK~\cite{Swebok}. During the article sorting process, certain articles were found to be equivocal. In such cases, we associated those attributes to the articles that received majority votes from the participants of this study. Please note that the knowledge areas mentioned in~SWEBOK were not strictly used in the categorization but merely employed as a defining factor to provide a high level abstraction of attributes that represented the set of articles. To get a better understanding, a graphical representation of the workflow starting from the attribute extraction process leading to the classification scheme is shown in Fig.~\ref{CS}.

\textbf{8. Systematic map}
The construction of the map comprises of a series of discussions among the participants of this study, which lead to the careful association of the facets with the high level attributes of the articles. To get a better understanding of the systematic map, in the following we describe its main facets.

\begin{figure*}
\centering
\includegraphics[width=1.0\linewidth]{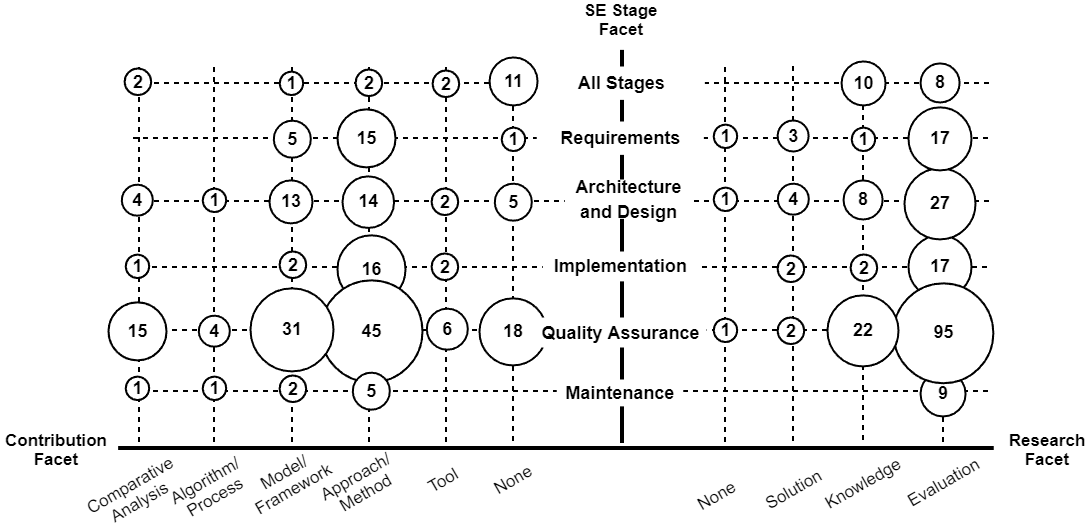}
% \centerline{}
\caption{Systematic Map - Association of Contribution/Research Facets with the SE Stage Facet}
\label{Systematic map}
\end{figure*}

\begin{enumerate}
	\item \textbf{SE Stage Facet:} The SE Stage Facet comprises of attributes on a higher level of abstraction showing partial relevance between knowledge areas of SWEBOK~\cite{Swebok} and the extracted attributes.
	\item \textbf{Contribution Facet:} The Contribution types, such as tools, approaches, or algorithms, are derived from the articles in a fashion similar to the ones described in~\cite{Petersen2008,Banerjee2013} and supplemented by our own perspective on the obtained set of articles.
	\item \textbf{Research Facet:} The Research types, such as evaluations and solutions, are derived from the work of Wieringa et al.~\cite{Wieringa2006}, where the type knowledge refers to the articles expressing experiences and opinions of the researchers.
\end{enumerate}
Fig. \ref{Systematic map} shows the resultant systematic map.

\subsection{Map Evaluation}
This section evaluates the systematic map by addressing the research questions discussed in Section~\ref{RQ}. In order to get a better understanding, the questions are answered in line.

\begin{figure}
\centering
\includegraphics[width=1.0\linewidth]{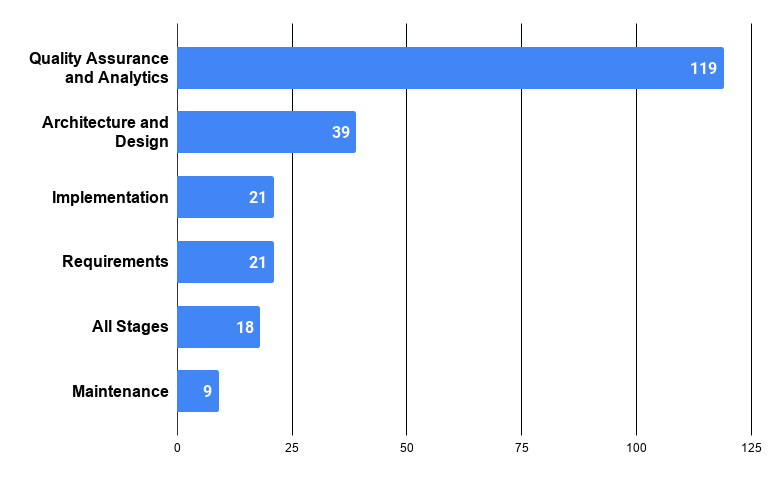}
% \centerline{}
\setlength{\belowcaptionskip}{-10pt}
\caption{Articles by SE life cycle Stages}
\label{Articles_by_SE_life_cycle_stages}
\end{figure}

\textbf{Q1.1} SE life cycle stages:
This question relates to our classification scheme, which is partially based on knowledge areas involved in traditional SE as mentioned in SWEBOK~\cite{Swebok}. 
The SE stages and articles that fall into the corresponding stage are addressed in Fig.~\ref{Articles_by_SE_life_cycle_stages}. 119 out of 227 (52\%) articles belong to quality assurance and analytics. 39 out of 227 (17\%) articles have focused on architecture and design. 21 out of 227 (9\%) articles have addressed the implementation and requirements engineering stage each. 9 (4\%) articles were focusing on the maintenance phase. Rest of the articles were not particularly focusing on any stage but were generally applicable to SE.

\begin{figure*}
\includegraphics[width=1.0\linewidth]{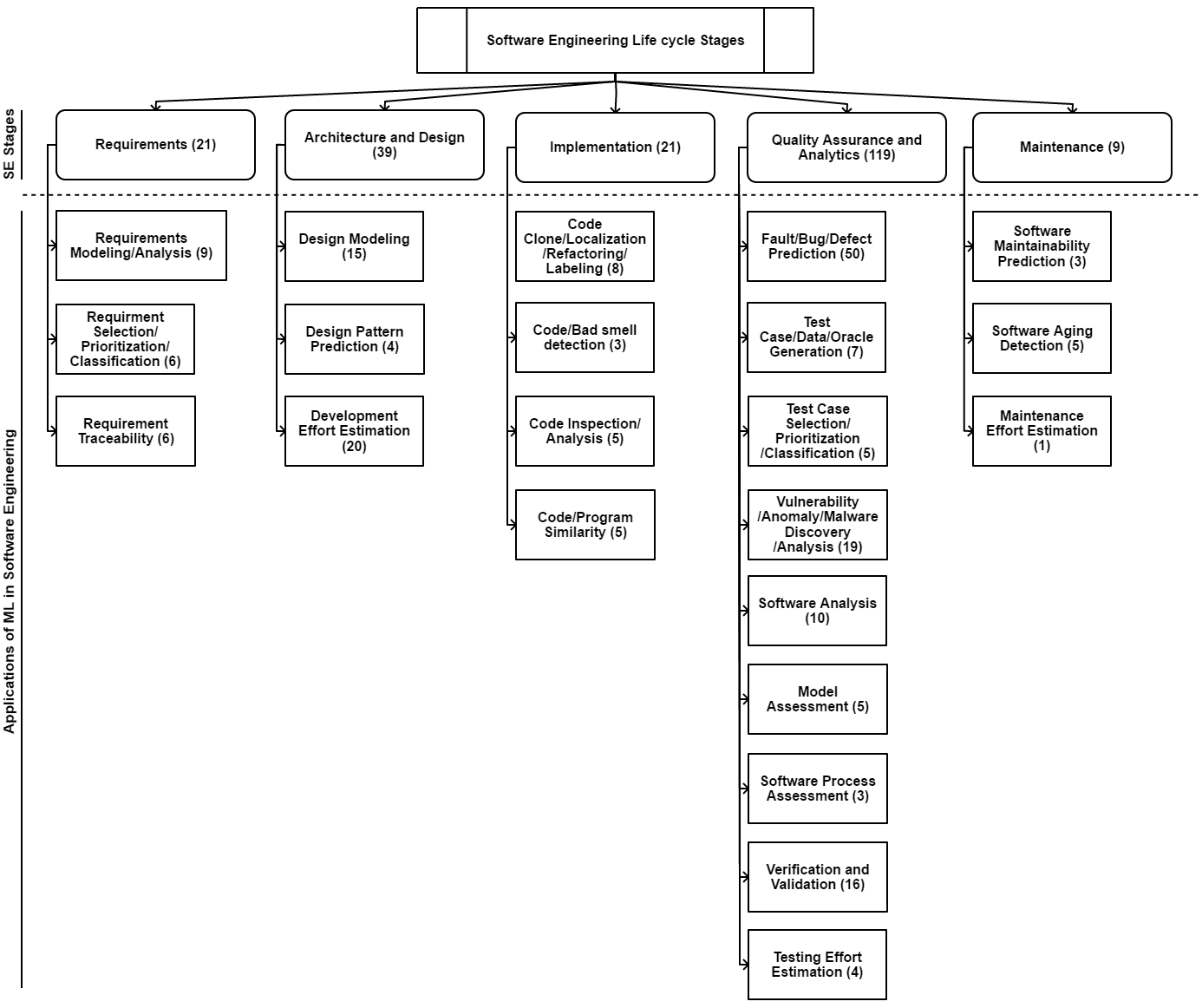}
\centerline{}
\caption{MLSE Taxonomy}
\label{Taxonomy}
\end{figure*}

\textbf{Q1.2} Applications of ML for SE:
To address this question, we have developed a taxonomy based on the identified applications of ML for SE in order to characterize the obtained articles into appropriate categories. We named the taxonomy as MLSE (machine learning for software engineering). The taxonomy was devised following the principles mentioned in~\cite{Cuevas2014,Usman2017}. As aforementioned, we have consulted the knowledge areas in SE from the SWEBOK~\cite{Swebok} and envisioned a hierarchical-based classification structure of the taxonomy. Each participant of the study analyzed the applications in their assigned set the articles and aggregated them based on the similarities as described in step 7 of Section~\ref{researchprotocol}. Subsequently, we have organized the applications of ML for SE as subbranches, which belong to five life cycle stages of SE (knowledge areas). The applications of ML for SE that come under corresponding SE life cycle stages along with the number of articles are briefly explained below. Table~\ref{Classification_by_Articles} shows the corresponding articles with respect to the classification proposed as a MLSE taxonomy as shown in Fig.~\ref{Taxonomy}. Following is a brief description of the elements of the MLSE taxonomy:

The Requirements stage comprises of three categories.
\begin{itemize}
	\item \textbf{Requirements Modeling and Analysis (9 (4\%) articles):} Requirement Modeling and Analysis contains articles that are focusing on distinguishing ambiguous requirements, resolving incompleteness, correctness of requirements, etc.
	\item \textbf{Requirements Selection\slash{}Prioritization\slash{}Classification (6 (3\%) articles):} Requirements Selection\slash{}Prioritization\slash{}Classification deals with articles proposing ML techniques that emphasize on automating prioritization of requirements or their classification.
	\item \textbf{Requirements Traceability (6 (3\%) articles):} Requirements traceability contains articles that refer to the ML approaches that assist in linking requirements to code or other artifacts.
\end{itemize}

The Architecture and Design stage consists of three categories.

\begin{itemize}
	\item \textbf{Design Modeling (15 (7\%) articles):} Design Modeling comprises of articles in which software process\slash{}services recommendation models have been proposed in order to facilitate the project managers in selection of the most suitable process model for their projects. Apart from this, model smells and re-factoring techniques of object-oriented structures using ML have also been proposed in the articles.
	\item \textbf{Design Pattern Prediction (4 (2\%) articles):} Design Pattern Prediction comprises of articles that primarily focus on recognizing design patterns in software through source code or user interface layout using ML techniques. 
	\item \textbf{Development Effort Estimation (20 (9\%) articles):} Development Effort Estimation refers to the effort estimation of software projects using ML techniques.
\end{itemize}
    
The Implementation stage has four categories. 

\begin{itemize}
	\item \textbf{Code Clone\slash{}Localization\slash{}Re-factoring\slash{}Labelling (8 (3\%) articles):} Code Clone\slash{}Localization\slash{}Re-factoring\slash{}Labelling comprises of articles that aim at finding code clones, specific piece of code in software, re-factoring of code or labelling of the code with the help of ML. 
	\item \textbf{Code\slash{}Bad smell detection (3 (1\%) articles):} Code\slash{}Bad smell detection contains articles that focus on applying ML in order to detect code and bad smells in software source code. 
	\item \textbf{Code Inspection\slash{}Analysis (5 (2\%) articles):} Code Inspection\slash{}Analysis contains articles in which a ML technique is employed for the purpose of code reviews. 
	\item \textbf{Code\slash{}Program similarity (5 (2\%) articles):} Code\slash{}Program similarity category refers to articles that identify specific pieces of code, which are similar between two or more software projects. Additionally, such articles distinguish between original and pirated/cracked software.
\end{itemize}

The Quality Assurance and Analytics stage has nine categories.

\begin{itemize}
	\item \textbf{Fault\slash{}Bug\slash{}Defect Prediction (50 (20\%) articles):} Fault\slash{}Bug\slash{}Defection Prediction category contains articles that revolve around the prediction of faults, bug or defects using ML techniques. 
	\item \textbf{Test Case\slash{}Data\slash{}Oracle Generation (7 (2\%) articles):} Test Case\slash{}Data\slash{}Oracle Generation surrounds articles that specifically propose ML techniques that help in generating test data, test oracle or entire test suite. 
	\item \textbf{Test Case Selection\slash{}Prioritization\slash{}Classification (5 (2\%) articles):} Test Case Selection\slash{}Prioritization\slash{} Classification deals with articles that particularly focus on test case prioritization or classification techniques using ML.
	\item \textbf{Vulnerability\slash{}Anomaly\slash{}Malware Discovery\slash{}Analysis (19 (8\%) articles):} Vulnerability\slash{}Anomaly\slash{}Malware Discovery\slash{}Analysis mostly concerns with the security aspect of software quality addressed through ML techniques.
	\item \textbf{Software Analysis (10 (4\%) articles), Technique Assessment (5 (2\%) articles), Software Process Assessment (3 (1\%) articles):} Software Analysis, Model Assessment and Software Process Assessment contain articles that come under assessment and analysis of software and ML models using existing ML techniques.
	\item \textbf{Verification and Validation (16 (7\%) articles):} Verification and validation category holds articles that specifically address prediction and verification of software reliability through ML. 
	\item \textbf{Testing Effort Estimation (4 (2\%) articles):} Testing Effort Estimation comprises of articles that address the amount of testing effort required in order to test a software system using ML techniques.
\end{itemize}

The Maintenance stage has three categories.

\begin{itemize}
	\item \textbf{Software Maintainability Prediction (3 (1\%) articles):} The category of Software Maintainability Prediction holds articles that employ ML technique in order to assist the prediction of maintainability metrics appropriate for specific software projects. 
	\item \textbf{Software Aging Detection (5 (2\%) articles):} Software Aging Detection comprises of articles that use ML in order to detect software maturity and its aging in terms of resource depletion such as memory leaks, high CPU usage, and overtime.
	\item \textbf{Maintenance Effort Estimation (1 (0.4\%) article):} Maintenance Effort Estimation contains articles that estimate effort required for the maintenance of a software system using ML.
\end{itemize}

\begin{table*}
\renewcommand\arraystretch{1.0}
\centering
\caption{Classification by Articles}
\begin{tabular}[t]{l>{\raggedright}p{0.35\linewidth}>{\raggedright\arraybackslash}p{0.4\linewidth}}
\toprule
SE Stages&Applications of ML for SE &Articles\\
\midrule
\multirow{3}{*}{Requirements}&Requirements Modeling and Analysis&{\cite{Ahsan2010,Chioasca2012,Jahan2019,Khosrowjerdi2018,Ramirez2011,Sanctis2019,Sharma2014,Singh2018,Turliuc2011}}\\
&Requirement Selection/Prioritization/Classification&{\cite{Ahsan2009,Licea2017,Moranna2018,Perini2013,Rahman2019,Rooijen2017}}\\
&Requirement Traceability&{\cite{Cleland-Huang2010,Guo2017,Mills2017,Nafi2018,Sultanov2013,Wieloch2013}}\\
\hline
\multirow{2}{*}{Architecture and Design}&Design Modeling&{\cite{Ahn2018,Castro-Lopez2018,Chollet2013,Dawoud2018,Factors2000,B2019,Muccini2019,Neal2017,Oberta2018,Sankaran2017,Shin2005,Sidhu2018,Song2016,White2015,WU2017}}\\
&Design Pattern Prediction&{\cite{Dwivedi2016,Nguyen2018,Thaller2019,B2019b}}\\
&Development Effort Estimation&{\cite{AlAsheeri2019,Ba2007,Banimustafa2018,Braga2007,Corazza2009,Cuadrado-gallego2010,ERTUGRUL2019,Huang2015,Idri2008,Ionescu2017,Iwata2016,Mizuno2010,Murillo-Morera2017,Satapathy2017,Sharma2017,Souza2003,Srinivasan1995,Vlv2017,Wen2012,Wright2019}}\\
\hline
\multirow{4}{*}{Implementation}&Code Clone/Localization/Refactoring/Labeling&{\cite{Alahmadi,Cambronero,Gelman2018,Mostaeen2019,Ott2018,Parr,White,Yang2012}}\\
&Code/Bad smell detection&{\cite{Q2018,Maneerat2011,Pecorelli2019}}\\
&Code Inspection/Analysis&{\cite{Ayesha2018,Chandra2016,Fouqu,Lal2017a,Madera2017a}}\\
&Code/Program Similarity&{\cite{Kim2015,Leclair2018,Tufano2018,Zhang2018a,Zhao2018}}\\
\hline
\multirow{9}{*}{Quality Assurance}&Fault/Bug/Defect Prediction&{\cite{Abbineni2018,Alshehri2018,B2019a,Bhandari2018b,Bhandari2018,Bharathi2019,Borg2019,Brun2004,Cerrada2017,Ceylan2006,Challagulla2005,Clemente2018,Dam2019,Dwarakanath2018,Ferzund2008,Hall2012,Heo2017,Huo2018a,Jindal,Kahles2019,Kalibhat2017,Karim,Kumar2019a,Lal2017b,Li2017a,Liang2019,Luo2010,Iru2017,Ni2017,Phan2017a,Phan2017,PhuongHa2019,Polisetty2019,Ghhs2017,Rana2003,Dhanda2019,Shepperd2014,Shukla2018,Singh2008,Singh2017,Sudharson2019,Sun2018,Tanaka2019,Tran2019,Wang2018,Wen2018,Xu2004,Yang2019,Zhang2018,Zhao2019}}\\
&Test Case/Data/Oracle Generation&{\cite{Babamir,Braga2018,Briand2008,Ma2018,Oster2005,Paduraru,Zhang2006}}\\
&Test Case Selection/Prioritization/Classification&{\cite{Gove,Hu2018,Lenz2013,Rosenfeld2017,Zheng2018}}\\
and Analytics&Vulnerability/Anomaly/Malware Discovery/Analysis&{\cite{Alonso2011,Alotaibi2019,Bisio2014,Dam2015,Dqga,Ghaffarian2017,Gowda2018,Han2017,Huang2019,Huch2018,Jie2016,Kronjee,Kuznetsov2019,Muzzammel2019,Niu2019,Ognawala2018,Paper,Rezende2017,Yan2017}}\\
&Software Analysis&{\cite{B2016,Dam2018,Fu2018,Lim2018,Liu2017,Moghadam,Quin2019,Rajbahadur2019,Sahin2019,Tofighi-Shirazi2019}}\\
&Technique Assessment&{\cite{Harandi1991,Lounis,Lounis2014,Sajnani2012,Simpson2016}}\\
&Software Process Assessment&{\cite{Chen2011,Lopez-martin2014,Rughetti2012}}\\
&Verification and Validation&{\cite{Al-jamimi2013,Baskiotis2006,Cummins2018,Eniser2018,Guo2018,Kanewala2016,Karpov2019a,Lounis2014,Masuda2018,Nakajima,Paul2007,Rana,DeSantiago2018,Shanthi2018,Tamura2016,Wang2018a}}\\
&Testing Effort Estimation&{\cite{Almeida1998,Cheatham,Ramasamy2017,Silva2010}}\\
\hline
\multirow{3}{*}{Maintenance}&Software Maintainability Prediction&{\cite{Al-Jamimi2013b,Kumar2019,Reddivari2019}}\\
&Software Aging Detection&{\cite{Alonso2010,Andrzejak2008,Huo2018,Lee2014,Yan2016}}\\
&Maintenance Effort Estimation&{\cite{Chandra2017}}\\
\hline
\multirow{1}{*}{All Stages}&N/A&{\cite{B2013a,B2018,Bruegge2009,Falcini2017,Harman2012,Eshuis2019,Kalles2016,Lin2018a,Meinke2018b,Michie1991,Nascimento2018,Park1997,Robbes2019,Schreck2018,Shen2019,Wan2019,White2015a,Xie2013b}}\\
\bottomrule
\end{tabular}
\label{Classification_by_Articles}
\end{table*}

\begin{figure}
\centering
\includegraphics[width=1.0\linewidth]{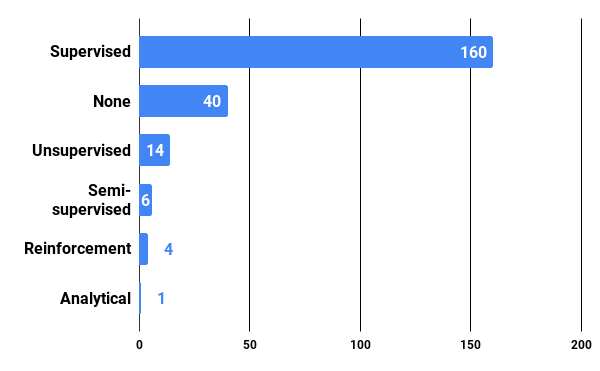}
% \centerline{}
\caption{Articles by ML Type}
\label{Articles_by_Machine_learning_Type}
\end{figure}

\begin{figure}
\centering
\includegraphics[width=1.05\linewidth]{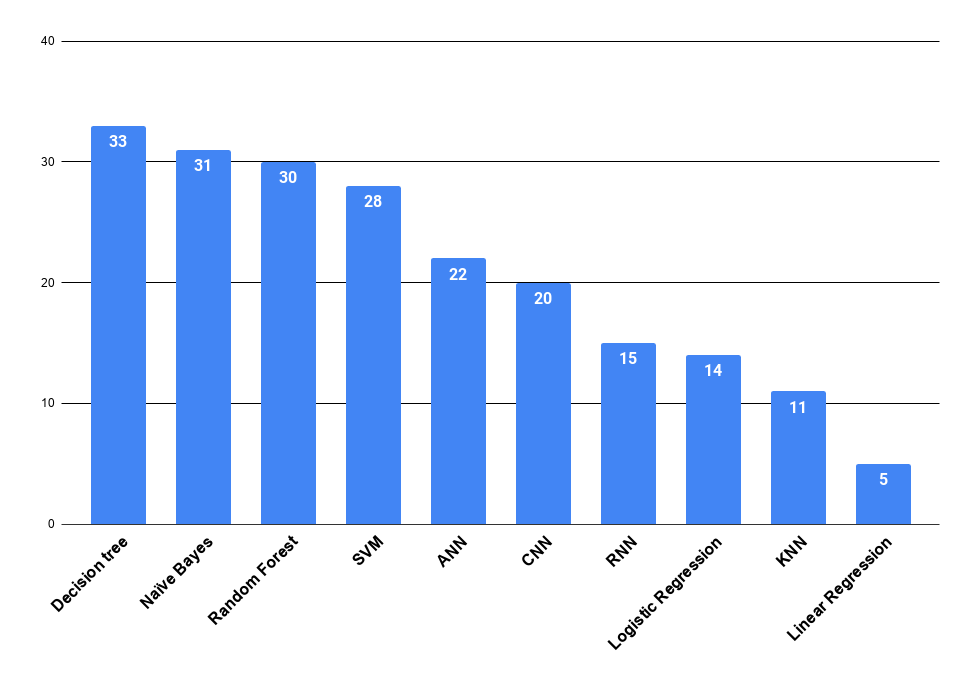}
% \centerline{}
\caption{Articles by Techniques}
\label{Articles_by_Techniques}
\end{figure}

\begin{figure}
\centering
\includegraphics[width=1.0\linewidth]{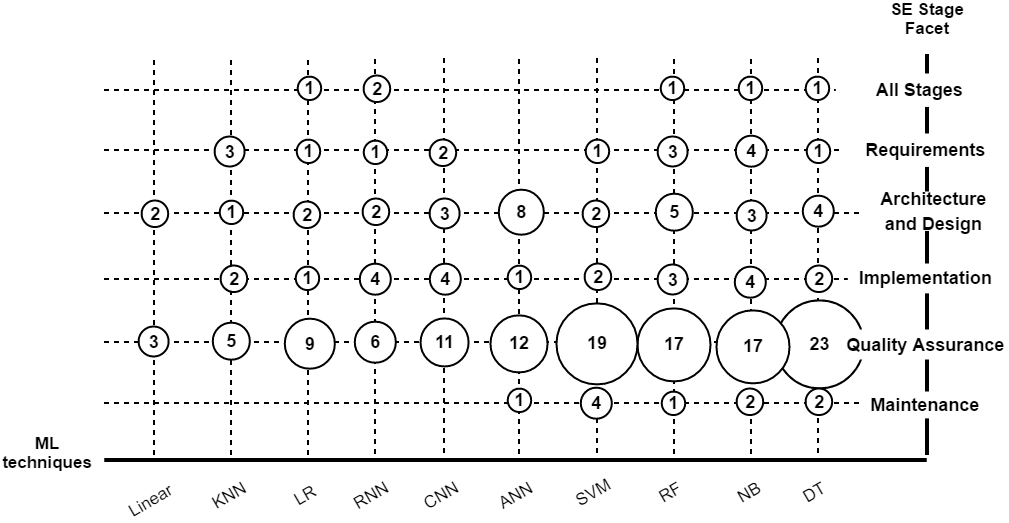}
% \centerline{}
\caption{ML techniques usage in SE}
\label{MLUsageinSE}
\end{figure}

\textbf{Q1.3} ML type and techniques:
The purpose of this question is to understand which types of ML are being employed in the selected articles. As shown in Fig.~\ref{Articles_by_Machine_learning_Type}, 162 out of 227 (71\%) articles employed supervised learning, whereas 14 out of 227 (6\%) articles employed unsupervised learning, and 6 out of 227 (3\%) articles employed semi-supervised learning. While, 4 out of 227 (2\%) articles addressed reinforcement learning, 1 out of 227 (0.4\%) focused analytical (inference based) learning, and the rest of the articles 40 out of 227 (18\%) reported none. The techniques being employed in those articles are shown in Fig.~\ref{Articles_by_Techniques}. The top 3 most commonly used techniques are Decision Trees, Naive Bayes and Random Forrest, respectively. While 33 out of 227 (15\%) articles employed Decision Trees, 31 out of 227 (14\%) articles have used Naive Bayes and 30 out of 227 (13\%) articles used Random Forest for model training. Moreover, the techniques targeting specific life cycle stages are shown in Fig.~\ref{MLUsageinSE}. As shown in Fig.~\ref{MLUsageinSE}, mostly ML techniques were employed to solve problems related to the Quality Assurance and Analytics stage. Decision Trees were again the most commonly used technique here (23 articles), followed by Support Vector Machine (19 articles). Random Forrest and Naive Bayes were next in line with 17 articles apiece. Artificial Neural Network, which was used in 12 articles in the Quality Assurance stage was also a subject of interest for the researchers working in the Architecture and Design stage (8 articles). Although, all the ML techniques have certain pros and cons but the selection of the most suitable technique depends on the type of dataset being constructed or employed. In general, decision trees appeared to be highly employed among the articles due to its simplicity and strong classification and regression capabilities~\cite{Almeida1998,Ferzund2008,Q2018}.

\begin{figure}
\centering
\includegraphics[width=1.05\linewidth]{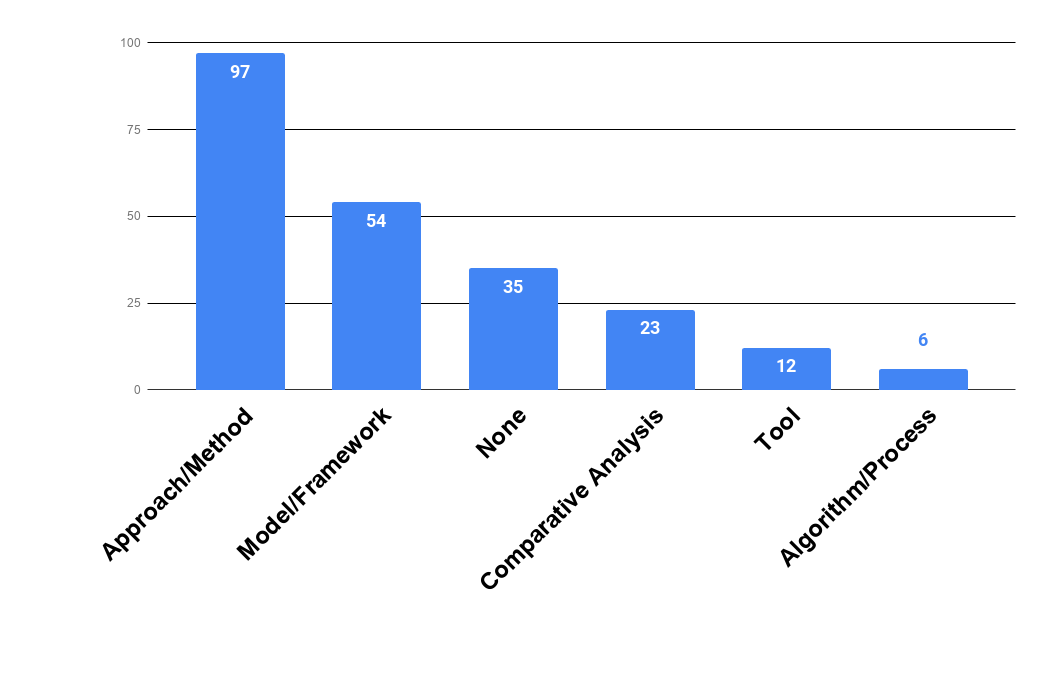}
% \centerline{}
\setlength{\abovecaptionskip}{-20pt}
\caption{Articles by Contribution Facet}
\label{Articles_by_Contribution_Facet}
\end{figure}

\textbf{Q2.1:} Contribution facet of the articles:
The contribution facet addresses the novel propositions of the articles. This represents the current state-of-the-art and enables researchers and industrial practitioners to get an overview of the existing tools and techniques in the literature. As shown in Fig.~\ref{Articles_by_Contribution_Facet}, 97 out of 227 (43\%) articles focused on approaches\slash{}methods, followed by 54 (24\%) articles proposing models\slash{}frameworks, 23 (10\%) articles focusing on comparative analysis of existing techniques, 12 (5\%) articles focusing on tools and 6 (3\%) articles focusing on algorithms/processes. Rest of the articles -- 35 out of 227 (15\%) -- reported no new propositions. These articles were either investigating existing approaches, performing comparative studies, discussing opinions, or reporting their experiences.

Table~\ref{Tools} shows the names of the propositions along with the contribution facet and references of the articles. %The types of propositions mentioned in Table~\ref{Tools} are explicitly stated by the authors in their articles. 
Interestingly, only 23 out of 227 (10\%) articles have explicitly named their propositions.

\begin{table}
\caption{Named propositions in the articles}
\begin{tabular}{cp{2.5cm}p{1.9cm}c}
\toprule
Sr. no. & Name& Contribution Facet & Article \\
\midrule
1& WIRECAML &Tool& \cite{Kronjee} \\
\hline
2& Trace-by-Classification &Approach& \cite{Wieloch2013} \\
\hline
3& SOA-based integrated software &Tool& \cite{Cerrada2017}\\
\hline
4& ProbPoly &Framework& \cite{Turliuc2011}\\
\hline
5& Modelware &Tool& \cite{WU2017}\\
\hline
6& Featuretools &Tool& \cite{Schreck2018}\\
\hline
7& Feature Maps &Algorithm& \cite{Thaller2019}\\
\hline
8& ExploitMeter &Framework& \cite{Yan2017}\\
\hline
9& DLFuzz &Framework& \cite{Guo2018}\\
\hline
10& DeepSim &Approach& \cite{Zhao2018}\\
\hline
11& DeepGauge &Process& \cite{Ma2018}\\
\hline
12& DARVIZ &Framework& \cite{Sankaran2017}\\
\hline
13& CroLSim &Model& \cite{Nafi2018}\\
\hline
14& Code-Buff &Tool& \cite{Parr}\\
\hline
15& CDGDroid &Approach& \cite{Xu2004}\\
\hline
16& AppFlow &Tool& \cite{Hu2018}\\
\hline
17& CloneCognition &Tool& \cite{Mostaeen2019}\\
\hline
18& ArchLearner &Tool& \cite{Muccini2019}\\
\hline
19& SZZ Unleashed &Tool& \cite{Borg2019}\\
\hline
20& Auto-sklearn &Tool& \cite{Tanaka2019}\\
\hline
21& SLDeep &Approach& \cite{Majd2020}\\
\hline
22& RIVER &Tool& \cite{Paduraru}\\
\hline
23& Seml &Framework& \cite{Liang2019}\\
\bottomrule
\end{tabular}
\label{Tools}
\end{table}

\begin{figure}
\centering
\includegraphics[width=1.0\linewidth]{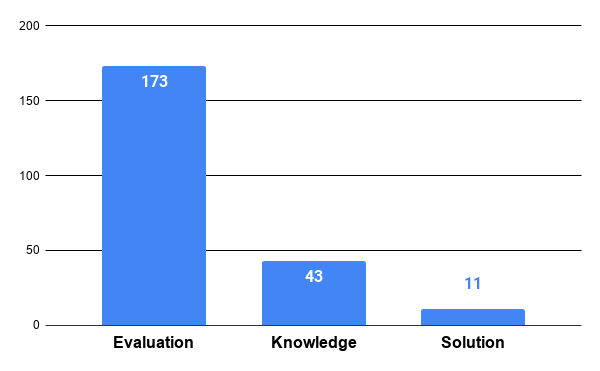}
% \centerline{}
\caption{Articles by Research Facet}
\label{Articles_by_Research_Facet}
\end{figure}

\textbf{Q2.2} Research facet of the articles:
The Research facet describes the nature of articles in terms of their purpose of conducting the research. %This reflects the overall goal of the research happening in this specific area.
Fig.~\ref{Articles_by_Research_Facet} shows the articles by the research facet. 173 out of 227 (76\%) articles have contributions with empirically evaluated propositions, whereas 43 out of 227 (19\%) articles are knowledge-based, 11 out of 227 (5\%) articles have proposed solutions without any empirical evaluation.

\begin{figure}
\centering
\includegraphics[width=1.0\linewidth]{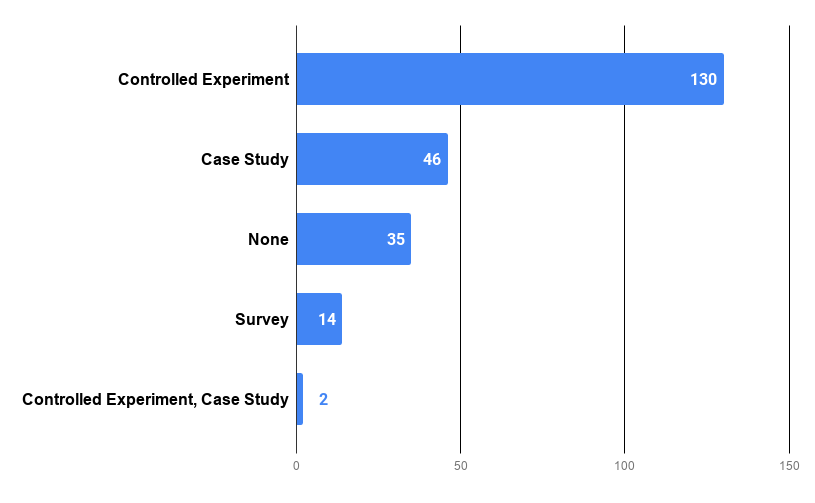}
% \centerline{}
\caption{Articles by Evaluation Facet}
\label{Articles_by_Evaluation_Facet}
\end{figure}

The evaluation facet represents the type of evaluation that has been performed in the articles in order to evaluate the propositions. The articles by the evaluation facet are shown in Fig.~\ref{Articles_by_Evaluation_Facet}. Controlled Experiments have been performed in 130 out 227 (57\%) articles followed by Case Studies in 46 out of 227 (20\%) articles and Surveys in 14 out of 227 (6\%) articles. 2 out of 227 (1\%) articles have employed both a controlled experiment and a case study for an empirical evaluation; whereas, rest of the articles -- 35 out of 227 (15\%) -- did not use any empirical method for evaluation purposes. Moreover, we found no article employing ethnography or action research as empirical methods for evaluation. Among the articles those performed control experiments, 63 articles proposed approaches\slash{}techniques\slash{}methods and 36 articles proposed models\slash{}frameworks.
% This shows the practicality of the propositions in the area is significantly large.
%We have also discussed the association of contribution and research facets with the evaluation facet. Fig. \ref{AI-SE_bubble} shows the association of Contribution/Research facet with the Evaluation facet. 

\begin{figure}
\centering
\includegraphics[width=1.0\linewidth]{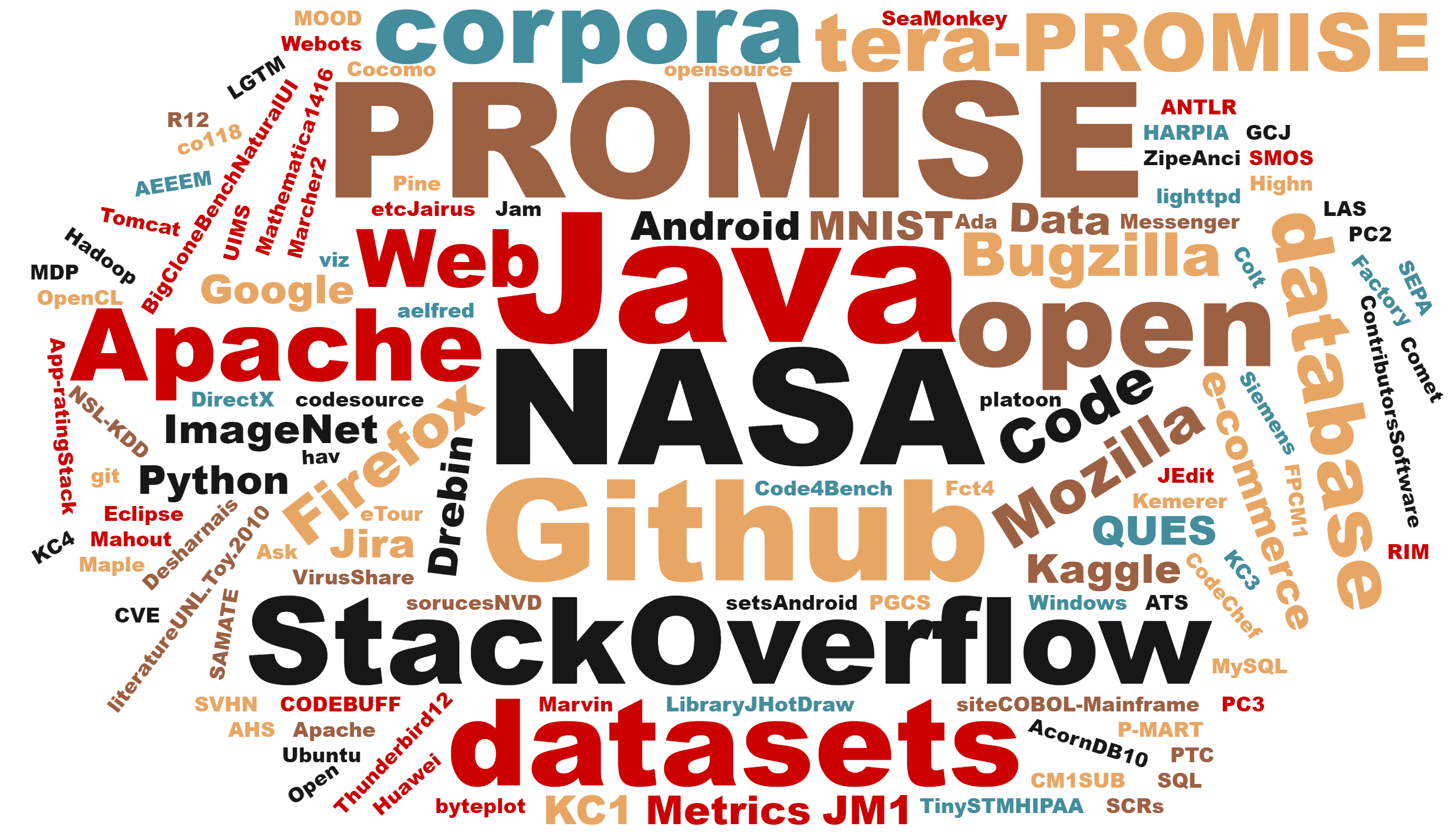}
% \centerline{}
\caption{Datasets Word Cloud}
\label{wordcloud}
\end{figure}

%\begin{figure*}[htbp]
%\includegraphics[width=0.8\linewidth]{AI-SE_bubble.png}
%\centerline{}
%\caption{Association of Contribution/Research Facet with Evaluation Facet as bubble plot}
%\label{AI-SE_bubble}
%\end{figure*}
\textbf{Q2.3} Datasets:
This question refers to the datasets that have been used in most of the articles in order to evaluate their proposed approaches or comparative studies. Evidently, wide spread of articles employed JAVA applications followed by repositories made publicly available by NASA\footnote{\url{https://data.nasa.gov/}}. StackOverflow\footnote{\url{https://archive.org/details/stackexchange}}, Github\footnote{\url{https://ghtorrent.org/}} and Promise\footnote{\url{http://promise.site.uottawa.ca/SERepository/datasets-page.html}} repositories have also been addressed in various studies. 
Fig.~\ref{wordcloud} shows the word cloud for datasets that have been most commonly used in the articles. The size of the terms indicates their frequency in the articles. The greater the size, the more number of occurrences (appearances) in the articles.

\begin{figure}
\centering
\includegraphics[width=1.0\linewidth]{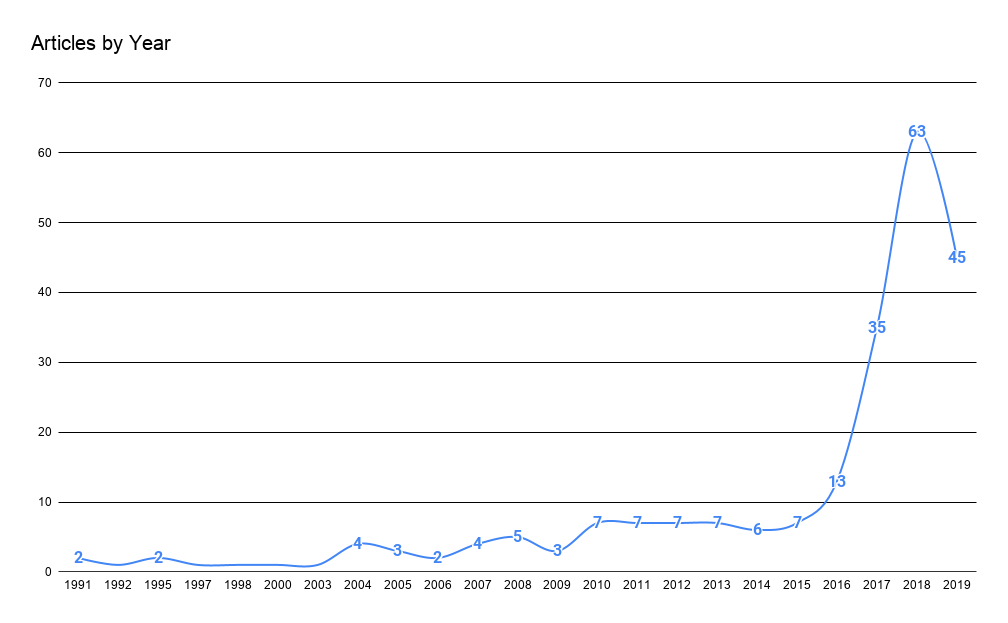}
% \centerline{}
\caption{Articles by Year}
\label{Articles_by_Year}
\end{figure}

\textbf{Q3.1} Trends in terms of year:
This refers to the trends in terms of publication years of articles. It shows the evolution of the adoption of ML for SE. As shown in Fig.~\ref{Articles_by_Year}, the use of ML for SE is consistently growing. One can also observe an exponential growth in this trend from 2016 - 2018, where 2018 proved to be the highest publication year with 63 (28\%) publications. In 2019, we recorded relatively less publications: 45 out of 227 (20\%). There could be two plausible reasons for that. Either some articles are still in press (as this study was conducted in the start of 2020) or like any hype cycle, the peak of inflated expectations regarding ML for SE was reached in 2018 and now the trend is slowly going towards the trough of disillusionment.   %followed by 2019 and 2017 with  and 35 out of 227 (15\%) articles respectively. Apparently. the reason for relatively less number of articles in year 2019 as compared to 2018 shows the hype cycle of ML for SE as it grows to maturity.
% study was conducted in the last quarter of 2019 therefore most of the articles were still under review and yet to be published as of this writing.

\begin{figure}
\centering
\includegraphics[width=1.0\linewidth]{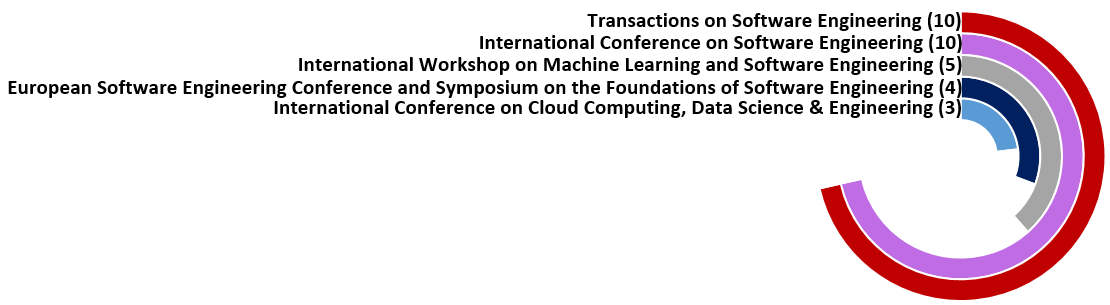}
% \centerline{}
\caption{Articles by Venues (Top 5)}
\label{Top_5_Venues}
\end{figure}

\begin{figure}
\centering
\includegraphics[width=1.0\linewidth]{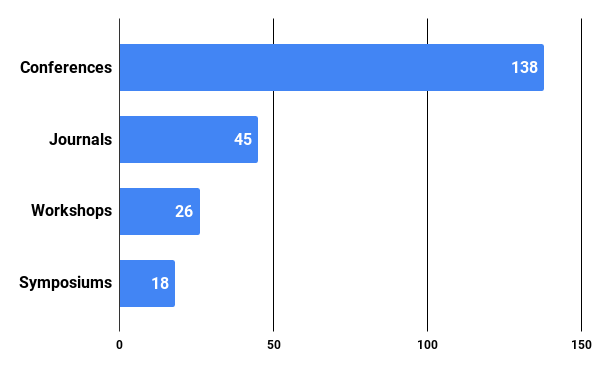}
% \centerline{}
\caption{Articles by publishing venues}
\label{PublishingVenues}
\end{figure}

\textbf{Q3.2} Venues with highest publications:
Fig.~\ref{Top_5_Venues} shows the top 5 venues where most researchers of the domain tend to publish. International Conference on Software Engineering (ICSE) and Transactions on Software Engineering (TSE) are leading by 10 out of 227 (4\%) articles each. They are followed by International Workshop on Machine Learning and Software Engineering, which featured 5 out of 227 (2\%) articles, European Software Engineering Conference and Symposium on the Foundations of Software Engineering (ESEC/FSE), which featured 4 out of 227 (2\%) articles, and International Conference on Cloud Computing, Data Science \& Engineering (Confluence), which featured 3 out of 227 (1\%) articles. Moreover, Fig.~\ref{PublishingVenues} shows the overall distribution of articles with respect to publishing venues. 138 out of 227 (61\%) articles have been published in conferences while 45 out of 227 (20\%) articles are published in journals whereas 26 out 227 (11\%) articles belong to workshops and 18 out of 227 (8\%) articles belong to symposiums.

\section{Discussion}\label{Discussion}
This section relates to the fourth goal of this study (G4) and deals with implications and analysis of the aforementioned articles. Here, we elaborate the challenges, limitations and future directions in this field.
%Traditional Software Engineering stages are being employed in major areas of software development, however, these stages require intense human intervention to complete. As humans are ultimately the adjudicators in achieving the correctness and completeness of the particular software engineering stage. This perception has become somewhat obsolete as machine learning has proven to be highly successful in constraint solving \cite{Briand2008,Oberta2018} e.g., generating test inputs and cases by matching pre-defined patterns, analysing requirement specifications \cite{Sharma2014}, generating requirement specification form source code \cite{Thummalapenta2009} and assisting humans in debugging \cite{Wang2012}.

Quality Assurance and Analytics (52\%), being the SE stage with the most number of ML-related articles, shows that software quality is of prime focus for the researchers, while Architecture and Design, Implementation, and Requirements stages being the second and third highest targeted stages, respectively. Quality Assurance, Design, and Requirements are indeed human-centric stages of the SE life cycle and the high number of articles highlight the fact that ML is able to address the problems in these area. %Consequently, the use ML for software testing~\cite{Durelli2019} and software estimation~\cite{Sharma2017,Wen2012} -- a subset of Quality Assurance -- has also greatly evolved over the years. Furthermore, ML has also greatly reduced the efforts of performing various SE stages as observed in~\cite{Harman2012,Park1997,Xie2013b}.
To get a better understanding of the distribution of articles, we classified them as a MLSE taxonomy. The proposed taxonomy helps in understanding the general categories, which encapsulate the applications of ML specifically aiming at facilitating SE stages in literature. This also shows which stages are being covered the most and which (might) need more exploration. As can be seen in Table~\ref{Classification_by_Articles}, Fault/Bug/Defect Prediction has been the major focus as most articles emphasized on it. %considering its utmost importance in safety critical systems. 
We believe that the rest of the stages need the same kind of attention by the researchers in order to collectively improve the entire life cycle of SE. Looking at the MLSE taxonomy as shown in Table~\ref{Classification_by_Articles}, one can figure out that the  Maintenance stage has been the least interesting area for the researchers. We encourage researchers to investigate how ML can be used to automate certain tasks in this area. We further encourage researchers to adopt combinations of ML techniques and use diverse datasets from different sources in order to train the ML models so that the applicability of the techniques can be generalized as also observed in~\cite{Liu2017,Mizuno2010,Shin2005,Singh2008}. 

%Having said that, there is a lot of grey area that needs to be explored in this particular merger, e.g., 
We figured out that only 4 out of 227 (2\%) articles used reinforcement learning as shown in Fig.~\ref{Articles_by_Machine_learning_Type}. This implies a little interest of researchers in the applications of reinforcement learning to SE. Reinforcement learning has proven to be beneficial in solving complex problems specially in healthcare, business and robotics~\cite{Francois-Lavet2018}. Thus, we believe it would be an interesting area to explore in terms of facilitating SE. Our findings also show that simple neural networks (39 out of 227 (17\%)) and shallow neural networks (containing one or more hidden layers) (35 out of 227 (15\%)) are the most widely used ML techniques in SE, in general. % which shows that deep learning has a moderate applicability in the area of SE.  
Moreover, Boosting, Naive Bayes (NB) and Case-Based Ranking (CBRank) techniques were popular in requirements engineering, particularly.

192 out of 227 (85\%) articles suggest that evidence-based research is a focus of researchers of this domain. Moreover, the high number of controlled experiments (130 out of 227 articles (57\%)) implies that the propositions are being compared to the benchmarks and overall the research is progressing evidently. The demographics also suggest that the interest of the researchers is rapidly growing in this area.

Addressing the fourth goal mentioned in Section~\ref{Goals}, many researchers also reported the uncertain and stochastic nature of their approaches, and the difference in the captured data and results, e.g., difference in the deep learning model output values when executing it multiple times over the same input data~\cite{Brun2004,Dwarakanath2018}. Researchers also found that the availability of  sufficiently labeled and structured dataset is quite a challenge~\cite{Lal2017a,Lal2017b,Rana2003}. Moreover, imbalanced sizes of software projects and datasets were also pointed out to be major obstacles in evaluating the techniques empirically~\cite{Ghaffarian2017,Thaller2019}.
Lack of generalizability and overfitting problems appeared to be the highest limitation in the articles as the ML models have shown fewer results when applied to diverse cross-project datasets~\cite{Madera2017a,Ni2017}.
Future directions include improvement of precision while maintaining recall in ML models~\cite{Ghaffarian2017}. Researchers also emphasized on improving prediction accuracy of the ML model by conducting more experiments using larger numbers of datasets and software applications~\cite{Liu2017,Mizuno2010,Shin2005,Singh2008}. Furthermore, evaluation of similar studies with alternate ML techniques are suggested by researchers, which can further strengthen the knowledge base in terms of prediction capability~\cite{Alonso2010,Clemente2018,Guo2017,Sharma2017}.
%In this study, we have identified the key aspects regarding the amalgamation of ML and SE that would allow the readers to understand the challenges and limitations of applying ML techniques in SE and motivate the academic as well as professional researchers to explore the solutions of these challenges and further the research by producing applicable and highly generalizable techniques and meaningful insights.

\section{Threats to Validity}\label{TTV}
Similar to other secondary studies, the study is also prone to some validity threats. The threats and their mitigation strategies are described in this section. 
\subsection{Internal Validity}
The extraction of articles and choice of repositories constitute a threat to internal validity. Moreover, the screening of articles and the risk of our bias also make the study prone to this type of validity threat. To overcome the internal validity threat, we ensured that our search strategy yielded relevant articles through an iterative refinement of the query.  Each article was reviewed by the first author of this study, which may lead to a threat to the reliability of the results. This threat was reduced by double checking the article by the second author. In order to prevent the risk of bias, the articles underwent our defined QAP in a randomly distributed fashion. The fewer additional articles found through snowballing suggest that we succeeded in devising a robust query.
\subsection{External Validity}
We believe that the wide scope of our query formulation and the stringent exclusion/inclusion criteria has yielded a wide variety of articles that represent a significant and sufficient part of the research area, thus eliminating the generalizability threat to a significant extent.
\subsection{Construct Validity}
The adopted research methodology and protocol, and the data extraction process followed in this study is entirely based on established secondary study guidelines, such as~\cite{KITCHENHAM20097,Petersen2008, Petersen2015}, which reduce the threat to construct validity.
%\subsection{Conclusion Validity}
%The comprehensive and extensive search strategy of the study makes the results highly reproducible and accurate as of date the study has been conducted thus overcoming the conclusion validity threat.
\section{Conclusion}\label{Conclusion}
The conclusion of the study is manifold. We have provided an overview of the state-of-the-art in the area of machine learning for software engineering by evaluating carefully selected studies. We also proposed a classification scheme in the form of the MLSE (machine learning for software engineering) taxonomy that highlights the overall applications of machine learning for software engineering in terms of SE life cycle stages. The taxonomy shows the primary focus of researchers towards specific stages. This observation is one of the major contributions of this study. This study also reveals that the quality of primary studies in the domain of ML and SE is evidence-based with respect to the techniques being empirically evaluated by the researchers. Although, this research area is still showing an upward trend in terms of number of publications, further primary studies need to be conducted to emphasize on other lesser explored SE life cycle stages such as requirements engineering, maintenance and cost estimation. 

The challenges faced by the researchers and reported in the articles should motivate and further guide researchers. These challenges also indicate the presence of known and unknown obstacles that researchers have come across or have not been able to solve while conducting their research – implying a potential in ML for SE with obstacles in term of usefulness. Limitations pointed out by the articles show an inclination of not having enough resources or not being able to overcome certain aspects present in the domain considering the domain is still in its infancy. We also believe that this study provides the necessary impetus and further motivation to explore areas, which have been given lesser attention till date.

% \appendix
% \section{Appendix A}
% \verb+\printcredits+ command is used after appendix sections to list 
% author credit taxonomy contribution roles tagged using \verb+\credit+ in frontmatter.

\printcredits

\bibliographystyle{cas-model2-names}

\bibliography{AI-SE}

\begin{thebibliography}{240}
\expandafter\ifx\csname natexlab\endcsname\relax\def\natexlab#1{#1}\fi
\providecommand{\url}[1]{\texttt{#1}}
\providecommand{\href}[2]{#2}
\providecommand{\path}[1]{#1}
\providecommand{\DOIprefix}{doi:}
\providecommand{\ArXivprefix}{arXiv:}
\providecommand{\URLprefix}{URL: }
\providecommand{\Pubmedprefix}{pmid:}
\providecommand{\doi}[1]{\href{http://dx.doi.org/#1}{\path{#1}}}
\providecommand{\Pubmed}[1]{\href{pmid:#1}{\path{#1}}}
\providecommand{\bibinfo}[2]{#2}
\ifx\xfnm\relax \def\xfnm[#1]{\unskip,\space#1}\fi
%Type = Inproceedings
\bibitem[{Abbineni and Thalluri(2018)}]{Abbineni2018}
\bibinfo{author}{Abbineni, J.}, \bibinfo{author}{Thalluri, O.},
  \bibinfo{year}{2018}.
\newblock \bibinfo{title}{{Software Defect Detection Using Machine Learning
  Techniques}}, in: \bibinfo{booktitle}{2nd International Conference on Trends
  in Electronics and Informatics}, \bibinfo{publisher}{IEEE}. pp.
  \bibinfo{pages}{471--475}.
%Type = Inproceedings
\bibitem[{Ahn and Kim(2018)}]{Ahn2018}
\bibinfo{author}{Ahn, S.}, \bibinfo{author}{Kim, J.}, \bibinfo{year}{2018}.
\newblock \bibinfo{title}{{Poster : A Novel Shared Memory Framework for
  Distributed Deep Learning in High-Performance Computing Architecture}}, in:
  \bibinfo{booktitle}{2018 ACM/IEEE 40th International Conference on Software
  Engineering: Companion Proceedings Poster:}, pp. \bibinfo{pages}{191--192}.
\newblock \href{http://arxiv.org/abs/arXiv:1711.05979v2}{\tt
  arXiv:arXiv:1711.05979v2}.
%Type = Inproceedings
\bibitem[{Ahsan et~al.(2009)Ahsan, Ferzund and Wotawa}]{Ahsan2009}
\bibinfo{author}{Ahsan, S.N.}, \bibinfo{author}{Ferzund, J.},
  \bibinfo{author}{Wotawa, F.}, \bibinfo{year}{2009}.
\newblock \bibinfo{title}{{Automatic Classification of Software Change Request
  Using Multi-Label Machine Learning Methods}}, in: \bibinfo{booktitle}{2009
  33rd Annual IEEE Software Engineering Workshop}, \bibinfo{publisher}{IEEE}.
\newblock \DOIprefix\doi{10.1109/SEW.2009.15}.
%Type = Inproceedings
\bibitem[{Ahsan and Wotawa(2010)}]{Ahsan2010}
\bibinfo{author}{Ahsan, S.N.}, \bibinfo{author}{Wotawa, F.},
  \bibinfo{year}{2010}.
\newblock \bibinfo{title}{{Impact analysis of SCRs using single and multi-label
  machine learning classification}}, in: \bibinfo{booktitle}{International
  Symposium on Empirical Software Engineering and Measurement}.
\newblock \DOIprefix\doi{10.1145/1852786.1852851}.
%Type = Inproceedings
\bibitem[{{Al Asheeri} and Hammad(2019)}]{AlAsheeri2019}
\bibinfo{author}{{Al Asheeri}, M.M.}, \bibinfo{author}{Hammad, M.},
  \bibinfo{year}{2019}.
\newblock \bibinfo{title}{{Machine learning models for software cost
  estimation}}, in: \bibinfo{booktitle}{2019 International Conference on
  Innovation and Intelligence for Informatics, Computing, and Technologies,
  3ICT 2019}, \bibinfo{publisher}{IEEE}. pp. \bibinfo{pages}{1--6}.
\newblock \DOIprefix\doi{10.1109/3ICT.2019.8910327}.
%Type = Inproceedings
\bibitem[{Al-jamimi and Ahmed(2013)}]{Al-jamimi2013}
\bibinfo{author}{Al-jamimi, H.A.}, \bibinfo{author}{Ahmed, M.},
  \bibinfo{year}{2013}.
\newblock \bibinfo{title}{{Machine Learning-based Software Quality Prediction
  Models : State of the Art}}, in: \bibinfo{booktitle}{2013 International
  Conference on Information Science and Applications (ICISA)},
  \bibinfo{publisher}{IEEE}. pp. \bibinfo{pages}{1--4}.
\newblock \DOIprefix\doi{10.1109/ICISA.2013.6579473}.
%Type = Article
\bibitem[{Al-Jamimi and Ahmed(2013)}]{Al-Jamimi2013b}
\bibinfo{author}{Al-Jamimi, H.A.}, \bibinfo{author}{Ahmed, M.A.},
  \bibinfo{year}{2013}.
\newblock \bibinfo{title}{{Machine learning approaches for predicting software
  maintainability: a fuzzy-based transparent model}}.
\newblock \bibinfo{journal}{IET Software} \bibinfo{volume}{7},
  \bibinfo{pages}{317--326}.
\newblock \DOIprefix\doi{10.1049/iet-sen.2013.0046}.
%Type = Inproceedings
\bibitem[{Alahmadi et~al.(2018)Alahmadi, Hassel, Parajuli, Haiduc and
  Kumar}]{Alahmadi}
\bibinfo{author}{Alahmadi, M.}, \bibinfo{author}{Hassel, J.},
  \bibinfo{author}{Parajuli, B.}, \bibinfo{author}{Haiduc, S.},
  \bibinfo{author}{Kumar, P.}, \bibinfo{year}{2018}.
\newblock \bibinfo{title}{{Accurately Predicting the Location of Code Fragments
  in Programming Video Tutorials Using Deep Learning}}, in:
  \bibinfo{booktitle}{Proceedings of the 14th International Conference on
  Predictive Models and Data Analytics in Software Engineering}.
%Type = Inproceedings
\bibitem[{Almeida et~al.(1998)Almeida, Mont, Box, Ave and On}]{Almeida1998}
\bibinfo{author}{Almeida, M.A.D.}, \bibinfo{author}{Mont, D.},
  \bibinfo{author}{Box, P.O.}, \bibinfo{author}{Ave, M.C.},
  \bibinfo{author}{On, K.I.N.}, \bibinfo{year}{1998}.
\newblock \bibinfo{title}{{An Investigation on the Use of Machine Learned
  Models for Estimating Correction Costs}}, in: \bibinfo{booktitle}{Proceedings
  of the 20th international conference on Software engineering}, pp.
  \bibinfo{pages}{473--476}.
%Type = Inproceedings
\bibitem[{Alonso(2011)}]{Alonso2011}
\bibinfo{author}{Alonso, J.}, \bibinfo{year}{2011}.
\newblock \bibinfo{title}{{Predicting Software Anomalies using Machine Learning
  Techniques}}, in: \bibinfo{booktitle}{IEEE International Symposium on Network
  Computing and Applications Predicting}.
\newblock \DOIprefix\doi{10.1109/NCA.2011.29}.
%Type = Inproceedings
\bibitem[{Alonso and Torres(2010)}]{Alonso2010}
\bibinfo{author}{Alonso, J.}, \bibinfo{author}{Torres, J.},
  \bibinfo{year}{2010}.
\newblock \bibinfo{title}{{Adaptive on-line software aging prediction based on
  Machine Learning}}, in: \bibinfo{booktitle}{IEEEIIFIP International
  Conference on Dependable Systems {\&} Networks}, pp.
  \bibinfo{pages}{507--516}.
%Type = Article
\bibitem[{Alotaibi(2019)}]{Alotaibi2019}
\bibinfo{author}{Alotaibi, A.}, \bibinfo{year}{2019}.
\newblock \bibinfo{title}{{Identifying Malicious Software Using Deep Residual
  Long-Short Term Memory}}.
\newblock \bibinfo{journal}{IEEE Access} \bibinfo{volume}{7},
  \bibinfo{pages}{163128--163137}.
\newblock \DOIprefix\doi{10.1109/ACCESS.2019.2951751}.
%Type = Inproceedings
\bibitem[{Alshehri et~al.(2018)Alshehri, Goseva-popstojanova, Dzielski, Devine
  and Virginia}]{Alshehri2018}
\bibinfo{author}{Alshehri, Y.A.}, \bibinfo{author}{Goseva-popstojanova, K.},
  \bibinfo{author}{Dzielski, D.G.}, \bibinfo{author}{Devine, T.},
  \bibinfo{author}{Virginia, W.}, \bibinfo{year}{2018}.
\newblock \bibinfo{title}{{Applying machine learning to predict software fault
  proneness using change metrics , static code metrics , and a combination of
  them}}, in: \bibinfo{booktitle}{SoutheastCon}, \bibinfo{publisher}{IEEE}. pp.
  \bibinfo{pages}{1--7}.
%Type = Inproceedings
\bibitem[{Andrzejak et~al.(2008)Andrzejak, Silva and
  Inform{\'{a}}tica}]{Andrzejak2008}
\bibinfo{author}{Andrzejak, A.}, \bibinfo{author}{Silva, L.},
  \bibinfo{author}{Inform{\'{a}}tica, D.E.}, \bibinfo{year}{2008}.
\newblock \bibinfo{title}{{Using Machine Learning for Non-Intrusive Modeling
  and Prediction of Software Aging}}, in: \bibinfo{booktitle}{NOMS 2008 - 2008
  IEEE Network Operations and Management Symposium}, \bibinfo{publisher}{IEEE}.
  pp. \bibinfo{pages}{25--32}.
\newblock \URLprefix \url{http://ieeexplore.ieee.org/document/4575113/},
  \DOIprefix\doi{10.1109/NOMS.2008.4575113}.
%Type = Inproceedings
\bibitem[{Ayesha and Yethiraj(2018)}]{Ayesha2018}
\bibinfo{author}{Ayesha, N.}, \bibinfo{author}{Yethiraj, N.G.},
  \bibinfo{year}{2018}.
\newblock \bibinfo{title}{{Review on Code Examination Proficient System in
  Software Engineering by Using Machine Learning Approach}}, in:
  \bibinfo{booktitle}{Proceedings of the International Conference on Inventive
  Research in Computing Applications}, \bibinfo{publisher}{IEEE}. pp.
  \bibinfo{pages}{324--327}.
%Type = Inproceedings
\bibitem[{{Azadi, Umbarto, Fontana, Arcelli, Francesca, Zanoni}(2018)}]{Q2018}
\bibinfo{author}{{Azadi, Umbarto, Fontana, Arcelli, Francesca, Zanoni}, M.},
  \bibinfo{year}{2018}.
\newblock \bibinfo{title}{{Poster: Machine Learning Based Code Smell Detection
  Through WekaNose}}, in: \bibinfo{booktitle}{ACM/IEEE 40th International
  Conference on Software Engineering: Companion Proceedings}, pp.
  \bibinfo{pages}{29--32}.
%Type = Inproceedings
\bibitem[{B et~al.(2016)B, Meinke and Rausch}]{B2016}
\bibinfo{author}{B, F.H.}, \bibinfo{author}{Meinke, K.},
  \bibinfo{author}{Rausch, A.}, \bibinfo{year}{2016}.
\newblock \bibinfo{title}{{Learning Systems : Machine-Learning in Software
  Products and Learning-Based Analysis of Software Systems}}, in:
  \bibinfo{booktitle}{International Symposium on Leveraging Applications of
  Formal Methods}, pp. \bibinfo{pages}{651--654}.
\newblock \DOIprefix\doi{10.1007/978-3-319-47169-3}.
%Type = Article
\bibitem[{B et~al.(2013)B, Raj, Bosch and Holmstr}]{B2013a}
\bibinfo{author}{B, L.E.L.}, \bibinfo{author}{Raj, A.}, \bibinfo{author}{Bosch,
  J.}, \bibinfo{author}{Holmstr, H.}, \bibinfo{year}{2013}.
\newblock \bibinfo{title}{{A Taxonomy of Software Engineering Challenges for
  Machine Learning Systems: An Empirical Investigation}}.
\newblock \bibinfo{journal}{International Conference on Agile Software
  Development} \bibinfo{volume}{149}, \bibinfo{pages}{227--243}.
\newblock \URLprefix \url{http://link.springer.com/10.1007/978-3-642-38314-4},
  \DOIprefix\doi{10.1007/978-3-642-38314-4}.
%Type = Inproceedings
\bibitem[{B et~al.(2019)B, Buchgeher, Klammer, Pfeiffer, Salomon, Thaller and
  Linsbauer}]{B2019a}
\bibinfo{author}{B, R.R.}, \bibinfo{author}{Buchgeher, G.},
  \bibinfo{author}{Klammer, C.}, \bibinfo{author}{Pfeiffer, M.},
  \bibinfo{author}{Salomon, C.}, \bibinfo{author}{Thaller, H.},
  \bibinfo{author}{Linsbauer, L.}, \bibinfo{year}{2019}.
\newblock \bibinfo{title}{{Improving Defect Localization by Classifying the
  Affected Asset Using Machine Learning}}, in:
  \bibinfo{booktitle}{International Conference on Software Quality},
  \bibinfo{publisher}{Springer International Publishing}. pp.
  \bibinfo{pages}{125--148}.
\newblock \URLprefix \url{http://link.springer.com/10.1007/978-3-030-05767-1},
  \DOIprefix\doi{10.1007/978-3-030-05767-1}.
%Type = Inproceedings
\bibitem[{Ba and Turhan(2007)}]{Ba2007}
\bibinfo{author}{Ba, B.}, \bibinfo{author}{Turhan, B.}, \bibinfo{year}{2007}.
\newblock \bibinfo{title}{{Software Effort Estimation Using Machine Learning
  Methods}}, in: \bibinfo{booktitle}{22nd international symposium on computer
  and information sciences}, \bibinfo{publisher}{IEEE}. pp.
  \bibinfo{pages}{1--6}.
%Type = Inproceedings
\bibitem[{Babamir et~al.(2010)Babamir, Hatamizadeh and Babamir}]{Babamir}
\bibinfo{author}{Babamir, F.S.}, \bibinfo{author}{Hatamizadeh, A.},
  \bibinfo{author}{Babamir, S.M.}, \bibinfo{year}{2010}.
\newblock \bibinfo{title}{{Application of Genetic Algorithm in Automatic
  Software Testing}}, in: \bibinfo{booktitle}{International Conference on
  Networked Digital Technologies}, pp. \bibinfo{pages}{545--552}.
%Type = Article
\bibitem[{Banerjee et~al.(2013)Banerjee, Nguyen, Garousi and
  Memon}]{Banerjee2013}
\bibinfo{author}{Banerjee, I.}, \bibinfo{author}{Nguyen, B.},
  \bibinfo{author}{Garousi, V.}, \bibinfo{author}{Memon, A.},
  \bibinfo{year}{2013}.
\newblock \bibinfo{title}{{Graphical user interface (GUI) testing: Systematic
  mapping and repository}}.
\newblock \bibinfo{journal}{Information and Software Technology}
  \bibinfo{volume}{55}, \bibinfo{pages}{1679--1694}.
\newblock \URLprefix \url{http://dx.doi.org/10.1016/j.infsof.2013.03.004},
  \DOIprefix\doi{10.1016/j.infsof.2013.03.004}.
%Type = Inproceedings
\bibitem[{Banimustafa(2018)}]{Banimustafa2018}
\bibinfo{author}{Banimustafa, A.}, \bibinfo{year}{2018}.
\newblock \bibinfo{title}{{Predicting Software Effort Estimation Using Machine
  Learning Techniques}}, in: \bibinfo{booktitle}{8th International Conference
  on Computer Science and Information Technology}, \bibinfo{publisher}{IEEE}.
  pp. \bibinfo{pages}{249--256}.
%Type = Article
\bibitem[{Basili et~al.(1994)Basili, Caldiera and Rombach}]{Basili}
\bibinfo{author}{Basili, V.R.}, \bibinfo{author}{Caldiera, G.},
  \bibinfo{author}{Rombach, H.D.}, \bibinfo{year}{1994}.
\newblock \bibinfo{title}{{The goal question metric approach}}.
\newblock \bibinfo{journal}{Encyclopedia of software engineering}
  \bibinfo{volume}{2}, \bibinfo{pages}{528--532.}
\newblock \URLprefix
  \url{http://fub-taslim.googlecode.com/svn/trunk/WEMSE/INSTICC{\_}Conference{\_}Latex/gqm.pdf}.
%Type = Inproceedings
\bibitem[{Baskiotis et~al.(2006)Baskiotis, Gaudel and Gouraud}]{Baskiotis2006}
\bibinfo{author}{Baskiotis, N.}, \bibinfo{author}{Gaudel, M.c.},
  \bibinfo{author}{Gouraud, S.}, \bibinfo{year}{2006}.
\newblock \bibinfo{title}{{A Machine Learning Approach for Statistical Software
  Testing}}, in: \bibinfo{booktitle}{IJCAI International Joint Conference on
  Artificial Intelligence}, pp. \bibinfo{pages}{2274--2279}.
%Type = Inproceedings
\bibitem[{Bhandari and Gupta(2018a)}]{Bhandari2018b}
\bibinfo{author}{Bhandari, G.P.}, \bibinfo{author}{Gupta, R.},
  \bibinfo{year}{2018}a.
\newblock \bibinfo{title}{{Machine learning based software fault prediction
  utilizing source code metrics}}, in: \bibinfo{booktitle}{IEEE 3rd
  International Conference on Computing, Communication and Security},
  \bibinfo{publisher}{IEEE}. pp. \bibinfo{pages}{40--45}.
%Type = Inproceedings
\bibitem[{Bhandari and Gupta(2018b)}]{Bhandari2018}
\bibinfo{author}{Bhandari, G.P.}, \bibinfo{author}{Gupta, R.},
  \bibinfo{year}{2018}b.
\newblock \bibinfo{title}{{Measuring the Fault Predictability of Software using
  Deep Learning Techniques with Software Metrics}}, in: \bibinfo{booktitle}{5th
  IEEE Uttar Pradesh Section International Conference on Electrical, Computer
  and Electronics}, \bibinfo{publisher}{IEEE}.
%Type = Article
\bibitem[{Bharathi and Selvarani(2019)}]{Bharathi2019}
\bibinfo{author}{Bharathi, R.}, \bibinfo{author}{Selvarani, R.},
  \bibinfo{year}{2019}.
\newblock \bibinfo{title}{{A Machine Learning Approach for Quantifying the
  Design Error Propagation in Safety Critical Software System}}.
\newblock \bibinfo{journal}{IETE Journal of Research} \bibinfo{volume}{0},
  \bibinfo{pages}{1--15}.
\newblock \URLprefix \url{https://doi.org/03772063.2019.1611490},
  \DOIprefix\doi{10.1080/03772063.2019.1611490}.
%Type = Inproceedings
\bibitem[{Bisio et~al.(2014)Bisio, Gastaldo, Zunino and Decherchi}]{Bisio2014}
\bibinfo{author}{Bisio, F.}, \bibinfo{author}{Gastaldo, P.},
  \bibinfo{author}{Zunino, R.}, \bibinfo{author}{Decherchi, S.},
  \bibinfo{year}{2014}.
\newblock \bibinfo{title}{{Semi-supervised machine learning approach for
  unknown malicious software detection}}, in: \bibinfo{booktitle}{International
  Symposium on Innovations in Intelligent Systems and Applications},
  \bibinfo{publisher}{IEEE}.
%Type = Inproceedings
\bibitem[{Borg et~al.(2019)Borg, Svensson, Berg and Hansson}]{Borg2019}
\bibinfo{author}{Borg, M.}, \bibinfo{author}{Svensson, O.},
  \bibinfo{author}{Berg, K.}, \bibinfo{author}{Hansson, D.},
  \bibinfo{year}{2019}.
\newblock \bibinfo{title}{{SZZ unleashed: an open implementation of the SZZ
  algorithm - featuring example usage in a study of just-in-time bug prediction
  for the Jenkins project}}, in: \bibinfo{booktitle}{3rd ACM SIGSOFT
  International Workshop on Machine Learning Techniques for Software Quality
  Evaluation}, pp. \bibinfo{pages}{7--12}.
\newblock \DOIprefix\doi{10.1145/3340482.3342742},
  \href{http://arxiv.org/abs/1903.01742}{\tt arXiv:1903.01742}.
%Type = Inproceedings
\bibitem[{Braga et~al.(2007)Braga, Oliveira and Meira}]{Braga2007}
\bibinfo{author}{Braga, P.L.}, \bibinfo{author}{Oliveira, A.L.I.},
  \bibinfo{author}{Meira, S.R.L.}, \bibinfo{year}{2007}.
\newblock \bibinfo{title}{{Software Effort Estimation using Machine Learning
  Techniques with Robust Confidence Intervals}}, in:
  \bibinfo{booktitle}{Seventh International Conference on Hybrid Intelligent
  Systems}, pp. \bibinfo{pages}{352--357}.
\newblock \DOIprefix\doi{10.1109/HIS.2007.56}.
%Type = Inproceedings
\bibitem[{Braga et~al.(2018)Braga, Neto, Rab{\^{e}}lo, Santiago and
  Souza}]{Braga2018}
\bibinfo{author}{Braga, R.}, \bibinfo{author}{Neto, P.S.},
  \bibinfo{author}{Rab{\^{e}}lo, R.}, \bibinfo{author}{Santiago, J.},
  \bibinfo{author}{Souza, M.}, \bibinfo{year}{2018}.
\newblock \bibinfo{title}{{A machine learning approach to generate test
  oracles}}, in: \bibinfo{booktitle}{XXXII BRAZILIAN SYMPOSIUM ONSOFTWARE
  ENGINEERING}, pp. \bibinfo{pages}{142--151}.
\newblock \DOIprefix\doi{10.1145/3266237.3266273}.
%Type = Inproceedings
\bibitem[{Briand(2008)}]{Briand2008}
\bibinfo{author}{Briand, L.C.}, \bibinfo{year}{2008}.
\newblock \bibinfo{title}{{Novel Applications of Machine Learning in Software
  Testing}}, in: \bibinfo{booktitle}{Eighth International Conference on Quality
  Software Novel}, \bibinfo{publisher}{IEEE}. pp. \bibinfo{pages}{3--10}.
\newblock \DOIprefix\doi{10.1109/QSIC.2008.29}.
%Type = Inproceedings
\bibitem[{Bruegge et~al.(2009)Bruegge, David, Helming and Koegel}]{Bruegge2009}
\bibinfo{author}{Bruegge, B.}, \bibinfo{author}{David, J.},
  \bibinfo{author}{Helming, J.}, \bibinfo{author}{Koegel, M.},
  \bibinfo{year}{2009}.
\newblock \bibinfo{title}{{Classification of tasks using machine learning}},
  in: \bibinfo{booktitle}{5th International Conference on Predictor Models in
  Software Engineering}, pp. \bibinfo{pages}{1--11}.
\newblock \DOIprefix\doi{10.1145/1540438.1540455}.
%Type = Inproceedings
\bibitem[{Brun and Ernst(2004)}]{Brun2004}
\bibinfo{author}{Brun, Y.}, \bibinfo{author}{Ernst, M.}, \bibinfo{year}{2004}.
\newblock \bibinfo{title}{{Finding latent code errors via machine learning over
  program executions}}, in: \bibinfo{booktitle}{26th International Conference
  on Software Engineering}, pp. \bibinfo{pages}{480--490}.
\newblock \DOIprefix\doi{10.1109/icse.2004.1317470}.
%Type = Inproceedings
\bibitem[{Cambronero et~al.(2019)Cambronero, Kim and Chandra}]{Cambronero}
\bibinfo{author}{Cambronero, J.}, \bibinfo{author}{Kim, S.},
  \bibinfo{author}{Chandra, S.}, \bibinfo{year}{2019}.
\newblock \bibinfo{title}{{When Deep Learning Met Code Search}}, in:
  \bibinfo{booktitle}{ESEC/FSE 2019 - Proceedings of the 2019 27th ACM Joint
  Meeting European Software Engineering Conference and Symposium on the
  Foundations of Software Engineering}, pp. \bibinfo{pages}{964--974}.
%Type = Inproceedings
\bibitem[{Castro-Lopez and Vega-Lopez(2018)}]{Castro-Lopez2018}
\bibinfo{author}{Castro-Lopez, O.}, \bibinfo{author}{Vega-Lopez, I.F.},
  \bibinfo{year}{2018}.
\newblock \bibinfo{title}{{Fast deployment and scoring of support vector
  machine models in CPU and GPU}}, in: \bibinfo{booktitle}{1st International
  Workshop on Machine Learning and Software Engineering in Symbiosis}, pp.
  \bibinfo{pages}{45--52}.
\newblock \DOIprefix\doi{10.1145/3243127.3243133}.
%Type = Inproceedings
\bibitem[{Cerrada et~al.(2017)Cerrada, Cabrera, Macancela, Lucero, Pacheco,
  Sanchez, Cabrera, Macancela and Lucero}]{Cerrada2017}
\bibinfo{author}{Cerrada, M.}, \bibinfo{author}{Cabrera, D.},
  \bibinfo{author}{Macancela, J.}, \bibinfo{author}{Lucero, P.},
  \bibinfo{author}{Pacheco, F.}, \bibinfo{author}{Sanchez, R.V.},
  \bibinfo{author}{Cabrera, D.}, \bibinfo{author}{Macancela, J.},
  \bibinfo{author}{Lucero, P.}, \bibinfo{year}{2017}.
\newblock \bibinfo{title}{{SOA Based Integrated Software to Develop Fault
  Diagnosis Models Using Machine Learning in Rotating Machinery}}, in:
  \bibinfo{booktitle}{Proceedings - 11th IEEE International Symposium on
  Service-Oriented System Engineering, SOSE 2017}, pp. \bibinfo{pages}{28--37}.
\newblock \DOIprefix\doi{10.1109/SOSE.2017.19}.
%Type = Inproceedings
\bibitem[{Ceylan et~al.(2006)Ceylan, Kutlubay and Bener}]{Ceylan2006}
\bibinfo{author}{Ceylan, E.}, \bibinfo{author}{Kutlubay, F.O.},
  \bibinfo{author}{Bener, B.}, \bibinfo{year}{2006}.
\newblock \bibinfo{title}{{Software Defect Identification Using Machine
  Learning Techniques}}, in: \bibinfo{booktitle}{32nd EUROMICRO Conference on
  Software Engineering and Advanced Applications}.
%Type = Inproceedings
\bibitem[{Challagulla et~al.(2005)Challagulla, Bastani, Yen and
  Paul}]{Challagulla2005}
\bibinfo{author}{Challagulla, V.U.B.}, \bibinfo{author}{Bastani, F.B.},
  \bibinfo{author}{Yen, I.l.}, \bibinfo{author}{Paul, R.A.},
  \bibinfo{year}{2005}.
\newblock \bibinfo{title}{{Empirical Assessment of Machine Learning based
  Software Defect Prediction Techniques}}, in: \bibinfo{booktitle}{10th IEEE
  International Workshop on Object-Oriented Real-Time Dependable Systems}.
%Type = Inproceedings
\bibitem[{Chandra and Choudhary(2017)}]{Chandra2017}
\bibinfo{author}{Chandra, D.}, \bibinfo{author}{Choudhary, M.},
  \bibinfo{year}{2017}.
\newblock \bibinfo{title}{{Prophecy of Software Maintenance Effort with
  Univariate and Multivariate approach}}, in: \bibinfo{booktitle}{International
  Conference on Computing, Communication and Automation}, pp.
  \bibinfo{pages}{876--880}.
%Type = Inproceedings
\bibitem[{Chandra et~al.(2016)Chandra, Kapoor, Kohli and Gupta}]{Chandra2016}
\bibinfo{author}{Chandra, K.}, \bibinfo{author}{Kapoor, G.},
  \bibinfo{author}{Kohli, R.}, \bibinfo{author}{Gupta, A.},
  \bibinfo{year}{2016}.
\newblock \bibinfo{title}{{IMPROVING SOFTWARE QUALITY USING MACHINE LEARNING}},
  in: \bibinfo{booktitle}{1st International Conference on Innovation and
  Challenges in Cyber Security}, \bibinfo{publisher}{IEEE}. pp.
  \bibinfo{pages}{115--118}.
%Type = Inproceedings
\bibitem[{Cheatham and Wahl(1995)}]{Cheatham}
\bibinfo{author}{Cheatham, T.J.}, \bibinfo{author}{Wahl, N.J.},
  \bibinfo{year}{1995}.
\newblock \bibinfo{title}{{Software Testing : A Machine Learning}}, in:
  \bibinfo{booktitle}{23rd annual conference on Computer science}.
%Type = Inproceedings
\bibitem[{Chen et~al.(2011)Chen, Hoi and Xiao}]{Chen2011}
\bibinfo{author}{Chen, N.}, \bibinfo{author}{Hoi, S.C.H.},
  \bibinfo{author}{Xiao, X.}, \bibinfo{year}{2011}.
\newblock \bibinfo{title}{{Software Process Evaluation : A Machine Learning
  Approach}}, in: \bibinfo{booktitle}{26th IEEE/ACM International Conference on
  Automated Software Engineering}, \bibinfo{publisher}{IEEE}. pp.
  \bibinfo{pages}{333--342}.
%Type = Article
\bibitem[{Chioaşcǎ(2012)}]{Chioasca2012}
\bibinfo{author}{Chioaşcǎ, E.V.}, \bibinfo{year}{2012}.
\newblock \bibinfo{title}{{Using machine learning to enhance automated
  requirements model transformation}}.
\newblock \bibinfo{journal}{Proceedings - International Conference on Software
  Engineering} ,
  \bibinfo{pages}{1487--1490}\DOIprefix\doi{10.1109/ICSE.2012.6227055}.
%Type = Inproceedings
\bibitem[{Chollet et~al.(2013)Chollet, Lalanda and Bardin}]{Chollet2013}
\bibinfo{author}{Chollet, S.}, \bibinfo{author}{Lalanda, P.},
  \bibinfo{author}{Bardin, J.}, \bibinfo{year}{2013}.
\newblock \bibinfo{title}{{Service-Oriented Computing}}, in:
  \bibinfo{booktitle}{International Conference on Service-Oriented Computing},
  \bibinfo{publisher}{Springer International Publishing}. pp.
  \bibinfo{pages}{1--20}.
\newblock \URLprefix \url{http://dx.doi.org/10.1007/978-3-030-03596-9{\_}28},
  \DOIprefix\doi{10.4018/978-1-61350-159-7.ch001}.
%Type = Inproceedings
\bibitem[{Cleland-Huang et~al.(2010)Cleland-Huang, Czauderna, Gibiec and
  Emenecker}]{Cleland-Huang2010}
\bibinfo{author}{Cleland-Huang, J.}, \bibinfo{author}{Czauderna, A.},
  \bibinfo{author}{Gibiec, M.}, \bibinfo{author}{Emenecker, J.},
  \bibinfo{year}{2010}.
\newblock \bibinfo{title}{{A machine learning approach for tracing regulatory
  codes to product specific requirements}}, in: \bibinfo{booktitle}{32nd
  ACM/IEEE International Conference on Software Engineering}, p.
  \bibinfo{pages}{155}.
\newblock \DOIprefix\doi{10.1145/1806799.1806825}.
%Type = Inproceedings
\bibitem[{Clemente et~al.(2018)Clemente, Jaafar and Malik}]{Clemente2018}
\bibinfo{author}{Clemente, C.J.}, \bibinfo{author}{Jaafar, F.},
  \bibinfo{author}{Malik, Y.}, \bibinfo{year}{2018}.
\newblock \bibinfo{title}{{Is Predicting Software Security Bugs using Deep
  Learning Better than the Traditional Machine Learning Algorithms ?}}, in:
  \bibinfo{booktitle}{IEEE International Conference on Software Quality,
  Reliability and Security Is}, \bibinfo{publisher}{IEEE}.
\newblock \DOIprefix\doi{10.1109/QRS.2018.00023}.
%Type = Article
\bibitem[{Corazza et~al.(2009)Corazza, {Di Martino}, Ferrucci, Gravino and
  Mendes}]{Corazza2009}
\bibinfo{author}{Corazza, A.}, \bibinfo{author}{{Di Martino}, S.},
  \bibinfo{author}{Ferrucci, F.}, \bibinfo{author}{Gravino, C.},
  \bibinfo{author}{Mendes, E.}, \bibinfo{year}{2009}.
\newblock \bibinfo{title}{{Using Support Vector Regression for web development
  effort estimation}}.
\newblock \bibinfo{journal}{Lecture Notes in Computer Science (including
  subseries Lecture Notes in Artificial Intelligence and Lecture Notes in
  Bioinformatics)} \bibinfo{volume}{5891 LNCS}, \bibinfo{pages}{255--271}.
\newblock \DOIprefix\doi{10.1007/978-3-642-05415-0_19}.
%Type = Inproceedings
\bibitem[{Cuadrado-gallego et~al.(2010)Cuadrado-gallego, Rodr{\'{i}}guez-soria
  and Mart{\'{i}}n-herrera}]{Cuadrado-gallego2010}
\bibinfo{author}{Cuadrado-gallego, J.J.},
  \bibinfo{author}{Rodr{\'{i}}guez-soria, P.},
  \bibinfo{author}{Mart{\'{i}}n-herrera, B.}, \bibinfo{year}{2010}.
\newblock \bibinfo{title}{{Analogies and differences between Machine Learning
  and Expert based Software Project Effort Estimation}}, in:
  \bibinfo{booktitle}{11th ACIS International Conference on Software
  Engineering, Artificial Intelligence, Networking and Parallel/Distributed
  Computing}, pp. \bibinfo{pages}{269--276}.
\newblock \DOIprefix\doi{10.1109/SNPD.2010.47}.
%Type = Article
\bibitem[{Cuevas and San-feliu(2014)}]{Cuevas2014}
\bibinfo{author}{Cuevas, J.A.C.m.G.}, \bibinfo{author}{San-feliu, T.},
  \bibinfo{year}{2014}.
\newblock \bibinfo{title}{{Critical success factors taxonomy for software
  process deployment}}.
\newblock \bibinfo{journal}{Software Qual J} \bibinfo{volume}{22},
  \bibinfo{pages}{21--48}.
\newblock \DOIprefix\doi{10.1007/s11219-012-9190-y}.
%Type = Article
\bibitem[{Cummins et~al.(2018)Cummins, Petoumenos, Murray and
  Leather}]{Cummins2018}
\bibinfo{author}{Cummins, C.}, \bibinfo{author}{Petoumenos, P.},
  \bibinfo{author}{Murray, A.}, \bibinfo{author}{Leather, H.},
  \bibinfo{year}{2018}.
\newblock \bibinfo{title}{{Compiler fuzzing through deep learning}}.
\newblock \bibinfo{journal}{Proceedings of the 27th ACM SIGSOFT International
  Symposium on Software Testing and Analysis - ISSTA 2018} ,
  \bibinfo{pages}{95--105}\URLprefix
  \url{http://dl.acm.org/citation.cfm?doid=3213846.3213848},
  \DOIprefix\doi{10.1145/3213846.3213848}.
%Type = Inproceedings
\bibitem[{Dam(2018)}]{Dam2018}
\bibinfo{author}{Dam, H.K.}, \bibinfo{year}{2018}.
\newblock \bibinfo{title}{{Explainable Software Analytics}}, in:
  \bibinfo{booktitle}{40th International Conference on Software Engineering:
  New Ideas and Emerging Results}, \bibinfo{publisher}{ACM}. pp.
  \bibinfo{pages}{53--56}.
%Type = Inproceedings
\bibitem[{Dam et~al.(2019)Dam, Pham, Ng, Tran, Grundy, Ghose, Kim and
  Kim}]{Dam2019}
\bibinfo{author}{Dam, H.K.}, \bibinfo{author}{Pham, T.}, \bibinfo{author}{Ng,
  S.W.}, \bibinfo{author}{Tran, T.}, \bibinfo{author}{Grundy, J.},
  \bibinfo{author}{Ghose, A.}, \bibinfo{author}{Kim, T.}, \bibinfo{author}{Kim,
  C.J.}, \bibinfo{year}{2019}.
\newblock \bibinfo{title}{{Lessons learned from using a deep tree-based model
  for software defect prediction in practice}}, in: \bibinfo{booktitle}{IEEE
  International Working Conference on Mining Software Repositories},
  \bibinfo{publisher}{IEEE}. pp. \bibinfo{pages}{46--57}.
\newblock \DOIprefix\doi{10.1109/MSR.2019.00017}.
%Type = Article
\bibitem[{Dam et~al.(2015)Dam, Tran, Pham, Ng, Grundy and Ghose}]{Dam2015}
\bibinfo{author}{Dam, H.K.}, \bibinfo{author}{Tran, T.}, \bibinfo{author}{Pham,
  T.}, \bibinfo{author}{Ng, S.W.}, \bibinfo{author}{Grundy, J.},
  \bibinfo{author}{Ghose, A.}, \bibinfo{year}{2015}.
\newblock \bibinfo{title}{{Automatic feature learning for predicting vulnerable
  software components}}.
\newblock \bibinfo{journal}{IEEE Transactions on Software Engineering}
  \bibinfo{volume}{14}, \bibinfo{pages}{1--19}.
%Type = Article
\bibitem[{Dawoud et~al.(2018)Dawoud, Shahristani and Raun}]{Dawoud2018}
\bibinfo{author}{Dawoud, A.}, \bibinfo{author}{Shahristani, S.},
  \bibinfo{author}{Raun, C.}, \bibinfo{year}{2018}.
\newblock \bibinfo{title}{{Internet of Things Deep learning and
  software-defined networks : Towards secure IoT architecture}}.
\newblock \bibinfo{journal}{Internet of Things} \bibinfo{volume}{3-4},
  \bibinfo{pages}{82--89}.
\newblock \URLprefix \url{https://doi.org/10.1016/j.iot.2018.09.003},
  \DOIprefix\doi{10.1016/j.iot.2018.09.003}.
%Type = Inproceedings
\bibitem[{Dqg et~al.(2018)Dqg, Vdpudw, Frp, Dkrr, Dqg and Jpdlo}]{Dqga}
\bibinfo{author}{Dqg, X.H.}, \bibinfo{author}{Vdpudw, V.},
  \bibinfo{author}{Frp, J.}, \bibinfo{author}{Dkrr, P.}, \bibinfo{author}{Dqg,
  F.R.P.}, \bibinfo{author}{Jpdlo, P.}, \bibinfo{year}{2018}.
\newblock \bibinfo{title}{{Detection of Flow Based Anomaly in OpenFlow
  Controller : Machine Learning Approach in Software Defined Networking}}, in:
  \bibinfo{booktitle}{4th International Conference on Electrical Engineering
  and Information {\&} Communication Technology}.
%Type = Article
\bibitem[{Durelli et~al.(2019)Durelli, Durelli, Borges, Endo, Eler, Dias,
  Guimar and Guimaraes}]{Durelli2019}
\bibinfo{author}{Durelli, V.H.S.}, \bibinfo{author}{Durelli, R.S.},
  \bibinfo{author}{Borges, S.S.}, \bibinfo{author}{Endo, A.T.},
  \bibinfo{author}{Eler, M.M.}, \bibinfo{author}{Dias, D.R.C.},
  \bibinfo{author}{Guimar, M.P.}, \bibinfo{author}{Guimaraes, M.P.},
  \bibinfo{year}{2019}.
\newblock \bibinfo{title}{{Machine Learning Applied to Software Testing : A
  Systematic Mapping Study}}.
\newblock \bibinfo{journal}{IEEE Transactions on Reliability} ,
  \bibinfo{pages}{1--24}\DOIprefix\doi{10.1109/tr.2019.2892517}.
%Type = Inproceedings
\bibitem[{Dwarakanath et~al.(2018)Dwarakanath, Ahuja, Sikand, Rao, Bose, Dubash
  and Podder}]{Dwarakanath2018}
\bibinfo{author}{Dwarakanath, A.}, \bibinfo{author}{Ahuja, M.},
  \bibinfo{author}{Sikand, S.}, \bibinfo{author}{Rao, R.M.},
  \bibinfo{author}{Bose, R.P.J.C.}, \bibinfo{author}{Dubash, N.},
  \bibinfo{author}{Podder, S.}, \bibinfo{year}{2018}.
\newblock \bibinfo{title}{{Identifying Implementation Bugs in Machine Learning
  based Image Classifiers using Metamorphic Testing}}, in:
  \bibinfo{booktitle}{27th ACM SIGSOFT International Symposium on Software
  Testing and Analysis}, pp. \bibinfo{pages}{118--128}.
\newblock \URLprefix
  \url{http://arxiv.org/abs/1808.05353{\%}0Ahttp://dx.doi.org/10.1145/3213846.3213858},
  \DOIprefix\doi{10.1145/3213846.3213858},
  \href{http://arxiv.org/abs/1808.05353}{\tt arXiv:1808.05353}.
%Type = Inproceedings
\bibitem[{Dwivedi et~al.(2016)Dwivedi, Tirkey, Ray and Rath}]{Dwivedi2016}
\bibinfo{author}{Dwivedi, A.K.}, \bibinfo{author}{Tirkey, A.},
  \bibinfo{author}{Ray, R.B.}, \bibinfo{author}{Rath, S.K.},
  \bibinfo{year}{2016}.
\newblock \bibinfo{title}{{Software Design Pattern Recognition using Machine
  Learning Techniques}}, in: \bibinfo{booktitle}{2016 IEEE Region 10 Conference
  (TENCON)}, \bibinfo{publisher}{IEEE}. pp. \bibinfo{pages}{222--227}.
\newblock \DOIprefix\doi{10.1109/TENCON.2016.7847994}.
%Type = Inproceedings
\bibitem[{Enişer and Sen(2018)}]{Eniser2018}
\bibinfo{author}{Enişer, H.F.}, \bibinfo{author}{Sen, A.},
  \bibinfo{year}{2018}.
\newblock \bibinfo{title}{{Testing service oriented architectures using
  stateful service visualization via machine learning}}, in:
  \bibinfo{booktitle}{ACM/IEEE 13th International Workshop on Automation of
  Software Test Testing}, pp. \bibinfo{pages}{9--15}.
\newblock \DOIprefix\doi{10.1145/3194733.3194737}.
%Type = Article
\bibitem[{ERTUĞRUL et~al.(2019)ERTUĞRUL, BAYTAR, {\c{C}}ATAL and
  MURATLI}]{ERTUGRUL2019}
\bibinfo{author}{ERTUĞRUL, E.}, \bibinfo{author}{BAYTAR, Z.},
  \bibinfo{author}{{\c{C}}ATAL, {\c{C}}.}, \bibinfo{author}{MURATLI,
  {\"{O}}.C.}, \bibinfo{year}{2019}.
\newblock \bibinfo{title}{{Performance tuning for machine learning-based
  software development effort prediction models}}.
\newblock \bibinfo{journal}{TURKISH JOURNAL OF ELECTRICAL ENGINEERING {\&}
  COMPUTER SCIENCES} , \bibinfo{pages}{1308--1324}\URLprefix
  \url{http://online.journals.tubitak.gov.tr/openDoiPdf.htm?mKodu=elk-1809-129},
  \DOIprefix\doi{10.3906/elk-1809-129}.
%Type = Inproceedings
\bibitem[{Factors et~al.(2000)Factors, Revised, Hern{\'{a}}ndez-orallo and
  Ram{\'{i}}rez-quintana}]{Factors2000}
\bibinfo{author}{Factors, Q.}, \bibinfo{author}{Revised, L.c.},
  \bibinfo{author}{Hern{\'{a}}ndez-orallo, J.},
  \bibinfo{author}{Ram{\'{i}}rez-quintana, M.J.}, \bibinfo{year}{2000}.
\newblock \bibinfo{title}{{Software as Learning :}}, in:
  \bibinfo{booktitle}{International Conference on Fundamental Approaches to
  Software Engineering}, pp. \bibinfo{pages}{147--162}.
%Type = Article
\bibitem[{Falcini et~al.(2017)Falcini, Lami, Science, Costanza and
  Automobiles}]{Falcini2017}
\bibinfo{author}{Falcini, F.}, \bibinfo{author}{Lami, G.},
  \bibinfo{author}{Science, I.}, \bibinfo{author}{Costanza, A.M.},
  \bibinfo{author}{Automobiles, F.C.}, \bibinfo{year}{2017}.
\newblock \bibinfo{title}{{Deep Learning in Automotive Software}}.
\newblock \bibinfo{journal}{IEEE Software} .
%Type = Article
\bibitem[{Ferzund et~al.(2008)Ferzund, Ahsan and Wotawa}]{Ferzund2008}
\bibinfo{author}{Ferzund, J.}, \bibinfo{author}{Ahsan, S.N.},
  \bibinfo{author}{Wotawa, F.}, \bibinfo{year}{2008}.
\newblock \bibinfo{title}{{Analysing bug prediction capabilities of static code
  metrics in open source software}}.
\newblock \bibinfo{journal}{Lecture Notes in Computer Science (including
  subseries Lecture Notes in Artificial Intelligence and Lecture Notes in
  Bioinformatics)} \bibinfo{volume}{5338 LNCS}, \bibinfo{pages}{331--343}.
\newblock \DOIprefix\doi{10.1007/978-3-540-89403-2-27}.
%Type = Inproceedings
\bibitem[{Fouqu and Vrain(1992)}]{Fouqu}
\bibinfo{author}{Fouqu, G.}, \bibinfo{author}{Vrain, C.}, \bibinfo{year}{1992}.
\newblock \bibinfo{title}{{Building a Tool for Software Code Analysis A Machine
  Learning Approach structures}}, in: \bibinfo{booktitle}{International
  Conference on Advanced Information Systems Engineering}.
%Type = Book
\bibitem[{Fran{\c{c}}ois-Lavet et~al.(2018)Fran{\c{c}}ois-Lavet, Henderson,
  Islam, Bellemare and Pineau}]{Francois-Lavet2018}
\bibinfo{author}{Fran{\c{c}}ois-Lavet, V.}, \bibinfo{author}{Henderson, P.},
  \bibinfo{author}{Islam, R.}, \bibinfo{author}{Bellemare, M.G.},
  \bibinfo{author}{Pineau, J.}, \bibinfo{year}{2018}.
\newblock \bibinfo{title}{{An introduction to deep reinforcement learning}}.
  volume~\bibinfo{volume}{11}.
\newblock \DOIprefix\doi{10.1561/2200000071}.
%Type = Inproceedings
\bibitem[{Fu(2018)}]{Fu2018}
\bibinfo{author}{Fu, C.}, \bibinfo{year}{2018}.
\newblock \bibinfo{title}{{Estimating Software Energy Consumption with Machine
  Learning Approach by Software}}, in: \bibinfo{booktitle}{IEEE Confs on
  Internet of Things, Green Computing and Communications, Cyber, Physical and
  Social Computing, Smart Data, Blockchain, Computer and Information
  Technology, Congress on Cybermatics}, \bibinfo{publisher}{IEEE}. pp.
  \bibinfo{pages}{490--496}.
\newblock \DOIprefix\doi{10.1109/Cybermatics}.
%Type = Inproceedings
\bibitem[{Gelman et~al.(2018)Gelman, Hoyle, Moore, Saxe and
  Slater}]{Gelman2018}
\bibinfo{author}{Gelman, B.}, \bibinfo{author}{Hoyle, B.},
  \bibinfo{author}{Moore, J.}, \bibinfo{author}{Saxe, J.},
  \bibinfo{author}{Slater, D.}, \bibinfo{year}{2018}.
\newblock \bibinfo{title}{{A language-agnostic model for semantic source code
  labeling}}, in: \bibinfo{booktitle}{1st International Workshop on Machine
  Learning and Software Engineering in Symbiosis}, pp. \bibinfo{pages}{36--44}.
\newblock \DOIprefix\doi{10.1145/3243127.3243132}.
%Type = Article
\bibitem[{Ghaffarian and Shahriari(2017)}]{Ghaffarian2017}
\bibinfo{author}{Ghaffarian, S.M.}, \bibinfo{author}{Shahriari, H.R.},
  \bibinfo{year}{2017}.
\newblock \bibinfo{title}{{Software Vulnerability Analysis and Discovery Using
  Machine-Learning and Data-Mining Techniques : A Survey}}.
\newblock \bibinfo{journal}{ACM Computing Surveys} \bibinfo{volume}{50}.
%Type = Article
\bibitem[{Gove and Faytong(2012)}]{Gove}
\bibinfo{author}{Gove, R.}, \bibinfo{author}{Faytong, J.},
  \bibinfo{year}{2012}.
\newblock \bibinfo{title}{{Machine Learning and Event-Based Software Testing :
  Classifiers for Identifying Infeasible GUI Event Sequences}}.
\newblock \bibinfo{journal}{Advances in Computers} \bibinfo{volume}{86},
  \bibinfo{pages}{109--135}.
\newblock \URLprefix
  \url{http://dx.doi.org/10.1016/B978-0-12-396535-6.00004-1},
  \DOIprefix\doi{10.1016/B978-0-12-396535-6.00004-1}.
%Type = Inproceedings
\bibitem[{Gowda et~al.(2018)Gowda, Prajapati, Singh and Gadre}]{Gowda2018}
\bibinfo{author}{Gowda, S.}, \bibinfo{author}{Prajapati, D.},
  \bibinfo{author}{Singh, R.}, \bibinfo{author}{Gadre, S.S.},
  \bibinfo{year}{2018}.
\newblock \bibinfo{title}{{False Positive Analysis of software vulnerabilities
  using Machine learning}}, in: \bibinfo{booktitle}{2018 IEEE International
  Conference on Cloud Computing in Emerging Markets (CCEM)},
  \bibinfo{publisher}{IEEE}. pp. \bibinfo{pages}{3--6}.
\newblock \URLprefix \url{https://ieeexplore.ieee.org/document/8648633/},
  \DOIprefix\doi{10.1109/CCEM.2018.00010}.
%Type = Inproceedings
\bibitem[{Guo et~al.(2017)Guo, Cheng and Cleland-huang}]{Guo2017}
\bibinfo{author}{Guo, J.}, \bibinfo{author}{Cheng, J.},
  \bibinfo{author}{Cleland-huang, J.}, \bibinfo{year}{2017}.
\newblock \bibinfo{title}{{Semantically Enhanced Software Traceability Using
  Deep Learning Techniques}}, in: \bibinfo{booktitle}{IEEE/ACM 39th
  International Conference on Software Engineering Semantically},
  \bibinfo{publisher}{IEEE}.
\newblock \DOIprefix\doi{10.1109/ICSE.2017.9}.
%Type = Inproceedings
\bibitem[{Guo et~al.(2018)Guo, Jiang, Zhao, Chen and Sun}]{Guo2018}
\bibinfo{author}{Guo, J.}, \bibinfo{author}{Jiang, Y.}, \bibinfo{author}{Zhao,
  Y.}, \bibinfo{author}{Chen, Q.}, \bibinfo{author}{Sun, J.},
  \bibinfo{year}{2018}.
\newblock \bibinfo{title}{{DLFuzz: Differential Fuzzing Testing of Deep
  Learning Systems}}, in: \bibinfo{booktitle}{26th ACM Joint European Software
  Engineering Conference and Symposium on the Foundations of Software
  Engineering}, pp. \bibinfo{pages}{739--743}.
\newblock \URLprefix
  \url{http://arxiv.org/abs/1808.09413{\%}0Ahttp://dx.doi.org/10.1145/3236024.3264835},
  \DOIprefix\doi{10.1145/3236024.3264835},
  \href{http://arxiv.org/abs/1808.09413}{\tt arXiv:1808.09413}.
%Type = Inproceedings
\bibitem[{Hall and Bowes(2012)}]{Hall2012}
\bibinfo{author}{Hall, T.}, \bibinfo{author}{Bowes, D.}, \bibinfo{year}{2012}.
\newblock \bibinfo{title}{{The State of Machine Learning Methodology in
  Software Fault Prediction}}, in: \bibinfo{booktitle}{11th International
  Conference on Machine Learning and Applications}, \bibinfo{publisher}{IEEE}.
  pp. \bibinfo{pages}{308--313}.
\newblock \DOIprefix\doi{10.1109/ICMLA.2012.226}.
%Type = Inproceedings
\bibitem[{Han et~al.(2017)Han, Li, Xing, Liu and Feng}]{Han2017}
\bibinfo{author}{Han, Z.}, \bibinfo{author}{Li, X.}, \bibinfo{author}{Xing,
  Z.}, \bibinfo{author}{Liu, H.}, \bibinfo{author}{Feng, Z.},
  \bibinfo{year}{2017}.
\newblock \bibinfo{title}{{Learning to Predict Severity of Software
  Vulnerability Using Only Vulnerability Description}}, in:
  \bibinfo{booktitle}{IEEE International Conference on Software Maintenance and
  Evolution Learning}.
\newblock \DOIprefix\doi{10.1109/ICSME.2017.52}.
%Type = Inproceedings
\bibitem[{Harandi and Lee(1991)}]{Harandi1991}
\bibinfo{author}{Harandi, M.T.}, \bibinfo{author}{Lee, H.y.},
  \bibinfo{year}{1991}.
\newblock \bibinfo{title}{{Acquiring Software Design Schemas : A Machine
  Learning Perspective}}, in: \bibinfo{booktitle}{6th International Conference
  on Knowledge-Based Software Engineering}.
%Type = Inproceedings
\bibitem[{Harman(2012)}]{Harman2012}
\bibinfo{author}{Harman, M.}, \bibinfo{year}{2012}.
\newblock \bibinfo{title}{{The Role of Artificial Intelligence in Software
  Engineering}}, in: \bibinfo{booktitle}{2012 First International Workshop on
  Realizing AI Synergies in Software Engineering (RAISE)},
  \bibinfo{publisher}{IEEE}. pp. \bibinfo{pages}{1--6}.
\newblock \DOIprefix\doi{10.1109/RAISE.2012.6227961}.
%Type = Article
\bibitem[{Heo et~al.(2017)Heo, Oh and Yi}]{Heo2017}
\bibinfo{author}{Heo, K.}, \bibinfo{author}{Oh, H.}, \bibinfo{author}{Yi, K.},
  \bibinfo{year}{2017}.
\newblock \bibinfo{title}{{Machine-Learning-Guided Selectively Unsound Static
  Analysis}}.
\newblock \bibinfo{journal}{Proceedings - 2017 IEEE/ACM 39th International
  Conference on Software Engineering, ICSE 2017} ,
  \bibinfo{pages}{519--529}\DOIprefix\doi{10.1109/ICSE.2017.54}.
%Type = Inproceedings
\bibitem[{{Hoa Khanh Dam}(2019)}]{Eshuis2019}
\bibinfo{author}{{Hoa Khanh Dam}}, \bibinfo{year}{2019}.
\newblock \bibinfo{title}{{Empowering Software Engineering with Artificial
  Intelligence}}, in: \bibinfo{booktitle}{Australian Symposium on Service
  Research and Innovation}, \bibinfo{publisher}{Springer International
  Publishing}. pp. \bibinfo{pages}{3--12}.
\newblock \URLprefix \url{http://dx.doi.org/10.1007/978-3-030-32242-7{\_}1},
  \DOIprefix\doi{10.1007/978-3-030-32242-7}.
%Type = Inproceedings
\bibitem[{Hu et~al.(2018)Hu, Zhu and Yang}]{Hu2018}
\bibinfo{author}{Hu, G.}, \bibinfo{author}{Zhu, L.}, \bibinfo{author}{Yang,
  J.}, \bibinfo{year}{2018}.
\newblock \bibinfo{title}{{AppFlow: using machine learning to synthesize
  robust, reusable UI tests}}, in: \bibinfo{booktitle}{26th ACM Joint European
  Software Engineering Conference and Symposium on the Foundations of Software
  Engineering}, pp. \bibinfo{pages}{269--282}.
\newblock \DOIprefix\doi{10.1145/3236024.3236055}.
%Type = Article
\bibitem[{Huang et~al.(2019)Huang, Li, Wang, Ren, Cheng and Zhao}]{Huang2019}
\bibinfo{author}{Huang, G.}, \bibinfo{author}{Li, Y.}, \bibinfo{author}{Wang,
  Q.}, \bibinfo{author}{Ren, J.}, \bibinfo{author}{Cheng, Y.},
  \bibinfo{author}{Zhao, X.}, \bibinfo{year}{2019}.
\newblock \bibinfo{title}{{Automatic classification method for software
  vulnerability based on deep neural network}}.
\newblock \bibinfo{journal}{IEEE Access} \bibinfo{volume}{7},
  \bibinfo{pages}{28291--28298}.
\newblock \DOIprefix\doi{10.1109/ACCESS.2019.2900462}.
%Type = Article
\bibitem[{Huang et~al.(2015)Huang, Li and Xie}]{Huang2015}
\bibinfo{author}{Huang, J.}, \bibinfo{author}{Li, Y.F.}, \bibinfo{author}{Xie,
  M.}, \bibinfo{year}{2015}.
\newblock \bibinfo{title}{{An empirical analysis of data preprocessing for
  machine learning-based software cost estimation}}.
\newblock \bibinfo{journal}{INFORMATION AND SOFTWARE TECHNOLOGY}
  \bibinfo{volume}{67}, \bibinfo{pages}{108--127}.
\newblock \URLprefix \url{http://dx.doi.org/10.1016/j.infsof.2015.07.004
  https://linkinghub.elsevier.com/retrieve/pii/S0950584915001275},
  \DOIprefix\doi{10.1016/j.infsof.2015.07.004}.
%Type = Inproceedings
\bibitem[{Huch et~al.(2018)Huch, Golagha, Petrovska and Krauss}]{Huch2018}
\bibinfo{author}{Huch, F.}, \bibinfo{author}{Golagha, M.},
  \bibinfo{author}{Petrovska, A.}, \bibinfo{author}{Krauss, A.},
  \bibinfo{year}{2018}.
\newblock \bibinfo{title}{{Machine Learning-Based Run-Time Anomaly Detection in
  Software Systems : An Industrial Evaluation}}, in: \bibinfo{booktitle}{IEEE
  Workshop on Machine Learning Techniques for Software Quality Evaluation},
  \bibinfo{publisher}{IEEE}. pp. \bibinfo{pages}{13--18}.
%Type = Inproceedings
\bibitem[{Huo et~al.(2018a)Huo, Zhao, Liu, Xiang, Zhong and Yu}]{Huo2018}
\bibinfo{author}{Huo, S.}, \bibinfo{author}{Zhao, D.}, \bibinfo{author}{Liu,
  X.}, \bibinfo{author}{Xiang, J.}, \bibinfo{author}{Zhong, Y.},
  \bibinfo{author}{Yu, H.}, \bibinfo{year}{2018}a.
\newblock \bibinfo{title}{{Using Machine Learning for Software Aging Detection
  in Android System}}, in: \bibinfo{booktitle}{Tenth International Conference
  on Advanced Computational Intelligence}, \bibinfo{publisher}{IEEE}. pp.
  \bibinfo{pages}{741--746}.
%Type = Inproceedings
\bibitem[{Huo et~al.(2018b)Huo, Yang, Li and Zhan}]{Huo2018a}
\bibinfo{author}{Huo, X.}, \bibinfo{author}{Yang, Y.}, \bibinfo{author}{Li,
  M.}, \bibinfo{author}{Zhan, D.c.}, \bibinfo{year}{2018}b.
\newblock \bibinfo{title}{{Learning Semantic Features for Software Defect
  Prediction by Code Comments Embedding}}, in: \bibinfo{booktitle}{IEEE
  International Conference on Data Mining}, \bibinfo{publisher}{IEEE}.
\newblock \DOIprefix\doi{10.1109/ICDM.2018.00133}.
%Type = Article
\bibitem[{Idri et~al.(2008)Idri, Zahi, Mendes and Zakrani}]{Idri2008}
\bibinfo{author}{Idri, A.}, \bibinfo{author}{Zahi, A.},
  \bibinfo{author}{Mendes, E.}, \bibinfo{author}{Zakrani, A.},
  \bibinfo{year}{2008}.
\newblock \bibinfo{title}{{Software cost estimation models using radial basis
  function neural networks}}.
\newblock \bibinfo{journal}{Lecture Notes in Computer Science (including
  subseries Lecture Notes in Artificial Intelligence and Lecture Notes in
  Bioinformatics)} \bibinfo{volume}{4895 LNCS}, \bibinfo{pages}{21--31}.
\newblock \DOIprefix\doi{10.1007/978-3-540-85553-8_2}.
%Type = Inproceedings
\bibitem[{Ionescu(2017)}]{Ionescu2017}
\bibinfo{author}{Ionescu, V.s.}, \bibinfo{year}{2017}.
\newblock \bibinfo{title}{{An approach to software development effort
  estimation using machine learning}}, in: \bibinfo{booktitle}{13th IEEE
  International Conference on Intelligent Computer Communication and
  Processing}, pp. \bibinfo{pages}{197--203}.
%Type = Inproceedings
\bibitem[{Iwata et~al.(2016)Iwata, Nakashima, Anan and Ishii}]{Iwata2016}
\bibinfo{author}{Iwata, K.}, \bibinfo{author}{Nakashima, T.},
  \bibinfo{author}{Anan, Y.}, \bibinfo{author}{Ishii, N.},
  \bibinfo{year}{2016}.
\newblock \bibinfo{title}{{Effort Estimation for Embedded Software Development
  Projects by Combining Machine Learning with Classification}}, in:
  \bibinfo{booktitle}{4th Intl Conf on Applied Computing and Information
  Technology/3rd Intl Conf on Computational Science/Intelligence and Applied
  Informatics/1st Intl Conf on Big Data, Cloud Computing, Data Science {\&}
  Engineering}, \bibinfo{publisher}{IEEE}.
\newblock \DOIprefix\doi{10.1109/ACIT-CSII-BCD.2016.57}.
%Type = Article
\bibitem[{Jahan et~al.(2019)Jahan, Abad and Far}]{Jahan2019}
\bibinfo{author}{Jahan, M.}, \bibinfo{author}{Abad, Z.S.H.},
  \bibinfo{author}{Far, B.}, \bibinfo{year}{2019}.
\newblock \bibinfo{title}{{Detecting emergent behaviors and implied scenarios
  in scenario-based specifications: A machine learning approach}}.
\newblock \bibinfo{journal}{Proceedings - 2019 IEEE/ACM 11th International
  Workshop on Modelling in Software Engineering, MiSE 2019} ,
  \bibinfo{pages}{8--14}\DOIprefix\doi{10.1109/MiSE.2019.00009}.
%Type = Inproceedings
\bibitem[{Jie et~al.(2016)Jie, Xiao-Hui and Qiang}]{Jie2016}
\bibinfo{author}{Jie, G.}, \bibinfo{author}{Xiao-Hui, K.},
  \bibinfo{author}{Qiang, L.}, \bibinfo{year}{2016}.
\newblock \bibinfo{title}{{Survey on Software Vulnerability Analysis method
  based on Machine Learning}}, in: \bibinfo{booktitle}{2016 IEEE First
  International Conference on Data Science in Cyberspace (DSC)},
  \bibinfo{publisher}{IEEE}. pp. \bibinfo{pages}{642--647}.
\newblock \URLprefix \url{http://ieeexplore.ieee.org/document/7866201/},
  \DOIprefix\doi{10.1109/DSC.2016.33}.
%Type = Inproceedings
\bibitem[{Jindal(2018)}]{Jindal}
\bibinfo{author}{Jindal, V.}, \bibinfo{year}{2018}.
\newblock \bibinfo{title}{{Towards an Intelligent Fault Prediction Code Editor
  to Improve Software Quality using Deep Learning}}, in:
  \bibinfo{booktitle}{2nd International Conference on the Art, Science, and
  Engineering of Programming}, pp. \bibinfo{pages}{222--223}.
%Type = Inproceedings
\bibitem[{Kahles and Jung(2019)}]{Kahles2019}
\bibinfo{author}{Kahles, J.}, \bibinfo{author}{Jung, A.}, \bibinfo{year}{2019}.
\newblock \bibinfo{title}{{Automating Root Cause Analysis via Machine Learning
  in Agile Software Testing Environments}}, in: \bibinfo{booktitle}{12th IEEE
  Conference on Software Testing, Validation and Verification},
  \bibinfo{publisher}{IEEE}.
\newblock \DOIprefix\doi{10.1109/ICST.2019.00047}.
%Type = Inproceedings
\bibitem[{Kalibhat and Varshini(2017)}]{Kalibhat2017}
\bibinfo{author}{Kalibhat, N.M.}, \bibinfo{author}{Varshini, S.},
  \bibinfo{year}{2017}.
\newblock \bibinfo{title}{{Software Troubleshooting using Machine Learning}},
  in: \bibinfo{booktitle}{24th International Conference on High Performance
  Computing Workshops}.
\newblock \DOIprefix\doi{10.1109/HiPCW.2017.00010}.
%Type = Inproceedings
\bibitem[{Kalles(2016)}]{Kalles2016}
\bibinfo{author}{Kalles, D.}, \bibinfo{year}{2016}.
\newblock \bibinfo{title}{{Artificial Intelligence meets Software Engineering
  in Computing Education}}, in: \bibinfo{booktitle}{9th Hellenic Conference on
  Artificial Intelligence}, pp. \bibinfo{pages}{1--5}.
\newblock \DOIprefix\doi{10.1145/2903220.2903223}.
%Type = Article
\bibitem[{Kanewala et~al.(2016)Kanewala, Bieman and Ben-Hur}]{Kanewala2016}
\bibinfo{author}{Kanewala, U.}, \bibinfo{author}{Bieman, J.M.},
  \bibinfo{author}{Ben-Hur, A.}, \bibinfo{year}{2016}.
\newblock \bibinfo{title}{{Predicting metamorphic relations for testing
  scientific software: a machine learning approach using graph kernels}}.
\newblock \bibinfo{journal}{Software Testing, Verification and Reliability}
  \bibinfo{volume}{26}, \bibinfo{pages}{245--269}.
\newblock \URLprefix \url{http://doi.wiley.com/10.1002/stvr.1594},
  \DOIprefix\doi{10.1002/stvr.1594}.
%Type = Inproceedings
\bibitem[{Karim et~al.(2017)Karim, Leslie, Spits, Abdurachman and
  Soewito}]{Karim}
\bibinfo{author}{Karim, S.}, \bibinfo{author}{Leslie, H.},
  \bibinfo{author}{Spits, H.}, \bibinfo{author}{Abdurachman, E.},
  \bibinfo{author}{Soewito, B.}, \bibinfo{year}{2017}.
\newblock \bibinfo{title}{{Software Metrics for Fault Prediction Using Machine
  Learning Approaches}}, in: \bibinfo{booktitle}{IEEE International Conference
  on Cybernetics and Computational Intelligence}, pp. \bibinfo{pages}{19--23}.
%Type = Article
\bibitem[{Karpov et~al.(2019)Karpov, Smetanin and Karpov}]{Karpov2019a}
\bibinfo{author}{Karpov, Y.}, \bibinfo{author}{Smetanin, Y.},
  \bibinfo{author}{Karpov, L.}, \bibinfo{year}{2019}.
\newblock \bibinfo{title}{{Adaptation of general software testing concepts to
  neural networks}}.
\newblock \bibinfo{journal}{Programming and Computer Software}
  \bibinfo{volume}{44}, \bibinfo{pages}{43--56}.
\newblock \DOIprefix\doi{10.31857/s013234740001214-0}.
%Type = Inproceedings
\bibitem[{Khosrowjerdi and Meinke(2018)}]{Khosrowjerdi2018}
\bibinfo{author}{Khosrowjerdi, H.}, \bibinfo{author}{Meinke, K.},
  \bibinfo{year}{2018}.
\newblock \bibinfo{title}{{Learning-based testing for autonomous systems using
  spatial and temporal requirements}}, in: \bibinfo{booktitle}{1st
  International Workshop on Machine Learning and Software Engineering in
  Symbiosis}, pp. \bibinfo{pages}{6--15}.
\newblock \DOIprefix\doi{10.1145/3243127.3243129}.
%Type = Inproceedings
\bibitem[{Kim et~al.(2015)Kim, Park and Park}]{Kim2015}
\bibinfo{author}{Kim, Y.}, \bibinfo{author}{Park, J.}, \bibinfo{author}{Park,
  M.}, \bibinfo{year}{2015}.
\newblock \bibinfo{title}{{Machine Learning-based Software Classification
  Scheme for Efficient Program Similarity Analysis}}, in:
  \bibinfo{booktitle}{Conference on research in adaptive and convergent
  systems}, pp. \bibinfo{pages}{114--118}.
%Type = Article
\bibitem[{Kitchenham et~al.(2009)Kitchenham, {Pearl Brereton}, Budgen, Turner,
  Bailey and Linkman}]{KITCHENHAM20097}
\bibinfo{author}{Kitchenham, B.}, \bibinfo{author}{{Pearl Brereton}, O.},
  \bibinfo{author}{Budgen, D.}, \bibinfo{author}{Turner, M.},
  \bibinfo{author}{Bailey, J.}, \bibinfo{author}{Linkman, S.},
  \bibinfo{year}{2009}.
\newblock \bibinfo{title}{{Systematic literature reviews in software
  engineering - A systematic literature review}}.
\newblock \bibinfo{journal}{Information and Software Technology}
  \bibinfo{volume}{51}, \bibinfo{pages}{7--15}.
\newblock \URLprefix
  \url{http://www.sciencedirect.com/science/article/pii/S0950584908001390},
  \DOIprefix\doi{10.1016/j.infsof.2008.09.009}.
%Type = Inproceedings
\bibitem[{Kronjee and Vranken(2018)}]{Kronjee}
\bibinfo{author}{Kronjee, J.}, \bibinfo{author}{Vranken, H.},
  \bibinfo{year}{2018}.
\newblock \bibinfo{title}{{Discovering software vulnerabilities using data-flow
  analysis and machine learning}}, in: \bibinfo{booktitle}{International
  Conference on Availability, Reliability and Security}.
%Type = Inproceedings
\bibitem[{Kumar and Bansal(2019)}]{Kumar2019a}
\bibinfo{author}{Kumar, A.}, \bibinfo{author}{Bansal, A.},
  \bibinfo{year}{2019}.
\newblock \bibinfo{title}{{Software Fault Proneness Prediction Using Genetic
  Based Machine Learning Techniques}}, in: \bibinfo{booktitle}{Proceedings -
  2019 4th International Conference on Internet of Things: Smart Innovation and
  Usages, IoT-SIU 2019}, \bibinfo{publisher}{IEEE}. pp. \bibinfo{pages}{1--5}.
\newblock \DOIprefix\doi{10.1109/IoT-SIU.2019.8777494}.
%Type = Article
\bibitem[{Kumar et~al.(2019)Kumar, Son, Abdel-basset, Priyadarshini, Sharma and
  Long}]{Kumar2019}
\bibinfo{author}{Kumar, R.}, \bibinfo{author}{Son, L.E.H.},
  \bibinfo{author}{Abdel-basset, M.}, \bibinfo{author}{Priyadarshini, I.},
  \bibinfo{author}{Sharma, R.}, \bibinfo{author}{Long, H.V.},
  \bibinfo{year}{2019}.
\newblock \bibinfo{title}{{Deep Learning Approach for Software Maintainability
  Metrics Prediction}}.
\newblock \bibinfo{journal}{IEEE Access} \bibinfo{volume}{7},
  \bibinfo{pages}{2169--3536}.
%Type = Inproceedings
\bibitem[{Kuznetsov et~al.(2019)Kuznetsov, Yeromin, Shapoval, Chernov, Popova
  and Serdukov}]{Kuznetsov2019}
\bibinfo{author}{Kuznetsov, A.}, \bibinfo{author}{Yeromin, Y.},
  \bibinfo{author}{Shapoval, O.}, \bibinfo{author}{Chernov, K.},
  \bibinfo{author}{Popova, M.}, \bibinfo{author}{Serdukov, K.},
  \bibinfo{year}{2019}.
\newblock \bibinfo{title}{{Automated software vulnerability testing using deep
  learning methods}}, in: \bibinfo{booktitle}{2019 IEEE 2nd Ukraine Conference
  on Electrical and Computer Engineering, UKRCON 2019 - Proceedings},
  \bibinfo{publisher}{IEEE}. pp. \bibinfo{pages}{837--841}.
\newblock \DOIprefix\doi{10.1109/UKRCON.2019.8879997}.
%Type = Inproceedings
\bibitem[{Lal(2017a)}]{Lal2017a}
\bibinfo{author}{Lal, H.}, \bibinfo{year}{2017}a.
\newblock \bibinfo{title}{{Code Review Analysis of Software System using
  Machine Learning Techniques}}, in: \bibinfo{booktitle}{2017 11th
  International Conference on Intelligent Systems and Control (ISCO)},
  \bibinfo{publisher}{IEEE}. pp. \bibinfo{pages}{8--13}.
\newblock \DOIprefix\doi{10.1109/ISCO.2017.7855962}.
%Type = Inproceedings
\bibitem[{Lal(2017b)}]{Lal2017b}
\bibinfo{author}{Lal, H.}, \bibinfo{year}{2017}b.
\newblock \bibinfo{title}{{Root Cause Analysis of Software Bugs using Machine
  Learning Techniques}}, in: \bibinfo{booktitle}{2017 7th International
  Conference on Cloud Computing, Data Science {\&} Engineering - Confluence},
  \bibinfo{publisher}{IEEE}. pp. \bibinfo{pages}{105--111}.
\newblock \DOIprefix\doi{10.1109/CONFLUENCE.2017.7943132}.
%Type = Inproceedings
\bibitem[{Leclair et~al.(2018)Leclair, Eberhart and Mcmillan}]{Leclair2018}
\bibinfo{author}{Leclair, A.}, \bibinfo{author}{Eberhart, Z.},
  \bibinfo{author}{Mcmillan, C.}, \bibinfo{year}{2018}.
\newblock \bibinfo{title}{{Adapting Neural Text Classification for Improved
  Software Categorization}}, in: \bibinfo{booktitle}{IEEE International
  Conference on Software Maintenance and Evolution}, \bibinfo{publisher}{IEEE}.
\newblock \DOIprefix\doi{10.1109/ICSME.2018.00056}.
%Type = Inproceedings
\bibitem[{Lee et~al.(2014)Lee, Jung and Pande}]{Lee2014}
\bibinfo{author}{Lee, S.}, \bibinfo{author}{Jung, C.}, \bibinfo{author}{Pande,
  S.}, \bibinfo{year}{2014}.
\newblock \bibinfo{title}{{Detecting memory leaks through introspective dynamic
  behavior modelling using machine learning}}, in: \bibinfo{booktitle}{36th
  International Conference on Software Engineering}, pp.
  \bibinfo{pages}{814--824}.
\newblock \DOIprefix\doi{10.1145/2568225.2568307}.
%Type = Article
\bibitem[{Lenz et~al.(2013)Lenz, Pozo and Regina}]{Lenz2013}
\bibinfo{author}{Lenz, A.R.}, \bibinfo{author}{Pozo, A.},
  \bibinfo{author}{Regina, S.}, \bibinfo{year}{2013}.
\newblock \bibinfo{title}{{Engineering Applications of Arti fi cial
  Intelligence Linking software testing results with a machine learning
  approach}}.
\newblock \bibinfo{journal}{Engineering Applications of Artificial
  Intelligence} \bibinfo{volume}{26}, \bibinfo{pages}{1631--1640}.
\newblock \URLprefix \url{http://dx.doi.org/10.1016/j.engappai.2013.01.008},
  \DOIprefix\doi{10.1016/j.engappai.2013.01.008}.
%Type = Inproceedings
\bibitem[{Li et~al.(2017)Li, He, Zhu and Lyu}]{Li2017a}
\bibinfo{author}{Li, J.}, \bibinfo{author}{He, P.}, \bibinfo{author}{Zhu, J.},
  \bibinfo{author}{Lyu, M.R.}, \bibinfo{year}{2017}.
\newblock \bibinfo{title}{{Software Defect Prediction via Convolutional Neural
  Network}}, in: \bibinfo{booktitle}{IEEE International Conference on Software
  Quality, Reliability and Security}, pp. \bibinfo{pages}{318--328}.
\newblock \DOIprefix\doi{10.1109/QRS.2017.42}.
%Type = Article
\bibitem[{Liang et~al.(2019)Liang, Yu, Jiang and Xie}]{Liang2019}
\bibinfo{author}{Liang, H.}, \bibinfo{author}{Yu, Y.}, \bibinfo{author}{Jiang,
  L.}, \bibinfo{author}{Xie, Z.}, \bibinfo{year}{2019}.
\newblock \bibinfo{title}{{Seml: A Semantic LSTM Model for Software Defect
  Prediction}}.
\newblock \bibinfo{journal}{IEEE Access} \bibinfo{volume}{7},
  \bibinfo{pages}{83812--83824}.
\newblock \DOIprefix\doi{10.1109/ACCESS.2019.2925313}.
%Type = Inproceedings
\bibitem[{Licea(2017)}]{Licea2017}
\bibinfo{author}{Licea, G.}, \bibinfo{year}{2017}.
\newblock \bibinfo{title}{{Towards supporting Software Engineering using Deep
  Learning : A case of Software Requirements Classification}}, in:
  \bibinfo{booktitle}{5th International Conference in Software Engineering
  Research and Innovation}.
\newblock \DOIprefix\doi{10.1109/CONISOFT.2017.00021}.
%Type = Inproceedings
\bibitem[{Lim and Vector(2018)}]{Lim2018}
\bibinfo{author}{Lim, H.i.}, \bibinfo{author}{Vector, A.C.},
  \bibinfo{year}{2018}.
\newblock \bibinfo{title}{{Applying Code Vectors for Presenting Software
  Features in Machine Learning}}, in: \bibinfo{booktitle}{42nd IEEE
  International Conference on Computer Software {\&} Applications Applying},
  \bibinfo{publisher}{IEEE}. pp. \bibinfo{pages}{803--804}.
\newblock \DOIprefix\doi{10.1109/COMPSAC.2018.00128}.
%Type = Inproceedings
\bibitem[{Lin et~al.(2018)Lin, Bavota, Penta and Lanza}]{Lin2018a}
\bibinfo{author}{Lin, B.}, \bibinfo{author}{Bavota, G.},
  \bibinfo{author}{Penta, M.D.}, \bibinfo{author}{Lanza, M.},
  \bibinfo{year}{2018}.
\newblock \bibinfo{title}{{Sentiment Analysis for Software Engineering : How
  Far Can We Go ?}}, in: \bibinfo{booktitle}{ACM/IEEE 40th International
  Conference on Software Engineering}, \bibinfo{publisher}{ACM}. pp.
  \bibinfo{pages}{94--104}.
%Type = Inproceedings
\bibitem[{Liu et~al.(2017)Liu, Li, Zhu and Fan}]{Liu2017}
\bibinfo{author}{Liu, Q.}, \bibinfo{author}{Li, X.}, \bibinfo{author}{Zhu, H.},
  \bibinfo{author}{Fan, H.}, \bibinfo{year}{2017}.
\newblock \bibinfo{title}{{Acquisition of Open Source Software Project Maturity
  Based on Time Series Machine Learning}}, in: \bibinfo{booktitle}{10th
  International Symposium on Computational Intelligence and Design}, pp.
  \bibinfo{pages}{10--13}.
\newblock \DOIprefix\doi{10.1109/ISCID.2017.20}.
%Type = Inproceedings
\bibitem[{Lopez-martin et~al.(2014)Lopez-martin, Chavoya and
  Meda-campa{\~{n}}a}]{Lopez-martin2014}
\bibinfo{author}{Lopez-martin, C.}, \bibinfo{author}{Chavoya, A.},
  \bibinfo{author}{Meda-campa{\~{n}}a, M.E.}, \bibinfo{year}{2014}.
\newblock \bibinfo{title}{{A Machine Learning Technique for Predicting the
  Productivity of Practitioners from Individually Developed Software
  Projects}}, in: \bibinfo{booktitle}{IEEE/ACIS International Conference on
  Software Engineering, Artificial Intelligence, Networking and
  Parallel/Distributed Computing}.
%Type = Inproceedings
\bibitem[{Lounis and Ait-mehedine(2004)}]{Lounis}
\bibinfo{author}{Lounis, H.}, \bibinfo{author}{Ait-mehedine, L.},
  \bibinfo{year}{2004}.
\newblock \bibinfo{title}{{Machine-Learning Techniques for Software Product
  Quality Assessment}}, in: \bibinfo{booktitle}{Fourth International Conference
  on Quality Software}, \bibinfo{publisher}{IEEE}.
%Type = Inproceedings
\bibitem[{Lounis et~al.(2014)Lounis, Gayed and Boukadoum}]{Lounis2014}
\bibinfo{author}{Lounis, H.}, \bibinfo{author}{Gayed, T.F.},
  \bibinfo{author}{Boukadoum, M.}, \bibinfo{year}{2014}.
\newblock \bibinfo{title}{{Machine-Learning Models for Software Quality : a
  Compromise Between Performance and Intelligibility Machine-Learning Models
  for Software Quality : a Compromise Between Performance and
  Intelligibility}}, in: \bibinfo{booktitle}{23rd International Conference on
  Tools with Artificial Intelligence}, pp. \bibinfo{pages}{64--67}.
\newblock \DOIprefix\doi{10.1109/ICTAI.2011.155}.
%Type = Article
\bibitem[{Luo et~al.(2010)Luo, Ben and Mi}]{Luo2010}
\bibinfo{author}{Luo, Y.}, \bibinfo{author}{Ben, K.}, \bibinfo{author}{Mi, L.},
  \bibinfo{year}{2010}.
\newblock \bibinfo{title}{{Software metrics reduction for fault-proneness
  prediction of software modules}}.
\newblock \bibinfo{journal}{Lecture Notes in Computer Science (including
  subseries Lecture Notes in Artificial Intelligence and Lecture Notes in
  Bioinformatics)} \bibinfo{volume}{6289 LNCS}, \bibinfo{pages}{432--441}.
\newblock \DOIprefix\doi{10.1007/978-3-642-15672-4_36}.
%Type = Inproceedings
\bibitem[{Ma et~al.(2018)Ma, Juefei-Xu, Zhang, Sun, Xue, Li, Chen, Su, Li, Liu,
  Zhao and Wang}]{Ma2018}
\bibinfo{author}{Ma, L.}, \bibinfo{author}{Juefei-Xu, F.},
  \bibinfo{author}{Zhang, F.}, \bibinfo{author}{Sun, J.}, \bibinfo{author}{Xue,
  M.}, \bibinfo{author}{Li, B.}, \bibinfo{author}{Chen, C.},
  \bibinfo{author}{Su, T.}, \bibinfo{author}{Li, L.}, \bibinfo{author}{Liu,
  Y.}, \bibinfo{author}{Zhao, J.}, \bibinfo{author}{Wang, Y.},
  \bibinfo{year}{2018}.
\newblock \bibinfo{title}{{DeepGauge: Multi-Granularity Testing Criteria for
  Deep Learning Systems}}, in: \bibinfo{booktitle}{33rd ACM/IEEE International
  Conference on Automated Software Engineering}, pp. \bibinfo{pages}{120--131}.
\newblock \URLprefix
  \url{http://arxiv.org/abs/1803.07519{\%}0Ahttp://dx.doi.org/10.1145/3238147.3238202},
  \DOIprefix\doi{10.1145/3238147.3238202},
  \href{http://arxiv.org/abs/1803.07519}{\tt arXiv:1803.07519}.
%Type = Inproceedings
\bibitem[{Madera and Tomo{\'{n}}(2017)}]{Madera2017a}
\bibinfo{author}{Madera, M.}, \bibinfo{author}{Tomo{\'{n}}, R.},
  \bibinfo{year}{2017}.
\newblock \bibinfo{title}{{A case study on machine learning model for code
  review expert system in software engineering}}, in:
  \bibinfo{booktitle}{Proceedings of the 2017 Federated Conference on Computer
  Science and Information Systems}, pp. \bibinfo{pages}{1357--1363}.
\newblock \DOIprefix\doi{10.15439/2017f536}.
%Type = Article
\bibitem[{Majd et~al.(2020)Majd, Vahidi-Asl, Khalilian, Poorsarvi-Tehrani and
  Haghighi}]{Majd2020}
\bibinfo{author}{Majd, A.}, \bibinfo{author}{Vahidi-Asl, M.},
  \bibinfo{author}{Khalilian, A.}, \bibinfo{author}{Poorsarvi-Tehrani, P.},
  \bibinfo{author}{Haghighi, H.}, \bibinfo{year}{2020}.
\newblock \bibinfo{title}{{SLDeep: Statement-level software defect prediction
  using deep-learning model on static code features}}.
\newblock \bibinfo{journal}{Expert Systems with Applications}
  \bibinfo{volume}{147}, \bibinfo{pages}{113156}.
\newblock \URLprefix \url{https://doi.org/10.1016/j.eswa.2019.113156},
  \DOIprefix\doi{10.1016/j.eswa.2019.113156}.
%Type = Inproceedings
\bibitem[{{Malhotra, Ruchika; Bahl, Laavanye; Sehgal}(2017)}]{Iru2017}
\bibinfo{author}{{Malhotra, Ruchika; Bahl, Laavanye; Sehgal}, S.P.P.},
  \bibinfo{year}{2017}.
\newblock \bibinfo{title}{{Empirical Comparison of Machine Learning Algorithms
  for Bug Prediction in Open Source Software}}, in:
  \bibinfo{booktitle}{International Conference on Big Data Analytics and
  Computational Intelligence}, pp. \bibinfo{pages}{40--45}.
%Type = Inproceedings
\bibitem[{Maneerat(2011)}]{Maneerat2011}
\bibinfo{author}{Maneerat, N.}, \bibinfo{year}{2011}.
\newblock \bibinfo{title}{{Bad-smell Prediction from Software Design Model
  Using Machine Learning Techniques}}, in: \bibinfo{booktitle}{Eighth
  International Joint Conference on Computer Science and Software Engineering},
  \bibinfo{publisher}{IEEE}. pp. \bibinfo{pages}{331--336}.
%Type = Inproceedings
\bibitem[{Masuda et~al.(2018)Masuda, Ono, Yasue and Hosokawa}]{Masuda2018}
\bibinfo{author}{Masuda, S.}, \bibinfo{author}{Ono, K.},
  \bibinfo{author}{Yasue, T.}, \bibinfo{author}{Hosokawa, N.},
  \bibinfo{year}{2018}.
\newblock \bibinfo{title}{{A Survey of Software Quality for Machine Learning
  Applications}}, in: \bibinfo{booktitle}{International Conference on Software
  Testing, Verification and Validation Workshops}, \bibinfo{publisher}{IEEE}.
\newblock \DOIprefix\doi{10.1109/ICSTW.2018.00061}.
%Type = Inproceedings
\bibitem[{{Matthias Galster, Fabian Gilson}(2019)}]{Sanctis2019}
\bibinfo{author}{{Matthias Galster, Fabian Gilson}, F.G.},
  \bibinfo{year}{2019}.
\newblock \bibinfo{title}{{What Quality Attributes Can We Find in Product
  Backlogs? A Machine Learning Perspective}}, in: \bibinfo{booktitle}{European
  Conference on Software Architecture}, pp. \bibinfo{pages}{88--96}.
\newblock \DOIprefix\doi{10.1007/978-3-030-29983-5}.
%Type = Article
\bibitem[{Meinke and Bennaceur(2018)}]{Meinke2018b}
\bibinfo{author}{Meinke, K.}, \bibinfo{author}{Bennaceur, A.},
  \bibinfo{year}{2018}.
\newblock \bibinfo{title}{{Machine Learning for Software Engineering: Models,
  Methods, and Applications}}.
\newblock \bibinfo{journal}{Proceedings of the 40th International Conference on
  Software Engineering: Companion Proceeedings} ,
  \bibinfo{pages}{548--549}\URLprefix
  \url{http://doi.acm.org/10.1145/3183440.3183461},
  \DOIprefix\doi{10.1145/3183440.3183461}.
%Type = Article
\bibitem[{Michie(1991)}]{Michie1991}
\bibinfo{author}{Michie, D.}, \bibinfo{year}{1991}.
\newblock \bibinfo{title}{{Methodologies from Machine Learning in Data Analysis
  and Software}}.
\newblock \bibinfo{journal}{THE COMPUTER JOURNAL} \bibinfo{volume}{34}.
%Type = Article
\bibitem[{Mills and Haiduc(2017)}]{Mills2017}
\bibinfo{author}{Mills, C.}, \bibinfo{author}{Haiduc, S.},
  \bibinfo{year}{2017}.
\newblock \bibinfo{title}{{A machine learning approach for determining the
  validity of traceability links}}.
\newblock \bibinfo{journal}{Proceedings - 2017 IEEE/ACM 39th International
  Conference on Software Engineering Companion, ICSE-C 2017} ,
  \bibinfo{pages}{121--123}\DOIprefix\doi{10.1109/ICSE-C.2017.86}.
%Type = Inproceedings
\bibitem[{Mizuno et~al.(2010)Mizuno, Hamasaki, Takagi and Kikuno}]{Mizuno2010}
\bibinfo{author}{Mizuno, O.}, \bibinfo{author}{Hamasaki, T.},
  \bibinfo{author}{Takagi, Y.}, \bibinfo{author}{Kikuno, T.},
  \bibinfo{year}{2010}.
\newblock \bibinfo{title}{{An Empirical Evaluation of Predicting Runaway
  Software Projects Using Bayesian Classification}}, in:
  \bibinfo{booktitle}{International Conference on Product Focused Software
  Process Improvement}, pp. \bibinfo{pages}{263--273}.
\newblock \DOIprefix\doi{10.1007/978-3-540-24659-6_19}.
%Type = Inproceedings
\bibitem[{Moghadam(2019)}]{Moghadam}
\bibinfo{author}{Moghadam, M.H.}, \bibinfo{year}{2019}.
\newblock \bibinfo{title}{{Machine Learning-Assisted Performance Testing}}, in:
  \bibinfo{booktitle}{ESEC/FSE 2019 - Proceedings of the 2019 27th ACM Joint
  Meeting European Software Engineering Conference and Symposium on the
  Foundations of Software Engineering}, pp. \bibinfo{pages}{12--14}.
%Type = Inproceedings
\bibitem[{Moranna(2018)}]{Moranna2018}
\bibinfo{author}{Moranna, G.}, \bibinfo{year}{2018}.
\newblock \bibinfo{title}{{Natural Engineering}}, in: \bibinfo{booktitle}{1st
  International Workshop on Software Engineering for Cognitive Services}, pp.
  \bibinfo{pages}{25--28}.
%Type = Article
\bibitem[{Mostaeen et~al.(2019)Mostaeen, Svajlenko, Roy, Roy and
  Schneider}]{Mostaeen2019}
\bibinfo{author}{Mostaeen, G.}, \bibinfo{author}{Svajlenko, J.},
  \bibinfo{author}{Roy, B.}, \bibinfo{author}{Roy, C.K.},
  \bibinfo{author}{Schneider, K.A.}, \bibinfo{year}{2019}.
\newblock \bibinfo{title}{{CloneCognition: Machine learning based code clone
  validation tool}}.
\newblock \bibinfo{journal}{ESEC/FSE 2019 - Proceedings of the 2019 27th ACM
  Joint Meeting European Software Engineering Conference and Symposium on the
  Foundations of Software Engineering} ,
  \bibinfo{pages}{1105--1109}\DOIprefix\doi{10.1145/3338906.3341182}.
%Type = Book
\bibitem[{Mousouliotis; and {Loukas P. Petrou}(2019)}]{B2019}
\bibinfo{author}{Mousouliotis;, P.G.}, \bibinfo{author}{{Loukas P. Petrou}},
  \bibinfo{year}{2019}.
\newblock \bibinfo{title}{{Software-Defined FPGA Accelerator Design for Mobile
  Deep Learning Applications}}. volume \bibinfo{volume}{11444}.
\newblock \bibinfo{publisher}{Springer International Publishing}.
\newblock \URLprefix \url{http://link.springer.com/10.1007/978-3-030-17227-5},
  \DOIprefix\doi{10.1007/978-3-030-17227-5}.
%Type = Article
\bibitem[{Muccini and Vaidhyanathan(2019)}]{Muccini2019}
\bibinfo{author}{Muccini, H.}, \bibinfo{author}{Vaidhyanathan, K.},
  \bibinfo{year}{2019}.
\newblock \bibinfo{title}{{ArchLearner: Leveraging Machine-learning Techniques
  for Proactive Architectural Adaptation}}.
\newblock \bibinfo{journal}{Proceedings of the 13th European Conference on
  Software Architecture - Volume 2} , \bibinfo{pages}{38--41}\URLprefix
  \url{http://doi.acm.org/10.1145/3344948.3344962},
  \DOIprefix\doi{10.1145/3344948.3344962}.
%Type = Inproceedings
\bibitem[{{Muhammad Noman Khalid, Humera Farooq, Muhammad Iqbal, Muhammad Talha
  Alam} and Rasheed(2019)}]{Muzzammel2019}
\bibinfo{author}{{Muhammad Noman Khalid, Humera Farooq, Muhammad Iqbal,
  Muhammad Talha Alam}}, \bibinfo{author}{Rasheed, K.}, \bibinfo{year}{2019}.
\newblock \bibinfo{title}{{Predicting Web Vulnerabilities in Web Applications
  Based on Machine Learning Muhammad}}, in: \bibinfo{booktitle}{International
  Conference on Intelligent Technologies and Applications},
  \bibinfo{publisher}{Springer Singapore}. pp. \bibinfo{pages}{496--510}.
\newblock \URLprefix \url{http://link.springer.com/10.1007/978-981-13-6052-7},
  \DOIprefix\doi{10.1007/978-981-13-6052-7}.
%Type = Article
\bibitem[{Murillo-Morera et~al.(2017)Murillo-Morera, Quesada-L{\'{o}}pez,
  Castro-Herrera and Jenkins}]{Murillo-Morera2017}
\bibinfo{author}{Murillo-Morera, J.}, \bibinfo{author}{Quesada-L{\'{o}}pez,
  C.}, \bibinfo{author}{Castro-Herrera, C.}, \bibinfo{author}{Jenkins, M.},
  \bibinfo{year}{2017}.
\newblock \bibinfo{title}{{A genetic algorithm based framework for software
  effort prediction}}.
\newblock \bibinfo{journal}{Journal of Software Engineering Research and
  Development} \bibinfo{volume}{5}.
\newblock \DOIprefix\doi{10.1186/s40411-017-0037-x}.
%Type = Inproceedings
\bibitem[{Nafi et~al.(2018)Nafi, Roy, Roy and Schneider}]{Nafi2018}
\bibinfo{author}{Nafi, K.W.}, \bibinfo{author}{Roy, B.}, \bibinfo{author}{Roy,
  C.K.}, \bibinfo{author}{Schneider, K.A.}, \bibinfo{year}{2018}.
\newblock \bibinfo{title}{{CroLSim : Cross Language Software Similarity
  Detector using API documentation}}, in: \bibinfo{booktitle}{18th
  International Working Conference on Source Code Analysis and Manipulation
  CroLSim:}, \bibinfo{publisher}{IEEE}.
\newblock \DOIprefix\doi{10.1109/SCAM.2018.00023}.
%Type = Inproceedings
\bibitem[{Nakajima(2018)}]{Nakajima}
\bibinfo{author}{Nakajima, S.}, \bibinfo{year}{2018}.
\newblock \bibinfo{title}{{Quality Assurance of Machine Learning Software}},
  in: \bibinfo{booktitle}{7th Global Conference on Consumer Electronics},
  \bibinfo{publisher}{IEEE}. pp. \bibinfo{pages}{601--604}.
%Type = Inproceedings
\bibitem[{Nascimento(2018)}]{Nascimento2018}
\bibinfo{author}{Nascimento, N.}, \bibinfo{year}{2018}.
\newblock \bibinfo{title}{{Toward Human-in-the-Loop Collaboration Between
  Software Engineers and Machine Learning Algorithms}}, in:
  \bibinfo{booktitle}{IEEE International Conference on Big Data},
  \bibinfo{publisher}{IEEE}. pp. \bibinfo{pages}{3534--3540}.
%Type = Article
\bibitem[{Neal et~al.(2017)Neal, Brisk, Abousamra, Waters, Shriver and
  Corporation}]{Neal2017}
\bibinfo{author}{Neal, K.O.}, \bibinfo{author}{Brisk, P.},
  \bibinfo{author}{Abousamra, A.}, \bibinfo{author}{Waters, Z.},
  \bibinfo{author}{Shriver, E.}, \bibinfo{author}{Corporation, I.},
  \bibinfo{year}{2017}.
\newblock \bibinfo{title}{{GPU Performance Estimation using Software
  Rasterization}}.
\newblock \bibinfo{journal}{Transactions on Embedded Computing Systems}
  \bibinfo{volume}{16}.
%Type = Inproceedings
\bibitem[{Nguyen et~al.(2018)Nguyen, Vu, Pham and Nguyen}]{Nguyen2018}
\bibinfo{author}{Nguyen, T.T.}, \bibinfo{author}{Vu, P.M.},
  \bibinfo{author}{Pham, H.V.}, \bibinfo{author}{Nguyen, T.T.},
  \bibinfo{year}{2018}.
\newblock \bibinfo{title}{{Deep learning UI design patterns of mobile apps}},
  in: \bibinfo{booktitle}{40th International Conference on Software
  Engineering: New Ideas and Emerging Results Deep}, pp.
  \bibinfo{pages}{65--68}.
\newblock \DOIprefix\doi{10.1145/3183399.3183422}.
%Type = Article
\bibitem[{Ni et~al.(2017)Ni, Liu, Chen, Gu, Chen and Huang}]{Ni2017}
\bibinfo{author}{Ni, C.}, \bibinfo{author}{Liu, W.S.}, \bibinfo{author}{Chen,
  X.}, \bibinfo{author}{Gu, Q.}, \bibinfo{author}{Chen, D.X.},
  \bibinfo{author}{Huang, Q.G.}, \bibinfo{year}{2017}.
\newblock \bibinfo{title}{{A Cluster Based Feature Selection Method for
  Cross-Project Software Defect Prediction}}.
\newblock \bibinfo{journal}{Journal of Computer Science and Technology}
  \bibinfo{volume}{32}, \bibinfo{pages}{1090--1107}.
\newblock \DOIprefix\doi{10.1007/s11390-017-1785-0}.
%Type = Article
\bibitem[{Niu et~al.(2019)Niu, Zhang, Du, Zhao, Cao and Guizani}]{Niu2019}
\bibinfo{author}{Niu, W.}, \bibinfo{author}{Zhang, X.}, \bibinfo{author}{Du,
  X.}, \bibinfo{author}{Zhao, L.}, \bibinfo{author}{Cao, R.},
  \bibinfo{author}{Guizani, M.}, \bibinfo{year}{2019}.
\newblock \bibinfo{title}{{A Deep Learning Based Static Taint Analysis Approach
  for IoT Software Vulnerability Location}}.
\newblock \bibinfo{journal}{Measurement} \bibinfo{volume}{152},
  \bibinfo{pages}{107139}.
\newblock \URLprefix \url{https://doi.org/10.1016/j.measurement.2019.107139},
  \DOIprefix\doi{10.1016/j.measurement.2019.107139}.
%Type = Inproceedings
\bibitem[{Oberta et~al.(2018)Oberta, Uoc and Claris{\'{o}}}]{Oberta2018}
\bibinfo{author}{Oberta, U.}, \bibinfo{author}{Uoc, D.C.},
  \bibinfo{author}{Claris{\'{o}}, R.}, \bibinfo{year}{2018}.
\newblock \bibinfo{title}{{Applying Graph Kernels to Model-Driven Engineering
  Problems}}, in: \bibinfo{booktitle}{1st International Workshop on Machine
  Learning and Software Engineering in Symbiosis}, pp. \bibinfo{pages}{1--5}.
%Type = Inproceedings
\bibitem[{Ognawala et~al.(2018)Ognawala, Amato, Pretschner and
  Kulkarni}]{Ognawala2018}
\bibinfo{author}{Ognawala, S.}, \bibinfo{author}{Amato, R.N.},
  \bibinfo{author}{Pretschner, A.}, \bibinfo{author}{Kulkarni, P.},
  \bibinfo{year}{2018}.
\newblock \bibinfo{title}{{Automatically Assessing Vulnerabilities Discovered
  by Compositional Analysis}}, in: \bibinfo{booktitle}{1st International
  Workshop on Machine Learning and Software Engineering in Symbiosis}, pp.
  \bibinfo{pages}{16--25}.
\newblock \URLprefix \url{http://arxiv.org/abs/1807.09160},
  \href{http://arxiv.org/abs/1807.09160}{\tt arXiv:1807.09160}.
%Type = Article
\bibitem[{Oster(2005)}]{Oster2005}
\bibinfo{author}{Oster, N.}, \bibinfo{year}{2005}.
\newblock \bibinfo{title}{{Automated generation and evaluation of
  dataflow-based test data for object-oriented software}}.
\newblock \bibinfo{journal}{Lecture Notes in Computer Science (including
  subseries Lecture Notes in Artificial Intelligence and Lecture Notes in
  Bioinformatics)} \bibinfo{volume}{3712 LNCS}, \bibinfo{pages}{212--226}.
\newblock \DOIprefix\doi{10.1007/11558569_16}.
%Type = Inproceedings
\bibitem[{Ott et~al.(2018)Ott, Atchison, Harnack, Bergh and Linstead}]{Ott2018}
\bibinfo{author}{Ott, J.}, \bibinfo{author}{Atchison, A.},
  \bibinfo{author}{Harnack, P.}, \bibinfo{author}{Bergh, A.},
  \bibinfo{author}{Linstead, E.}, \bibinfo{year}{2018}.
\newblock \bibinfo{title}{{A Deep Learning Approach to Identifying Source Code
  in Images and Video}}, in: \bibinfo{booktitle}{15th International Conference
  on Mining Software Repositories}.
%Type = Inproceedings
\bibitem[{Paduraru et~al.(2019)Paduraru, Melemciuc and Paduraru}]{Paduraru}
\bibinfo{author}{Paduraru, C.}, \bibinfo{author}{Melemciuc, M.c.},
  \bibinfo{author}{Paduraru, M.}, \bibinfo{year}{2019}.
\newblock \bibinfo{title}{{Automatic Test Data Generation for a Given Set of
  Applications Using Recurrent Neural Networks}}, in:
  \bibinfo{booktitle}{International Conference on Software Technologies},
  \bibinfo{publisher}{Springer International Publishing}. pp.
  \bibinfo{pages}{307--326}.
\newblock \URLprefix \url{http://dx.doi.org/10.1007/978-3-030-29157-0{\_}14},
  \DOIprefix\doi{10.1007/978-3-030-29157-0}.
%Type = Inproceedings
\bibitem[{Paper(2016)}]{Paper}
\bibinfo{author}{Paper, I.}, \bibinfo{year}{2016}.
\newblock \bibinfo{title}{{Deep Learning Approach for Network Intrusion
  Detection in Software Defined Networking}}, in:
  \bibinfo{booktitle}{International Conference on Wireless Networks and Mobile
  Communications}, \bibinfo{publisher}{IEEE}. pp. \bibinfo{pages}{1--6}.
%Type = Article
\bibitem[{Park(1997)}]{Park1997}
\bibinfo{author}{Park, F.}, \bibinfo{year}{1997}.
\newblock \bibinfo{title}{{Artificial Intelligence and Software Engineering :
  Breaking the Toy Mold}}.
\newblock \bibinfo{journal}{Automated Software Engineering}
  \bibinfo{volume}{270}, \bibinfo{pages}{255--270}.
%Type = Inproceedings
\bibitem[{Parr(2016)}]{Parr}
\bibinfo{author}{Parr, T.}, \bibinfo{year}{2016}.
\newblock \bibinfo{title}{{Towards a Universal Code Formatter through Machine
  Learning}}, in: \bibinfo{booktitle}{SIGPLAN International Conference on
  Software Language Engineering}, pp. \bibinfo{pages}{137--151}.
%Type = Inproceedings
\bibitem[{Paul(2007)}]{Paul2007}
\bibinfo{author}{Paul, R.A.}, \bibinfo{year}{2007}.
\newblock \bibinfo{title}{{A Machine Learning-Based Reliability Assessment
  Model for Critical Software Systems}}, in: \bibinfo{booktitle}{31st Annual
  International Computer Software and Applications Conference}.
%Type = Inproceedings
\bibitem[{Pecorelli et~al.(2019)Pecorelli, {Di Nucci}, {De Roover} and {De
  Lucia}}]{Pecorelli2019}
\bibinfo{author}{Pecorelli, F.}, \bibinfo{author}{{Di Nucci}, D.},
  \bibinfo{author}{{De Roover}, C.}, \bibinfo{author}{{De Lucia}, A.},
  \bibinfo{year}{2019}.
\newblock \bibinfo{title}{{On the role of data balancing for machine
  learning-based code smell detection}}, in: \bibinfo{booktitle}{3rd ACM
  SIGSOFT International Workshop on Machine Learning Techniques for Software
  Quality Evaluation}, pp. \bibinfo{pages}{19--24}.
\newblock \DOIprefix\doi{10.1145/3340482.3342744}.
%Type = Article
\bibitem[{Perini et~al.(2013)Perini, Susi and Avesani}]{Perini2013}
\bibinfo{author}{Perini, A.}, \bibinfo{author}{Susi, A.},
  \bibinfo{author}{Avesani, P.}, \bibinfo{year}{2013}.
\newblock \bibinfo{title}{{A Machine Learning Approach to Software Requirements
  Prioritization}}.
\newblock \bibinfo{journal}{IEEE Transactions on Software Engineering}
  \bibinfo{volume}{39}, \bibinfo{pages}{445--461}.
\newblock \DOIprefix\doi{10.1109/TSE.2012.52}.
%Type = Article
\bibitem[{Petersen et~al.(2008)Petersen, Feldt, Mujtaba and
  Mattsson}]{Petersen2008}
\bibinfo{author}{Petersen, K.}, \bibinfo{author}{Feldt, R.},
  \bibinfo{author}{Mujtaba, S.}, \bibinfo{author}{Mattsson, M.},
  \bibinfo{year}{2008}.
\newblock \bibinfo{title}{{Systematic Mapping Studies in Software
  Engineering}}.
\newblock \bibinfo{journal}{12Th International Conference on Evaluation and
  Assessment in Software Engineering} \bibinfo{volume}{17},
  \bibinfo{pages}{10}.
\newblock \DOIprefix\doi{10.1142/S0218194007003112}.
%Type = Article
\bibitem[{Petersen et~al.(2015)Petersen, Vakkalanka and
  Kuzniarz}]{Petersen2015}
\bibinfo{author}{Petersen, K.}, \bibinfo{author}{Vakkalanka, S.},
  \bibinfo{author}{Kuzniarz, L.}, \bibinfo{year}{2015}.
\newblock \bibinfo{title}{{Guidelines for conducting systematic mapping studies
  in software engineering: An update}}.
\newblock \bibinfo{journal}{Information and Software Technology}
  \bibinfo{volume}{64}, \bibinfo{pages}{1--18}.
\newblock \URLprefix \url{http://dx.doi.org/10.1016/j.infsof.2015.03.007},
  \DOIprefix\doi{10.1016/j.infsof.2015.03.007}.
%Type = Inproceedings
\bibitem[{Phan and Nguyen(2017a)}]{Phan2017a}
\bibinfo{author}{Phan, A.V.}, \bibinfo{author}{Nguyen, M.L.},
  \bibinfo{year}{2017}a.
\newblock \bibinfo{title}{{Convolutional Neural Networks on Assembly Code for
  Predicting Software Defects}}, in: \bibinfo{booktitle}{21st Asia Pacific
  Symposium on Intelligent and Evolutionary Systems}.
%Type = Inproceedings
\bibitem[{Phan and Nguyen(2017b)}]{Phan2017}
\bibinfo{author}{Phan, A.V.}, \bibinfo{author}{Nguyen, M.L.},
  \bibinfo{year}{2017}b.
\newblock \bibinfo{title}{{Convolutional Neural Networks over Control Flow
  Graphs for Software Defect Prediction}}, in:
  \bibinfo{booktitle}{International Conference on Tools with Artificial
  Intelligence Convolutional}.
\newblock \DOIprefix\doi{10.1109/ICTAI.2017.00019}.
%Type = Inproceedings
\bibitem[{{Phuong Ha} et~al.(2019){Phuong Ha}, {Hung Tran}, {My Hanh} and
  {Thanh Binh}}]{PhuongHa2019}
\bibinfo{author}{{Phuong Ha}, T.M.}, \bibinfo{author}{{Hung Tran}, D.},
  \bibinfo{author}{{My Hanh}, L.T.}, \bibinfo{author}{{Thanh Binh}, N.},
  \bibinfo{year}{2019}.
\newblock \bibinfo{title}{{Experimental study on software fault prediction
  using machine learning model}}, in: \bibinfo{booktitle}{Proceedings of 2019
  11th International Conference on Knowledge and Systems Engineering, KSE
  2019}, \bibinfo{publisher}{IEEE}. pp. \bibinfo{pages}{1--5}.
\newblock \DOIprefix\doi{10.1109/KSE.2019.8919429}.
%Type = Inproceedings
\bibitem[{Polisetty et~al.(2019)Polisetty, Miranskyy and
  Başar}]{Polisetty2019}
\bibinfo{author}{Polisetty, S.}, \bibinfo{author}{Miranskyy, A.},
  \bibinfo{author}{Başar, A.}, \bibinfo{year}{2019}.
\newblock \bibinfo{title}{{On Usefulness of the Deep-Learning-Based Bug
  Localization Models to Practitioners}}, in: \bibinfo{booktitle}{Fifteenth
  International Conference on Predictive Models and Data Analytics in Software
  Engineering}, pp. \bibinfo{pages}{16--25}.
\newblock \DOIprefix\doi{10.1145/3345629.3345632},
  \href{http://arxiv.org/abs/1907.08588}{\tt arXiv:1907.08588}.
%Type = Inproceedings
\bibitem[{{Praman Deep Singh}(2017)}]{Ghhs2017}
\bibinfo{author}{{Praman Deep Singh}, A.C.}, \bibinfo{year}{2017}.
\newblock \bibinfo{title}{{Software Defect Prediction Analysis Using Machine
  Learning Algorithms}}, in: \bibinfo{booktitle}{7th International Conference
  on Cloud Computing, Data Science {\&} Engineering-Confluence}, pp.
  \bibinfo{pages}{775--781}.
\newblock \DOIprefix\doi{10.1109/CONFLUENCE.2017.7943255}.
%Type = Inproceedings
\bibitem[{Quin et~al.(2019)Quin, Weyns, Bamelis, Sarpreet and
  Michiels}]{Quin2019}
\bibinfo{author}{Quin, F.}, \bibinfo{author}{Weyns, D.},
  \bibinfo{author}{Bamelis, T.}, \bibinfo{author}{Sarpreet, S.B.},
  \bibinfo{author}{Michiels, S.}, \bibinfo{year}{2019}.
\newblock \bibinfo{title}{{Efficient analysis of large adaptation spaces in
  self-adaptive systems using machine learning}}, in: \bibinfo{booktitle}{ICSE
  Workshop on Software Engineering for Adaptive and Self-Managing Systems}, pp.
  \bibinfo{pages}{1--12}.
\newblock \DOIprefix\doi{10.1109/SEAMS.2019.00011}.
%Type = Inproceedings
\bibitem[{Rahman et~al.(2019)Rahman, Haque, Tawhid and Siddik}]{Rahman2019}
\bibinfo{author}{Rahman, M.A.}, \bibinfo{author}{Haque, M.A.},
  \bibinfo{author}{Tawhid, M.N.A.}, \bibinfo{author}{Siddik, M.S.},
  \bibinfo{year}{2019}.
\newblock \bibinfo{title}{{Classifying non-functional requirements using RNN
  variants for quality software development}}, in: \bibinfo{booktitle}{3rd ACM
  SIGSOFT International Workshop on Machine Learning Techniques for Software
  Quality Evaluation}, pp. \bibinfo{pages}{25--30}.
\newblock \DOIprefix\doi{10.1145/3340482.3342745}.
%Type = Article
\bibitem[{Rajbahadur et~al.(2019)Rajbahadur, Wang, Kamei and
  Hassan}]{Rajbahadur2019}
\bibinfo{author}{Rajbahadur, G.K.}, \bibinfo{author}{Wang, S.},
  \bibinfo{author}{Kamei, Y.}, \bibinfo{author}{Hassan, A.E.},
  \bibinfo{year}{2019}.
\newblock \bibinfo{title}{{Impact of Discretization Noise of the Dependent
  variable on Machine Learning Classifiers in Software Engineering}}.
\newblock \bibinfo{journal}{IEEE Transactions on Software Engineering}
  \DOIprefix\doi{10.1109/TSE.2019.2924371}.
%Type = Article
\bibitem[{Ramasamy and Lakshmanan(2017)}]{Ramasamy2017}
\bibinfo{author}{Ramasamy, S.}, \bibinfo{author}{Lakshmanan, I.},
  \bibinfo{year}{2017}.
\newblock \bibinfo{title}{{Machine Learning Approach for Software Reliability
  Growth Modeling with Infinite Testing Effort Function}}.
\newblock \bibinfo{journal}{Mathematical Problems in Engineering}
  \bibinfo{volume}{2017}.
%Type = Article
\bibitem[{Ramirez and Cheng(2011)}]{Ramirez2011}
\bibinfo{author}{Ramirez, A.J.}, \bibinfo{author}{Cheng, B.H.},
  \bibinfo{year}{2011}.
\newblock \bibinfo{title}{{Automatic derivation of utility functions for
  monitoring software requirements}}.
\newblock \bibinfo{journal}{Lecture Notes in Computer Science (including
  subseries Lecture Notes in Artificial Intelligence and Lecture Notes in
  Bioinformatics)} \bibinfo{volume}{6981 LNCS}, \bibinfo{pages}{501--516}.
\newblock \DOIprefix\doi{10.1007/978-3-642-24485-8_37}.
%Type = Inproceedings
\bibitem[{Rana and Staron(2015)}]{Rana}
\bibinfo{author}{Rana, R.}, \bibinfo{author}{Staron, M.}, \bibinfo{year}{2015}.
\newblock \bibinfo{title}{{Machine Learning Approach for Quality Assessment and
  Prediction in Large Software Organizations}}, in: \bibinfo{booktitle}{6th
  IEEE International Conference on Software Engineering and Service Science},
  \bibinfo{publisher}{IEEE}.
%Type = Inproceedings
\bibitem[{Rana et~al.(2003)Rana, Staron, Hansson, Nilsson and
  Meding}]{Rana2003}
\bibinfo{author}{Rana, R.}, \bibinfo{author}{Staron, M.},
  \bibinfo{author}{Hansson, J.}, \bibinfo{author}{Nilsson, M.},
  \bibinfo{author}{Meding, W.}, \bibinfo{year}{2003}.
\newblock \bibinfo{title}{{A Framework for Adoption of Machine Learning in
  Industry for Software Defect Prediction}}, in: \bibinfo{booktitle}{9th
  International Conference on Software Engineering and Applications}.
%Type = Inproceedings
\bibitem[{Reddivari and Raman(2019)}]{Reddivari2019}
\bibinfo{author}{Reddivari, S.}, \bibinfo{author}{Raman, J.},
  \bibinfo{year}{2019}.
\newblock \bibinfo{title}{{Software quality prediction: An investigation based
  on machine learning}}, in: \bibinfo{booktitle}{Proceedings - 2019 IEEE 20th
  International Conference on Information Reuse and Integration for Data
  Science, IRI 2019}, \bibinfo{publisher}{IEEE}. pp. \bibinfo{pages}{115--122}.
\newblock \DOIprefix\doi{10.1109/IRI.2019.00030}.
%Type = Inproceedings
\bibitem[{Rezende et~al.(2017)Rezende, Ruppert, Carvalho, Ramos and
  Geus}]{Rezende2017}
\bibinfo{author}{Rezende, E.}, \bibinfo{author}{Ruppert, G.},
  \bibinfo{author}{Carvalho, T.}, \bibinfo{author}{Ramos, F.},
  \bibinfo{author}{Geus, P.D.}, \bibinfo{year}{2017}.
\newblock \bibinfo{title}{{Malicious Software Classification using Transfer
  Learning of ResNet-50 Deep Neural Network}}, in: \bibinfo{booktitle}{16th
  IEEE International Conference on Machine Learning and Applications
  Malicious}.
\newblock \DOIprefix\doi{10.1109/ICMLA.2017.00-19}.
%Type = Inproceedings
\bibitem[{Robbes and Janes(2019)}]{Robbes2019}
\bibinfo{author}{Robbes, R.}, \bibinfo{author}{Janes, A.},
  \bibinfo{year}{2019}.
\newblock \bibinfo{title}{{Leveraging small software engineering data sets with
  pre-trained neural networks}}, in: \bibinfo{booktitle}{Proceedings - 2019
  IEEE/ACM 41st International Conference on Software Engineering: New Ideas and
  Emerging Results, ICSE-NIER 2019}, \bibinfo{publisher}{IEEE}. pp.
  \bibinfo{pages}{29--32}.
\newblock \DOIprefix\doi{10.1109/ICSE-NIER.2019.00016}.
%Type = Inproceedings
\bibitem[{Rooijen et~al.(2017)Rooijen, B, Platenius, Geierhos, Hamann and
  Engels}]{Rooijen2017}
\bibinfo{author}{Rooijen, L.V.}, \bibinfo{author}{B, F.S.},
  \bibinfo{author}{Platenius, M.C.}, \bibinfo{author}{Geierhos, M.},
  \bibinfo{author}{Hamann, H.}, \bibinfo{author}{Engels, G.},
  \bibinfo{year}{2017}.
\newblock \bibinfo{title}{{From User Demand to Software Service : Using Machine
  Learning to Automate the Requirements Specification Process}}, in:
  \bibinfo{booktitle}{25th International Requirements Engineering Conference
  Workshops}, pp. \bibinfo{pages}{379--385}.
\newblock \DOIprefix\doi{10.1109/REW.2017.26}.
%Type = Inproceedings
\bibitem[{Rosenfeld et~al.(2017)Rosenfeld, Kardashov and Zang}]{Rosenfeld2017}
\bibinfo{author}{Rosenfeld, A.}, \bibinfo{author}{Kardashov, O.},
  \bibinfo{author}{Zang, O.}, \bibinfo{year}{2017}.
\newblock \bibinfo{title}{{Automation of Android Applications Testing Using
  Machine Learning Activities Classification}}, in: \bibinfo{booktitle}{5th
  International Conference on Mobile Software Engineering and Systems
  Automation}, pp. \bibinfo{pages}{122--132}.
\newblock \URLprefix \url{http://arxiv.org/abs/1709.00928},
  \href{http://arxiv.org/abs/1709.00928}{\tt arXiv:1709.00928}.
%Type = Inproceedings
\bibitem[{Rughetti et~al.(2012)Rughetti, Sanzo, Ciciani and
  Quaglia}]{Rughetti2012}
\bibinfo{author}{Rughetti, D.}, \bibinfo{author}{Sanzo, P.D.},
  \bibinfo{author}{Ciciani, B.}, \bibinfo{author}{Quaglia, F.},
  \bibinfo{year}{2012}.
\newblock \bibinfo{title}{{Machine Learning-based Self-adjusting Concurrency in
  Software Transactional Memory Systems}}, in: \bibinfo{booktitle}{20th
  International Symposium on Modeling, Analysis and Simulation of Computer and
  Telecommunication Systems}, \bibinfo{publisher}{IEEE}. pp.
  \bibinfo{pages}{278--285}.
\newblock \DOIprefix\doi{10.1109/MASCOTS.2012.40}.
%Type = Inproceedings
\bibitem[{{S. Delphine Immaculate, M. Farida Begam}(2019)}]{Dhanda2019}
\bibinfo{author}{{S. Delphine Immaculate, M. Farida Begam}, M.F.},
  \bibinfo{year}{2019}.
\newblock \bibinfo{title}{{Software Bug Prediction Using Supervised Machine
  Learning Algorithms}}, in: \bibinfo{booktitle}{2019 International Conference
  on Data Science and Communication (IconDSC)}, \bibinfo{publisher}{IEEE}. pp.
  \bibinfo{pages}{1--7}.
\newblock \DOIprefix\doi{10.4018/978-1-5225-7955-7.ch009}.
%Type = Inproceedings
\bibitem[{Sahin et~al.(2019)Sahin, Karpat and Tosun}]{Sahin2019}
\bibinfo{author}{Sahin, S.E.}, \bibinfo{author}{Karpat, K.},
  \bibinfo{author}{Tosun, A.}, \bibinfo{year}{2019}.
\newblock \bibinfo{title}{{Predicting Popularity of Open Source Projects Using
  Recurrent Neural Networks}}, in: \bibinfo{booktitle}{IFIP International
  Conference on Open Source Systems}, \bibinfo{publisher}{Springer
  International Publishing}. pp. \bibinfo{pages}{80--90}.
\newblock \URLprefix \url{http://dx.doi.org/10.1007/978-3-030-20883-7{\_}8},
  \DOIprefix\doi{10.1007/978-3-030-20883-7_8}.
%Type = Inproceedings
\bibitem[{Sajnani(2012)}]{Sajnani2012}
\bibinfo{author}{Sajnani, H.}, \bibinfo{year}{2012}.
\newblock \bibinfo{title}{{Automatic Software Architecture Recovery : A Machine
  Learning Approach}}, in: \bibinfo{booktitle}{20th IEEE International
  Conference on Program Comprehension}, \bibinfo{publisher}{IEEE}. pp.
  \bibinfo{pages}{265--268}.
%Type = Inproceedings
\bibitem[{Sangwan(2017)}]{Vlv2017}
\bibinfo{author}{Sangwan, O.P.}, \bibinfo{year}{2017}.
\newblock \bibinfo{title}{{Software Effort Estimation using Machine Learning
  Techniques}}, in: \bibinfo{booktitle}{7th International Conference on Cloud
  Computing, Data Science {\&} Engineering-Confluence}, pp.
  \bibinfo{pages}{92--98}.
%Type = Inproceedings
\bibitem[{Sankaran et~al.(2017)Sankaran, Aralikatte, Mani, Khare, Panwar and
  Gantayat}]{Sankaran2017}
\bibinfo{author}{Sankaran, A.}, \bibinfo{author}{Aralikatte, R.},
  \bibinfo{author}{Mani, S.}, \bibinfo{author}{Khare, S.},
  \bibinfo{author}{Panwar, N.}, \bibinfo{author}{Gantayat, N.},
  \bibinfo{year}{2017}.
\newblock \bibinfo{title}{{DARVIZ : Deep Abstract Representation ,
  Visualization , and Verification of Deep Learning Models}}, in:
  \bibinfo{booktitle}{International Conference on Software Engineering: New
  Ideas and Emerging Technologies Result}, pp. \bibinfo{pages}{47--50}.
\newblock \DOIprefix\doi{10.1109/ICSE-NIER.2017.13}.
%Type = Inproceedings
\bibitem[{de~Santiago et~al.(2018)de~Santiago, da~Silva and {de Andrade
  Neto}}]{DeSantiago2018}
\bibinfo{author}{de~Santiago, V.A.}, \bibinfo{author}{da~Silva, L.A.R.},
  \bibinfo{author}{{de Andrade Neto}, P.R.}, \bibinfo{year}{2018}.
\newblock \bibinfo{title}{{Testing Environmental Models supported by Machine
  Learning}}, in: \bibinfo{booktitle}{III Brazilian Symposium on Systematic and
  Automated Software Testing}, pp. \bibinfo{pages}{3--12}.
\newblock \DOIprefix\doi{10.1145/3266003.3266004}.
%Type = Inproceedings
\bibitem[{Satapathy and Rath(2017)}]{Satapathy2017}
\bibinfo{author}{Satapathy, S.M.}, \bibinfo{author}{Rath, S.K.},
  \bibinfo{year}{2017}.
\newblock \bibinfo{title}{{Empirical Assessment of Machine Learning Models for
  Effort Estimation of Web-based Applications}}, in: \bibinfo{booktitle}{10th
  Innovations in Software Engineering Conference}, pp. \bibinfo{pages}{74--84}.
\newblock \DOIprefix\doi{10.1145/3021460.3021468}.
%Type = Inproceedings
\bibitem[{Schreck et~al.(2018)Schreck, Mallapur, Damle and James}]{Schreck2018}
\bibinfo{author}{Schreck, B.}, \bibinfo{author}{Mallapur, S.},
  \bibinfo{author}{Damle, S.}, \bibinfo{author}{James, N.J.},
  \bibinfo{year}{2018}.
\newblock \bibinfo{title}{{Augmenting Software Project Managers with
  Predictions from Machine Learning}}, in: \bibinfo{booktitle}{IEEE
  International Conference on Big Data}, \bibinfo{publisher}{IEEE}. pp.
  \bibinfo{pages}{2004--2011}.
%Type = Inproceedings
\bibitem[{Shanthi et~al.(2018)Shanthi, Mohanty and Narsimha}]{Shanthi2018}
\bibinfo{author}{Shanthi, D.}, \bibinfo{author}{Mohanty, R.K.},
  \bibinfo{author}{Narsimha, G.}, \bibinfo{year}{2018}.
\newblock \bibinfo{title}{{Application of Machine Learning Reliability Data
  Sets}}, in: \bibinfo{booktitle}{2018 Second International Conference on
  Intelligent Computing and Control Systems (ICICCS)},
  \bibinfo{publisher}{IEEE}. pp. \bibinfo{pages}{1472--1474}.
%Type = Inproceedings
\bibitem[{Sharma(2017)}]{Sharma2017}
\bibinfo{author}{Sharma, P.}, \bibinfo{year}{2017}.
\newblock \bibinfo{title}{{Systematic Literature Review on Software Effort
  Estimation Using Machine Learning Approaches}}, in:
  \bibinfo{booktitle}{International Conference on Next Generation Computing and
  Information Systems}, \bibinfo{publisher}{IEEE}.
\newblock \DOIprefix\doi{10.1109/ICNGCIS.2017.33}.
%Type = Inproceedings
\bibitem[{Sharma et~al.(2014)Sharma, Bhatia and Biswas}]{Sharma2014}
\bibinfo{author}{Sharma, R.}, \bibinfo{author}{Bhatia, J.},
  \bibinfo{author}{Biswas, K.K.}, \bibinfo{year}{2014}.
\newblock \bibinfo{title}{{Machine learning for constituency test of
  coordinating conjunctions in requirements specifications}}, in:
  \bibinfo{booktitle}{International Workshop on Realizing Artificial
  Intelligence Synergies in Software Engineering}, \bibinfo{publisher}{ACM}.
  pp. \bibinfo{pages}{25--31}.
\newblock \DOIprefix\doi{10.1145/2593801.2593806}.
%Type = Inproceedings
\bibitem[{Shen et~al.(2019)Shen, Baysal and Shafiq}]{Shen2019}
\bibinfo{author}{Shen, J.}, \bibinfo{author}{Baysal, O.},
  \bibinfo{author}{Shafiq, M.O.}, \bibinfo{year}{2019}.
\newblock \bibinfo{title}{{Evaluating the performance of machine learning
  sentiment analysis algorithms in software engineering}}, in:
  \bibinfo{booktitle}{Proceedings - IEEE 17th International Conference on
  Dependable, Autonomic and Secure Computing, IEEE 17th International
  Conference on Pervasive Intelligence and Computing, IEEE 5th International
  Conference on Cloud and Big Data Computing, 4th Cyber Scienc},
  \bibinfo{publisher}{IEEE}. pp. \bibinfo{pages}{1023--1030}.
\newblock \DOIprefix\doi{10.1109/DASC/PiCom/CBDCom/CyberSciTech.2019.00185}.
%Type = Article
\bibitem[{Shepperd et~al.(2014)Shepperd, Hall, Bowes and Hall}]{Shepperd2014}
\bibinfo{author}{Shepperd, M.}, \bibinfo{author}{Hall, T.},
  \bibinfo{author}{Bowes, D.}, \bibinfo{author}{Hall, T.},
  \bibinfo{year}{2014}.
\newblock \bibinfo{title}{{Researcher Bias : The Use of Machine Learning in
  Software Defect Prediction}}.
\newblock \bibinfo{journal}{TRANSACTIONS ON SOFTWARE ENGINEERING}
  \bibinfo{volume}{40}, \bibinfo{pages}{603--616}.
%Type = Inproceedings
\bibitem[{Shin and Goel(2005)}]{Shin2005}
\bibinfo{author}{Shin, M.}, \bibinfo{author}{Goel, A.L.}, \bibinfo{year}{2005}.
\newblock \bibinfo{title}{{Modeling Software Component Criticality Using a
  Machine Learning Approach}}, in: \bibinfo{booktitle}{International Conference
  on AI, Simulation, and Planning in High Autonomy Systems}, pp.
  \bibinfo{pages}{440--448}.
%Type = Inproceedings
\bibitem[{Shukla et~al.(2018)Shukla, Behera, Misra and Rath}]{Shukla2018}
\bibinfo{author}{Shukla, S.}, \bibinfo{author}{Behera, R.K.},
  \bibinfo{author}{Misra, S.}, \bibinfo{author}{Rath, S.K.},
  \bibinfo{year}{2018}.
\newblock \bibinfo{title}{{Software Reliability Assessment Using Deep Learning
  Technique}}, in: \bibinfo{booktitle}{International Conference on
  Computational Science and Its Applications}, \bibinfo{publisher}{Springer
  Singapore}.
\newblock \URLprefix \url{http://link.springer.com/10.1007/978-981-13-2348-5},
  \DOIprefix\doi{10.1007/978-981-13-2348-5}.
%Type = Inproceedings
\bibitem[{Sidhu et~al.(2018)Sidhu, Singh and Sharma}]{Sidhu2018}
\bibinfo{author}{Sidhu, B.K.}, \bibinfo{author}{Singh, K.},
  \bibinfo{author}{Sharma, N.}, \bibinfo{year}{2018}.
\newblock \bibinfo{title}{{A Catalogue of Model Smells and Refactoring
  Operations for Object - Oriented Software}}, in: \bibinfo{booktitle}{2nd
  International Conference on Inventive Communication and Computational
  Technologies}, \bibinfo{publisher}{IEEE}. pp. \bibinfo{pages}{313--319}.
%Type = Inproceedings
\bibitem[{Silva et~al.(2010)Silva, Jino and Abreu}]{Silva2010}
\bibinfo{author}{Silva, D.G.}, \bibinfo{author}{Jino, M.},
  \bibinfo{author}{Abreu, B.T.D.}, \bibinfo{year}{2010}.
\newblock \bibinfo{title}{{Machine learning methods and asymmetric cost
  function to estimate execution effort of software testing}}, in:
  \bibinfo{booktitle}{Third International Conference on Software Testing,
  Verification and Validation}, \bibinfo{publisher}{IEEE}.
\newblock \DOIprefix\doi{10.1109/ICST.2010.46}.
%Type = Inproceedings
\bibitem[{Simpson(2016)}]{Simpson2016}
\bibinfo{author}{Simpson, M.C.}, \bibinfo{year}{2016}.
\newblock \bibinfo{title}{{Automatic Algorithm Selection in Computational
  Software Using Machine Learning}}, in: \bibinfo{booktitle}{15th IEEE
  International Conference on Machine Learning and Applications Automatic},
  \bibinfo{publisher}{IEEE}.
\newblock \DOIprefix\doi{10.1109/ICMLA.2016.47}.
%Type = Inproceedings
\bibitem[{Singh et~al.(2018)Singh, Anu, Walia and Goswami}]{Singh2018}
\bibinfo{author}{Singh, M.}, \bibinfo{author}{Anu, V.}, \bibinfo{author}{Walia,
  G.S.}, \bibinfo{author}{Goswami, A.}, \bibinfo{year}{2018}.
\newblock \bibinfo{title}{{Validating Requirements Reviews by Introducing
  Fault-Type Level Granularity}}, in: \bibinfo{booktitle}{SIGSOFT Innovations
  in Software Engineering Conference}, pp. \bibinfo{pages}{1--11}.
\newblock \DOIprefix\doi{10.1145/3172871.3172880}.
%Type = Inproceedings
\bibitem[{Singh and Malhotra(2017)}]{Singh2017}
\bibinfo{author}{Singh, P.}, \bibinfo{author}{Malhotra, R.},
  \bibinfo{year}{2017}.
\newblock \bibinfo{title}{{Assessment of Machine Learning Algorithms for
  Determining Defective Classes in an Object-Oriented Software}}, in:
  \bibinfo{booktitle}{6th International Conference on Reliability, Infocom
  Technologies and Optimization (Trends and Future Directions)}.
%Type = Article
\bibitem[{Singh et~al.(2008)Singh, Kaur and Malhotra}]{Singh2008}
\bibinfo{author}{Singh, Y.}, \bibinfo{author}{Kaur, A.},
  \bibinfo{author}{Malhotra, R.}, \bibinfo{year}{2008}.
\newblock \bibinfo{title}{{Predicting software fault proneness model using
  neural network}}.
\newblock \bibinfo{journal}{Lecture Notes in Computer Science (including
  subseries Lecture Notes in Artificial Intelligence and Lecture Notes in
  Bioinformatics)} \bibinfo{volume}{5089 LNCS}, \bibinfo{pages}{204--214}.
\newblock \DOIprefix\doi{10.1007/978-3-540-69566-0_18}.
%Type = Book
\bibitem[{Society et~al.(2014)Society, Bourque and Fairley}]{Swebok}
\bibinfo{author}{Society, I.C.}, \bibinfo{author}{Bourque, P.},
  \bibinfo{author}{Fairley, R.E.}, \bibinfo{year}{2014}.
\newblock \bibinfo{title}{{Guide to the Software Engineering Body of Knowledge
  (SWEBOK(R)): Version 3.0}}.
\newblock \bibinfo{edition}{3rd} ed., \bibinfo{publisher}{IEEE Computer Society
  Press}, \bibinfo{address}{Los Alamitos, CA, USA}.
%Type = Article
\bibitem[{Song et~al.(2016)Song, Zhu, Wang, Sun, Jiang and Xue}]{Song2016}
\bibinfo{author}{Song, Q.}, \bibinfo{author}{Zhu, X.}, \bibinfo{author}{Wang,
  G.}, \bibinfo{author}{Sun, H.}, \bibinfo{author}{Jiang, H.},
  \bibinfo{author}{Xue, C.}, \bibinfo{year}{2016}.
\newblock \bibinfo{title}{{The Journal of Systems and Software A machine
  learning based software process model recommendation method}}.
\newblock \bibinfo{journal}{Journal of Systems and Software}
  \bibinfo{volume}{118}, \bibinfo{pages}{85--100}.
\newblock \DOIprefix\doi{10.1016/j.jss.2016.05.002}.
%Type = Inproceedings
\bibitem[{Souza et~al.(2003)Souza, Pozo and Vergilio}]{Souza2003}
\bibinfo{author}{Souza, G.A.D.}, \bibinfo{author}{Pozo, A.R.T.},
  \bibinfo{author}{Vergilio, S.R.}, \bibinfo{year}{2003}.
\newblock \bibinfo{title}{{Exploring Machine Learning Techniques for Software
  Size Estimation}}, in: \bibinfo{booktitle}{XXIII International Conference of
  the Chilean Computer Science Society}.
%Type = Article
\bibitem[{Srinivasan and Fisher(1995)}]{Srinivasan1995}
\bibinfo{author}{Srinivasan, K.}, \bibinfo{author}{Fisher, D.},
  \bibinfo{year}{1995}.
\newblock \bibinfo{title}{{Machine Learning Approaches to Estimating Software
  Development Effort}}.
\newblock \bibinfo{journal}{TRANSACTIONS ON SOFTWARE ENGINEERING}
  \bibinfo{volume}{21}.
%Type = Article
\bibitem[{Sudharson and Prabha(2019)}]{Sudharson2019}
\bibinfo{author}{Sudharson, D.}, \bibinfo{author}{Prabha, D.},
  \bibinfo{year}{2019}.
\newblock \bibinfo{title}{{A novel machine learning approach for software
  reliability growth modelling with pareto distribution function}}.
\newblock \bibinfo{journal}{Soft Computing} \bibinfo{volume}{23},
  \bibinfo{pages}{8379--8387}.
\newblock \URLprefix \url{https://doi.org/10.1007/s00500-019-04047-7},
  \DOIprefix\doi{10.1007/s00500-019-04047-7}.
%Type = Inproceedings
\bibitem[{Sultanov and Hayes(2013)}]{Sultanov2013}
\bibinfo{author}{Sultanov, H.}, \bibinfo{author}{Hayes, J.H.},
  \bibinfo{year}{2013}.
\newblock \bibinfo{title}{{Application of reinforcement learning to
  requirements engineering: Requirements tracing}}, in:
  \bibinfo{booktitle}{2013 21st IEEE International Requirements Engineering
  Conference, RE 2013 - Proceedings}, \bibinfo{publisher}{IEEE}. pp.
  \bibinfo{pages}{52--61}.
\newblock \DOIprefix\doi{10.1109/RE.2013.6636705}.
%Type = Inproceedings
\bibitem[{Sun and Wang(2018)}]{Sun2018}
\bibinfo{author}{Sun, Y.}, \bibinfo{author}{Wang, Y.M.}, \bibinfo{year}{2018}.
\newblock \bibinfo{title}{{Utilizing Deep Architecture Networks of VAE in
  Software Fault Prediction}}, in: \bibinfo{booktitle}{Intl Conf on Parallel
  {\&} Distributed Processing with Applications, Ubiquitous Computing {\&}
  Communications, Big Data {\&} Cloud Computing, Social Computing {\&}
  Networking, Sustainable Computing {\&} Communications}, pp.
  \bibinfo{pages}{870--877}.
\newblock \DOIprefix\doi{10.1109/BDCloud.2018.00129}.
%Type = Inproceedings
\bibitem[{Tamura(2016)}]{Tamura2016}
\bibinfo{author}{Tamura, Y.}, \bibinfo{year}{2016}.
\newblock \bibinfo{title}{{Software Reliability Model Selection Based on Deep
  Learning}}, in: \bibinfo{booktitle}{2016 International Conference on
  Industrial Engineering, Management Science and Application (ICIMSA)},
  \bibinfo{publisher}{IEEE}. pp. \bibinfo{pages}{1--5}.
\newblock \DOIprefix\doi{10.1109/ICIMSA.2016.7504034}.
%Type = Inproceedings
\bibitem[{Tanaka et~al.(2019)Tanaka, Monden and Yucel}]{Tanaka2019}
\bibinfo{author}{Tanaka, K.}, \bibinfo{author}{Monden, A.},
  \bibinfo{author}{Yucel, Z.}, \bibinfo{year}{2019}.
\newblock \bibinfo{title}{{Prediction of Software Defects Using Automated
  Machine Learning}}, in: \bibinfo{booktitle}{Proceedings - 20th IEEE/ACIS
  International Conference on Software Engineering, Artificial Intelligence,
  Networking and Parallel/Distributed Computing, SNPD 2019}, pp.
  \bibinfo{pages}{490--494}.
\newblock \DOIprefix\doi{10.1109/SNPD.2019.8935839}.
%Type = Inproceedings
\bibitem[{Thaller et~al.(2019)Thaller, Linsbauer and Egyed}]{Thaller2019}
\bibinfo{author}{Thaller, H.}, \bibinfo{author}{Linsbauer, L.},
  \bibinfo{author}{Egyed, A.}, \bibinfo{year}{2019}.
\newblock \bibinfo{title}{{Feature Maps : A Comprehensible Software
  Representation for Design Pattern Detection}}, in: \bibinfo{booktitle}{26th
  International Conference on Software Analysis, Evolution and Reengineering},
  \bibinfo{publisher}{IEEE}. pp. \bibinfo{pages}{207--217}.
%Type = Inproceedings
\bibitem[{Tofighi-Shirazi et~al.(2019)Tofighi-Shirazi, Asavoae, Elbaz-Vincent
  and Le}]{Tofighi-Shirazi2019}
\bibinfo{author}{Tofighi-Shirazi, R.}, \bibinfo{author}{Asavoae, I.M.},
  \bibinfo{author}{Elbaz-Vincent, P.}, \bibinfo{author}{Le, T.H.},
  \bibinfo{year}{2019}.
\newblock \bibinfo{title}{{Defeating Opaque Predicates Statically through
  Machine Learning and Binary Analysis}}, in: \bibinfo{booktitle}{3rd Software
  Protection Workshop}, pp. \bibinfo{pages}{3--14}.
\newblock \DOIprefix\doi{10.1145/3338503.3357719},
  \href{http://arxiv.org/abs/1909.01640}{\tt arXiv:1909.01640}.
%Type = Inproceedings
\bibitem[{Tran et~al.(2019)Tran, Hanh and Binh}]{Tran2019}
\bibinfo{author}{Tran, H.D.}, \bibinfo{author}{Hanh, L.T.M.},
  \bibinfo{author}{Binh, N.T.}, \bibinfo{year}{2019}.
\newblock \bibinfo{title}{{Combining feature selection, feature learning and
  ensemble learning for software fault prediction}}, in:
  \bibinfo{booktitle}{Proceedings of 2019 11th International Conference on
  Knowledge and Systems Engineering, KSE 2019}, \bibinfo{publisher}{IEEE}. pp.
  \bibinfo{pages}{1--8}.
\newblock \DOIprefix\doi{10.1109/KSE.2019.8919292}.
%Type = Inproceedings
\bibitem[{Tufano et~al.(2018)Tufano, William, Watson, Bavota, Penta, White and
  Poshyvanyk}]{Tufano2018}
\bibinfo{author}{Tufano, M.}, \bibinfo{author}{William, C.},
  \bibinfo{author}{Watson, C.}, \bibinfo{author}{Bavota, G.},
  \bibinfo{author}{Penta, M.D.}, \bibinfo{author}{White, M.},
  \bibinfo{author}{Poshyvanyk, D.}, \bibinfo{year}{2018}.
\newblock \bibinfo{title}{{Deep Learning Similarities from Different
  Representations of Source Code}}, in: \bibinfo{booktitle}{15th International
  Conference on Mining Software Repositories}, pp. \bibinfo{pages}{542--553}.
%Type = Article
\bibitem[{Turliuc(2011)}]{Turliuc2011}
\bibinfo{author}{Turliuc, C.R.}, \bibinfo{year}{2011}.
\newblock \bibinfo{title}{{ProbPoly: A Probabilistic Inductive Logic
  Programming Framework with Application in Model Checking}}.
\newblock \bibinfo{journal}{Proceedings of the International Workshop on
  Machine Learning Technologies in Software Engineering} ,
  \bibinfo{pages}{43--50}\URLprefix
  \url{http://doi.acm.org/10.1145/2070821.2070827},
  \DOIprefix\doi{10.1145/2070821.2070827}.
%Type = Article
\bibitem[{Usman et~al.(2017)Usman, Britto, B{\"{o}}rstler and
  Mendes}]{Usman2017}
\bibinfo{author}{Usman, M.}, \bibinfo{author}{Britto, R.},
  \bibinfo{author}{B{\"{o}}rstler, J.}, \bibinfo{author}{Mendes, E.},
  \bibinfo{year}{2017}.
\newblock \bibinfo{title}{{Taxonomies in software engineering: A Systematic
  mapping study and a revised taxonomy development method}}.
\newblock \bibinfo{journal}{Information and Software Technology}
  \bibinfo{volume}{85}, \bibinfo{pages}{43--59}.
\newblock \URLprefix \url{http://dx.doi.org/10.1016/j.infsof.2017.01.006},
  \DOIprefix\doi{10.1016/j.infsof.2017.01.006}.
%Type = Inproceedings
\bibitem[{{Vasil Borozanov, Simon Hacks}(2019)}]{B2019b}
\bibinfo{author}{{Vasil Borozanov, Simon Hacks}, N.S.}, \bibinfo{year}{2019}.
\newblock \bibinfo{title}{{Using Machine Learning Techniques for Evaluating the
  Similarity of Enterprise Architecture Models}}, in:
  \bibinfo{booktitle}{International Conference on Advanced Information Systems
  Engineering}, \bibinfo{publisher}{Springer International Publishing}. pp.
  \bibinfo{pages}{563--578}.
\newblock \DOIprefix\doi{10.1007/978-3-030-21290-2}.
%Type = Article
\bibitem[{Wan et~al.(2019)Wan, Xia, Lo and Murphy}]{Wan2019}
\bibinfo{author}{Wan, Z.}, \bibinfo{author}{Xia, X.}, \bibinfo{author}{Lo, D.},
  \bibinfo{author}{Murphy, G.C.}, \bibinfo{year}{2019}.
\newblock \bibinfo{title}{{How does Machine Learning Change Software
  Development Practices?}}
\newblock \bibinfo{journal}{IEEE Transactions on Software Engineering}
  \bibinfo{volume}{PP}, \bibinfo{pages}{1--1}.
\newblock \DOIprefix\doi{10.1109/tse.2019.2937083}.
%Type = Article
\bibitem[{Wang and Zhang(2018)}]{Wang2018a}
\bibinfo{author}{Wang, J.}, \bibinfo{author}{Zhang, C.}, \bibinfo{year}{2018}.
\newblock \bibinfo{title}{{Software reliability prediction using a deep
  learning model based on the RNN encoder – decoder}}.
\newblock \bibinfo{journal}{Reliability Engineering {\&} System Safety}
  \bibinfo{volume}{170}, \bibinfo{pages}{73--82}.
\newblock \URLprefix
  \url{https://linkinghub.elsevier.com/retrieve/pii/S0951832017303538},
  \DOIprefix\doi{10.1016/j.ress.2017.10.019}.
%Type = Article
\bibitem[{Wang et~al.(2018)Wang, Liu, Nam and Tan}]{Wang2018}
\bibinfo{author}{Wang, S.}, \bibinfo{author}{Liu, T.}, \bibinfo{author}{Nam,
  J.}, \bibinfo{author}{Tan, L.}, \bibinfo{year}{2018}.
\newblock \bibinfo{title}{{Deep Semantic Feature Learning for Software Defect
  Prediction}}.
\newblock \bibinfo{journal}{Transactions on Software Engineering}
  \bibinfo{volume}{5589}, \bibinfo{pages}{1--26}.
\newblock \DOIprefix\doi{10.1109/TSE.2018.2877612}.
%Type = Article
\bibitem[{Wen et~al.(2012)Wen, Li, Lin, Hu and Huang}]{Wen2012}
\bibinfo{author}{Wen, J.}, \bibinfo{author}{Li, S.}, \bibinfo{author}{Lin, Z.},
  \bibinfo{author}{Hu, Y.}, \bibinfo{author}{Huang, C.}, \bibinfo{year}{2012}.
\newblock \bibinfo{title}{{Systematic literature review of machine learning
  based software development effort estimation models}}.
\newblock \bibinfo{journal}{Information and Software Technology}
  \bibinfo{volume}{54}, \bibinfo{pages}{41--59}.
\newblock \URLprefix
  \url{https://linkinghub.elsevier.com/retrieve/pii/S0950584911001832
  http://dx.doi.org/10.1016/j.infsof.2011.09.002},
  \DOIprefix\doi{10.1016/j.infsof.2011.09.002}.
%Type = Article
\bibitem[{Wen et~al.(2018)Wen, Wu and Cheung}]{Wen2018}
\bibinfo{author}{Wen, M.}, \bibinfo{author}{Wu, R.}, \bibinfo{author}{Cheung,
  S.c.}, \bibinfo{year}{2018}.
\newblock \bibinfo{title}{{How Well Do Change Sequences Predict Defects ?
  Sequence Learning from Software Changes}}.
\newblock \bibinfo{journal}{Transactions on Software Engineering}
  \bibinfo{volume}{5589}, \bibinfo{pages}{1--20}.
\newblock \DOIprefix\doi{10.1109/TSE.2018.2876256}.
%Type = Inproceedings
\bibitem[{White(2015)}]{White2015a}
\bibinfo{author}{White, M.}, \bibinfo{year}{2015}.
\newblock \bibinfo{title}{{Deep Representations for Software Engineering}}, in:
  \bibinfo{booktitle}{37th IEEE International Conference on Software
  Engineering}, \bibinfo{publisher}{IEEE}. pp. \bibinfo{pages}{781--783}.
\newblock \DOIprefix\doi{10.1109/ICSE.2015.248}.
%Type = Inproceedings
\bibitem[{White et~al.(2016)White, Tufano, Vendome and Poshyvanyk}]{White}
\bibinfo{author}{White, M.}, \bibinfo{author}{Tufano, M.},
  \bibinfo{author}{Vendome, C.}, \bibinfo{author}{Poshyvanyk, D.},
  \bibinfo{year}{2016}.
\newblock \bibinfo{title}{{Deep Learning Code Fragments for Code Clone
  Detection}}, in: \bibinfo{booktitle}{31st IEEE/ACM International Conference
  on Automated Software Engineering}, pp. \bibinfo{pages}{87--98}.
%Type = Inproceedings
\bibitem[{White et~al.(2015)White, Vendome, Linares-v and
  Poshyvanyk}]{White2015}
\bibinfo{author}{White, M.}, \bibinfo{author}{Vendome, C.},
  \bibinfo{author}{Linares-v, M.}, \bibinfo{author}{Poshyvanyk, D.},
  \bibinfo{year}{2015}.
\newblock \bibinfo{title}{{Toward Deep Learning Software Repositories}}, in:
  \bibinfo{booktitle}{12th Working Conference on Mining Software Repositories
  Toward}.
\newblock \DOIprefix\doi{10.1109/MSR.2015.38}.
%Type = Inproceedings
\bibitem[{Wieloch et~al.(2013)Wieloch, Amornborvornwong and
  Cleland-huang}]{Wieloch2013}
\bibinfo{author}{Wieloch, M.}, \bibinfo{author}{Amornborvornwong, S.},
  \bibinfo{author}{Cleland-huang, J.}, \bibinfo{year}{2013}.
\newblock \bibinfo{title}{{Trace-by-Classification : A Machine Learning
  Approach to Generate Trace Links for Frequently Occurring Software
  Artifacts}}, in: \bibinfo{booktitle}{7th International Workshop on
  Traceability in Emerging Forms of Software Engineering},
  \bibinfo{publisher}{IEEE}. pp. \bibinfo{pages}{110--114}.
%Type = Article
\bibitem[{Wieringa et~al.(2006)Wieringa, Maiden, Mead and
  Rolland}]{Wieringa2006}
\bibinfo{author}{Wieringa, R.}, \bibinfo{author}{Maiden, N.},
  \bibinfo{author}{Mead, N.}, \bibinfo{author}{Rolland, C.},
  \bibinfo{year}{2006}.
\newblock \bibinfo{title}{{Requirements engineering paper classification and
  evaluation criteria: a proposal and a discussion}}.
\newblock \bibinfo{journal}{Requirements Engineering} \bibinfo{volume}{11},
  \bibinfo{pages}{102--107}.
\newblock \DOIprefix\doi{10.1007/s00766-005-0021-6}.
%Type = Inproceedings
\bibitem[{Wohlin(2014)}]{Wohlin2014a}
\bibinfo{author}{Wohlin, C.}, \bibinfo{year}{2014}.
\newblock \bibinfo{title}{{Guidelines for Snowballing in Systematic Literature
  Studies and a Replication in Software Engineering}}, in:
  \bibinfo{booktitle}{18th international conference on evaluation and
  assessment in software engineering}.
%Type = Article
\bibitem[{Wright and Ziegler(2019)}]{Wright2019}
\bibinfo{author}{Wright, I.}, \bibinfo{author}{Ziegler, A.},
  \bibinfo{year}{2019}.
\newblock \bibinfo{title}{{The standard coder: A machine learning approach to
  measuring the effort required to produce source code change}}.
\newblock \bibinfo{journal}{Proceedings - 2019 IEEE/ACM 7th International
  Workshop on Realizing Artificial Intelligence Synergies in Software
  Engineering, RAISE 2019} ,
  \bibinfo{pages}{1--7}\DOIprefix\doi{10.1109/RAISE.2019.00009},
  \href{http://arxiv.org/abs/1903.02436}{\tt arXiv:1903.02436}.
%Type = Inproceedings
\bibitem[{WU and ZHOU(2017)}]{WU2017}
\bibinfo{author}{WU, X.}, \bibinfo{author}{ZHOU, Z.}, \bibinfo{year}{2017}.
\newblock \bibinfo{title}{{Model reuse with domain knowledge}}, in:
  \bibinfo{booktitle}{Proceedings of the 12th ACM/IEEE International Symposium
  on Empirical Software Engineering and Measurement}, pp.
  \bibinfo{pages}{1483--1492}.
\newblock \DOIprefix\doi{10.1360/n112017-00106}.
%Type = Inproceedings
\bibitem[{Xie(2013)}]{Xie2013b}
\bibinfo{author}{Xie, T.}, \bibinfo{year}{2013}.
\newblock \bibinfo{title}{{The Synergy of Human and Artificial Intelligence in
  Software Engineering}}, in: \bibinfo{booktitle}{2013 2nd International
  Workshop on Realizing Artificial Intelligence Synergies in Software
  Engineering (RAISE)}, \bibinfo{publisher}{IEEE}. pp. \bibinfo{pages}{4--6}.
\newblock \DOIprefix\doi{10.1109/RAISE.2013.6615197}.
%Type = Inproceedings
\bibitem[{Xie(2018)}]{B2018}
\bibinfo{author}{Xie, T.}, \bibinfo{year}{2018}.
\newblock \bibinfo{title}{{Dependable Software Engineering. Theories, Tools,
  and Applications}}, in: \bibinfo{booktitle}{International Symposium on
  Dependable Software Engineering: Theories, Tools, and Applications},
  \bibinfo{publisher}{Springer International Publishing}. pp.
  \bibinfo{pages}{3--7}.
\newblock \URLprefix \url{http://link.springer.com/10.1007/978-3-319-99933-3},
  \DOIprefix\doi{10.1007/978-3-319-99933-3}.
%Type = Inproceedings
\bibitem[{Xu et~al.(2004)Xu, Ren, Qin and Craciun}]{Xu2004}
\bibinfo{author}{Xu, Z.}, \bibinfo{author}{Ren, K.}, \bibinfo{author}{Qin, S.},
  \bibinfo{author}{Craciun, F.}, \bibinfo{year}{2004}.
\newblock \bibinfo{title}{{CDGDroid: Android Malware Detection Based on Deep
  Learning Using CFG and DFG}}, in: \bibinfo{booktitle}{International
  Conference on Formal Engineering Methods}, \bibinfo{publisher}{Springer
  International Publishing}. pp. \bibinfo{pages}{177--193}.
\newblock \URLprefix \url{http://link.springer.com/10.1007/b102837},
  \DOIprefix\doi{10.1007/b102837}.
%Type = Inproceedings
\bibitem[{Yan and Lu(2017)}]{Yan2017}
\bibinfo{author}{Yan, G.}, \bibinfo{author}{Lu, J.}, \bibinfo{year}{2017}.
\newblock \bibinfo{title}{{ExploitMeter : Combining Fuzzing with Machine
  Learning for Automated Evaluation of Software Exploitability}}, in:
  \bibinfo{booktitle}{2017 IEEE Symposium on Privacy-Aware Computing}.
\newblock \DOIprefix\doi{10.1109/PAC.2017.10}.
%Type = Article
\bibitem[{Yan and Guo(2016)}]{Yan2016}
\bibinfo{author}{Yan, Y.}, \bibinfo{author}{Guo, P.}, \bibinfo{year}{2016}.
\newblock \bibinfo{title}{{A Practice Guide of Software Aging Prediction in a
  Web Server Based on Machine Learning}}.
\newblock \bibinfo{journal}{SECURITY SCHEMES AND SOLUTIONS} ,
  \bibinfo{pages}{225--235}.
%Type = Inproceedings
\bibitem[{Yang et~al.(2012)Yang, Hotta, Higo, Igaki and Kusumoto}]{Yang2012}
\bibinfo{author}{Yang, J.}, \bibinfo{author}{Hotta, K.}, \bibinfo{author}{Higo,
  Y.}, \bibinfo{author}{Igaki, H.}, \bibinfo{author}{Kusumoto, S.},
  \bibinfo{year}{2012}.
\newblock \bibinfo{title}{{Filtering Clones for Individual User Based on
  Machine Learning Analysis}}, in: \bibinfo{booktitle}{6th International
  Workshop on Software Clones}, pp. \bibinfo{pages}{76--77}.
%Type = Inproceedings
\bibitem[{Yang et~al.(2019)Yang, Yang, Fang, Yu, Rui and Ma}]{Yang2019}
\bibinfo{author}{Yang, S.}, \bibinfo{author}{Yang, S.}, \bibinfo{author}{Fang,
  Z.}, \bibinfo{author}{Yu, X.}, \bibinfo{author}{Rui, L.},
  \bibinfo{author}{Ma, Y.}, \bibinfo{year}{2019}.
\newblock \bibinfo{title}{{Fault Prediction for Software System in Industrial
  Internet: A Deep Learning Algorithm via Effective Dimension Reduction}}, in:
  \bibinfo{booktitle}{International Conference on Cyber-Living, Cyber-Syndrome
  and Cyber-Health}, \bibinfo{publisher}{Springer Singapore}. pp.
  \bibinfo{pages}{572--580}.
\newblock \URLprefix \url{http://dx.doi.org/10.1007/978-981-15-1922-2{\_}40},
  \DOIprefix\doi{10.1007/978-981-15-1922-2_40}.
%Type = Inproceedings
\bibitem[{Zhang(2006)}]{Zhang2006}
\bibinfo{author}{Zhang, D.}, \bibinfo{year}{2006}.
\newblock \bibinfo{title}{{Machine Learning in Value-Based Software Test Data
  Generation}}, in: \bibinfo{booktitle}{18th IEEE International Conference on
  Tools with Artificial Intelligence}, \bibinfo{publisher}{IEEE}.
%Type = Inproceedings
\bibitem[{Zhang(2010)}]{Tsai2002}
\bibinfo{author}{Zhang, D.}, \bibinfo{year}{2010}.
\newblock \bibinfo{title}{{Machine learning and software development}}, in:
  \bibinfo{booktitle}{14th IEEE International Conference on Tools with
  Artificial Intelligence}, \bibinfo{publisher}{IEEE}. pp.
  \bibinfo{pages}{87--119}.
\newblock \DOIprefix\doi{10.1016/j.humov.2010.04.004}.
%Type = Inproceedings
\bibitem[{Zhang and Ben(2018)}]{Zhang2018a}
\bibinfo{author}{Zhang, X.}, \bibinfo{author}{Ben, K.}, \bibinfo{year}{2018}.
\newblock \bibinfo{title}{{A Neural Language Model with a Modified Attention}},
  in: \bibinfo{booktitle}{9th International Conference on Software Engineering
  and Service Science}, \bibinfo{publisher}{IEEE}. pp.
  \bibinfo{pages}{232--236}.
%Type = Inproceedings
\bibitem[{Zhang et~al.(2018)Zhang, Ben and Zeng}]{Zhang2018}
\bibinfo{author}{Zhang, X.}, \bibinfo{author}{Ben, K.}, \bibinfo{author}{Zeng,
  J.}, \bibinfo{year}{2018}.
\newblock \bibinfo{title}{{Cross-Entropy : A New Metric for Software Defect
  Prediction}}, in: \bibinfo{booktitle}{IEEE International Conference on
  Software Quality, Reliability and Security}, \bibinfo{publisher}{IEEE}.
\newblock \DOIprefix\doi{10.1109/QRS.2018.00025}.
%Type = Inproceedings
\bibitem[{Zhao(2018)}]{Zhao2018}
\bibinfo{author}{Zhao, G.}, \bibinfo{year}{2018}.
\newblock \bibinfo{title}{{DeepSim : Deep Learning Code Functional
  Similarity}}, in: \bibinfo{booktitle}{26th ACM Joint European Software
  Engineering Conference and Symposium on the Foundations of Software
  Engineering}, pp. \bibinfo{pages}{141--151}.
%Type = Article
\bibitem[{Zhao et~al.(2019)Zhao, Member and Shang}]{Zhao2019}
\bibinfo{author}{Zhao, L.}, \bibinfo{author}{Member, S.},
  \bibinfo{author}{Shang, Z.}, \bibinfo{year}{2019}.
\newblock \bibinfo{title}{{Siamese Dense Neural Network for Software Defect
  Prediction With Small Data}}.
\newblock \bibinfo{journal}{IEEE Access} \bibinfo{volume}{7},
  \bibinfo{pages}{7663--7677}.
%Type = Article
\bibitem[{Zheng et~al.(2018)Zheng, Bai and Che}]{Zheng2018}
\bibinfo{author}{Zheng, W.}, \bibinfo{author}{Bai, Y.}, \bibinfo{author}{Che,
  H.}, \bibinfo{year}{2018}.
\newblock \bibinfo{title}{{A computer-assisted instructional method based on
  machine learning in software testing class}}.
\newblock \bibinfo{journal}{Computer Applications in Engineering Education}
  \bibinfo{volume}{26}, \bibinfo{pages}{1150--1158}.
\newblock \URLprefix \url{http://doi.wiley.com/10.1002/cae.21962},
  \DOIprefix\doi{10.1002/cae.21962}.

\end{thebibliography}

\end{document}